\documentclass[aps,prc,reprint,qshowpacs,groupedaddress,superscriptaddress]{revtex4-1} 
\usepackage[latin1]{inputenc}
\usepackage{multirow}
\usepackage{amsmath,amssymb}
\usepackage[pdftex]{graphicx}
\pdfoutput=1
\usepackage{hyperref}

\usepackage{color}
\definecolor{dred}{rgb}{0.80,0.00,0.00} 
\definecolor{dgreen}{rgb}{0.00,0.60,0.00} 
\definecolor{dblue}{rgb}{0.00,0.00,0.80} 
\definecolor{dmagenta}{rgb}{0.80,0.00,0.80}

\begin{document} 
 
\title{The electric and magnetic form factors of the proton}

\newcommand{\kph}{\affiliation{Institut f\"ur Kernphysik, Johannes  
  Gutenberg-Universit\"at Mainz, 55099 Mainz, Germany.}}  
\newcommand{\clermont}{\affiliation{LPC-Clermont, Universit\'e Blaise Pascal,  
  CNRS/IN2P3, F-63177 Aubi\`{e}re Cedex, France.}}  
\newcommand{\zagreb}{\affiliation{Department of Physics,   
  University of Zagreb, 10002 Zagreb, Croatia.}}  
\newcommand{\cea}{\affiliation{CEA DAPNIA-SPhN, C.E. Saclay,   
  91191 Gif-sur-Yvette  Cedex, France.}}  
\newcommand{\stefan}{\affiliation{Jo\v zef Stefan Institute, Ljubljana,   
  Slovenia.}}

\author{J.\,C. Bernauer}
\altaffiliation{Current address: Laboratory for Nuclear Science, MIT, Cambridge,  Massachusetts 02139, USA}
\email{bernauer@mit.edu} \kph 
\author{M.\,O.~Distler}\email{distler@kph.uni-mainz.de}\kph  
\author{J.~Friedrich}\kph  
\author{Th.~Walcher}\kph
\author{P.~Achenbach}\kph      
\author{C.~{Ayerbe Gayoso}}\kph  
\author{R.~B\"ohm}\kph  
\author{D.~Bosnar}\zagreb      
\author{L.~Debenjak}\stefan  
\author{L.~Doria}\kph  
\author{A.~Esser}\kph  
\author{H.~Fonvieille}\clermont  
\author{M.~{G\'omez Rodr\'iguez de la Paz}}\kph  
\author{J.\,M.~Friedrich}\affiliation{CERN, CH-1211 Geneva 23,
  Switzerland, on leave of absence from Physik-Department,  
          Technische Universit\"at M\"unchen, 85748 Garching, Germany.}  
\author{M.~Makek}\zagreb  
\author{H.~Merkel}\kph  
\author{D.\,G.~Middleton}\kph  
\author{U.~M\"uller}\kph  
\author{L.~Nungesser}\kph      
\author{J.~Pochodzalla}\kph    
\author{M.~Potokar}\stefan  
\author{S.~{S\'anchez Majos}}\kph  
\author{B.\,S.~Schlimme}\kph   
\author{S.~\v{S}irca}\stefan \affiliation{Department of Physics, University of Ljubljana, Slovenia.}  
\author{M.~Weinriefer}\kph 
\collaboration{A1 Collaboration}\noaffiliation

\date{\today}  
 
\begin{abstract} 
  This paper describes a precise measurement of electron scattering
  off the proton at momentum transfers of $0.003 \lesssim Q^2 \lesssim
  1$\ GeV$^2$. The average point-to-point error of the cross sections
  in this experiment is $\sim$ 0.37\%. These data are used for a
  coherent new analysis together with all world data of unpolarized
  and polarized electron scattering from the very smallest to the
  highest momentum transfers so far measured. The extracted electric
  and magnetic form factors provide new insight into their exact
  shape, deviating from the classical dipole form, and of structure on
  top of this gross shape.  The data reaching very low $Q^2$ values
  are used for a new determination of the electric and magnetic
  radii. An empirical determination of the two-photon-exchange
  correction is presented. The implications of this correction on the
  radii and the question of a directly visible signal of the pion
  cloud are addressed.
\end{abstract}

\pacs{13.40.Gp, 14.20.Dh , 25.30.Bf} 
 
\maketitle  
 \section{Introduction}
\label{sec:intro}
Form factors of extended quantum systems are traditionally considered
as a means to access the distribution of charge, magnetism and weak
charge through their Fourier transforms. However, this traditional
interpretation is only approximately valid for a light system as the
proton.  More recently it was realized through the generalization of
quark-gluon structure functions that an interpretation on the
light-cone frame is not only mandatory, but also fruitful since it
offers better insight into the quark-gluon structure of the nucleon
(for a summary see Ref.\ \cite{Vanderhaeghen:2010nd}). However, for
such an application of form factors, good knowledge from the smallest
to the highest momentum transfers is needed. Though common fits of the
world data have been presented in recent years they were always
hampered by the insufficient knowledge at the small negative
four-momentum transfers $Q^2 \lessapprox 0.5$\ GeV$^2$
\cite{fw03,Arrington03,Arrington07,venkat11,Kelly:2004hm,alberico09}. This
lack of knowledge was recently remedied by a precise measurement of
the elastic electron scattering cross sections at the Mainz Microtron
(MAMI) \cite{Bernauer:2010wm}, and this paper presents a new
determination of the electric and magnetic form factors through a new
fit of these data together with the previous world data including the
results from polarization measurements. The precision was made
possible by the MAMI cw electron beam with energies up to 1600~MeV
with narrow halo and excellent energy definition. However, in this
measurement only electron energies up to 855~MeV were used since the
higher energies were not yet available at the time of the
experiment. Using the three high-resolution spectrometers of the A1
collaboration \cite{Blomqvist:1998xn} it was possible to measure the
elastic electron-proton scattering cross section and extract the form
factors up to a negative four-momentum transfer squared of 1~GeV$^2$
with an average total point-to-point error of the cross sections of
0.37\% \cite{Bernauer:2010wm}.

There are two established ways to extract form factors from cross
sections: the classic Rosenbluth separation and a direct fit of form
factor models to the measured cross sections. The first method is the
traditional way of analyzing and presenting experimental results, it
produces form factors without any model assumption. The second method
is often used in fits for at least a decade. It has many advantages,
especially when coupled with an experiment optimized for this style of
analysis

\begin{itemize}
\item The traditional Rosenbluth separation of the electric and
  magnetic form factors uses measurements of the cross section at
  constant $Q^2$ and varied values of the polarization parameter
  $\varepsilon$. This method, however, limits unnecessarily the
  kinematical range since the range of beam energy and scattering
  angle are larger than the constant-$Q^2$ domain.  Aiming the second
  approach, the experiment is not bound by this constraint. All data
  at any kinematical point can be used for the fit, even if the
  covered range of $\varepsilon$ at a given $Q^2$ is not wide enough to
  separate the form factors.
\item It is a notorious problem to measure an absolute cross section
  to better than about $\pm 2$\%.  This was already the case in the
  old fits of Hofstadter \cite{Hofstadter55} and is highly relevant
  for all older measurements since the uncertainty of the absolute
  normalization may have been as large as $\pm 5$\% mostly due to the
  uncertainty in the determination of the electron beam current,
  target thickness and solid angle of the spectrometers. A common fit
  to the world data has to account for the uncertainty in the cross
  section normalization of the different experiments. To this end one
  may introduce the normalization as parameters in the fits, as has
  been done in previous extractions (e.g.\ Refs.\
  \cite{Arrington03,Arrington07,venkat11,alberico09}).  We can employ
  the same technique to determine the normalizations. Having groups of
  measurements with good relative normalization internally given by
  the experiment, the fit can determine the relative normalization of
  the groups with regard to each other and the global normalization
  via the extrapolation to the known form factor values at $Q^2=0$.

\item While the classic Rosenbluth approach gives a correct error
  estimate for the form factor themselves, traditionally numbers for
  the anti-correlation between $G_E$ and $G_M$ are not given. Without
  this information, uncertainties of fits to the form factors, for
  example for the extraction of the radius, cannot be calculated
  correctly. A direct, global fit to cross sections does not have this
  problem.
\item We find that the robustness of the fits is increased (see
  Sec. \ref{chptextrosen}). This is, on the one hand, caused by
  the smoothing effect of the fit and, on the other hand, the fit allows
  us to separate the form factors even if the measurements do not
  coincide in $Q^2$. Therefore, the effective density of measurements
  in $Q^2$ can be increased.
 \end{itemize}

The experiment presented here, to our knowledge, is the first specifically optimized for
this method of analysis. This approach and the precise measurements
extending down to very low momentum transfer made some distinctive
improvements over previous extractions possible as follows:
 \begin{itemize}
 \item The form-factor normalization at $Q^2=0$: The momentum region
   covered by the new measurement at MAMI, the details of which are
   described in this paper, is 0.003\ GeV$^2 \lesssim Q^2 \lesssim$1\
   GeV$^2$. The small statistical error of 0.2\% at low $Q^2$ reduces
   dramatically the uncertainty of the normalization at $Q^2=0$.

 \item Normalization of different data sets: The new data presented
   here have excellent relative normalization in a large $Q^2$ range,
   tying together the normalizations of the different data sets in the
   overlap region. Together with the first point, the absolute
   normalization is fixed with small uncertainty for a broad range in
   $Q^2$.  Of course, the normalization factors extracted by the fit
   have to be independent of the specific fit model. The analysis
   presented here has this feature.

 \item Two-photon-exchange (TPE) correction: The cross section data
   and the asymmetry measurements with polarized electrons give
   inconsistent results for the form factors. This inconsistency is
   believed to be caused by the unconsidered TPE
   contribution which is deemed to be more important for the
   Rosenbluth formula, containing the electric and magnetic form
   factors $G_E^2$ and $G_M^2$ as a sum weighted by kinematical
   factors, than for the asymmetry formulas, which give the ratio
   $G_E/G_M$. Both methods are based on the one-photon-exchange
   approximation only \cite{Guichon03}.  We fit the cross-section
   data together with the polarization data using a simple empirical
   model to parametrize the inconsistency. This empirical ansatz can
   reconcile the measurements and can easily be compared to
   theoretical calculations of TPE.
 \end{itemize}
 
 Some of the results of this paper are topical theoretically and at
 the center of recent controversies as follows: 
 \begin{itemize}
 \item Electric and magnetic radii:  A very precise determination of the electric radius of the proton
 through the Lamb shift of muonic hydrogen \cite{pohl}  has given a
 4\% smaller value than both the CODATA value \cite{Mohr08} and the result
 of the present experiment \cite{Bernauer:2010wm}. The smaller muonic
 Lamb shift value has been confirmed recently with an updated result
 \cite{Antognini:1900ns}. This discrepancy is
a so far unresolved puzzle. A similar discrepancy existed for the
magnetic radius between the determination from the hyperfine splitting
of electric hydrogen \cite{volotka} and some of the fits of the
electron scattering world data  \cite{Ron:2011rd}. However, as will be discussed in Sec.\ \ref{sec:V.C}, this discrepancy disappears with the result of this experiment  \cite{Bernauer:2010wm}. 
 \item TPE correction:  This correction has been calculated by several
   groups but with diverging results (see Ref.\ \cite{Meziane:2010xc} and references
 therein). After a controversy
 \cite{Arrington:2011kv,Bernauer:2011zz,Ron:2011rd} we present an
 experimental method in this paper showing the size of the effect
 possibly assigned to TPE and make it directly accessible to a comparison with theoretical calculations.
\item Possible signal of a pion cloud: The common idea is that the
  proton form factors are smooth and show no narrow structure of small
  $Q^2$ scale. In an outdated fit of the pre-2003 data Friedrich and
  Walcher hypothesized the existence of a bump-dip structure which
  they attributed to the signal of a pion cloud \cite{fw03}. This
  bump-dip could not be confirmed by the results of this experiment
  \cite{Bernauer:2010wm}, which, however, shows some other similar and
  more significant structure. Though the idea of such structures is
  not very welcome \cite{Meissner:2007tp} as it is considered to be a
  ``popular fantasy'' on theoretical grounds, the fits presented here
  will show further evidence for it.
  \end{itemize}
 The resolution of these controversies is partially possible through the high-precision results for the form factors presented here and  as discussed later.
 
 The paper is organized as follows. In the second section we present
 the relevant details of the experiment at MAMI followed by the description of
 the theoretical basis in the next section. The fourth section
 describes the extraction of cross sections from the measured count rates.
The analysis of the cross section and some
 peculiar aspects of the statistics needed for the determination of
 errors and for the construction of confidence bands is described in
 the fifth section. It follows a
 discussion of the results in the frame work sketched above concluding
 with the new determination of the proton form factors in Sec.\ \ref{ffffitssec} and of the radii in Sec.\ \ref{secradii}.

 Because of the size of the data set, cross sections and tabulated
 fits are not included as an appendix in the paper but are available
 electronically as part of the Supplemental Material \cite{supp}, on
 the arXiv, and on request from the authors.

 \section{Experiment}
 \label{secexperiment}
In this section, we give an overview of the accelerator and detector facilities used in the experiment and describe the hydrogen target and the program of the measurement.

\subsection{Accelerator}
MAMI, the Mainz Microtron \cite{Herminghaus76,Jankowiak:2006yc,Kaiser:2008zz}, is a normal conducting continuous-wave electron accelerator. It consists of a cascade of three race-track microtrons (RTMs) and a fourth stage, a harmonic double-sided microtron (HDSM). 
The accelerator is equipped with two electron sources: a thermionic source, which can provide currents in excess of $100\ \mathrm{\mu A}$, and a polarized source that makes use of the photoelectric effect on a GaAs crystal using polarized light which can provide more than $30\ \mathrm{\mu A}$.\\
A linear accelerator injects the electrons with 3.97~MeV into the first RTM. Each of the three RTMs contains a normal conducting accelerator segment and two large high precision conventional magnets which recirculate the beam back into the accelerator segment. In the first RTM, the beam is recirculated 18 times, raising the electron energy to 14.86~MeV. The second RTM boosts this to 180~MeV in 51 turns.\\
The beam may now bypass the rest of the accelerator and may be directed to the different experimental sites. Alternatively, it can enter RTM 3, which can boost the energy up to 855~MeV in 90 turns. Every other recirculation path can be instrumented with a kicker magnet which deflects the beam to the exit beam line system. Thus, the energy can be selected in 15-MeV steps.\\
The beam may then be injected into the fourth stage. The HDSM stage comprises two anti-parallel accelerator segments, one of which is operated at the doubled frequency to suppress instabilities. The beam is recirculated by four magnets. The HDSM stage raises the energy up to 1.6 GeV in 43 recirculations.\\
The absolute beam energy uncertainty is 150~keV and the root-mean-square (rms) energy spread is 30~keV at 855~MeV and 110~keV at 1.5~GeV. For the measurements described in this work, an unpolarized beam with beam energies of 180, 315, 450, 585, 720 and 855~MeV was used.\\

\subsection{Detector setup}

The detector setup of the A1-collaboration at MAMI is called the three-spectrometer facility. The three high-resolution magnetic spectrometers, labeled A, B, and C, can be operated in single, double, or triple coincidence mode.  
A detailed description can be found in Ref.\ \cite{Blomqvist:1998xn}. The spectrometers can be rotated around a central pivot to measure at different scattering angles. The scattering angle can be read out with an absolute accuracy of 0.01$^\circ$, leading to an uncertainty in the cross-section of around 0.02\%.

The magnetic system of spectrometer A and C is composed of a quadrupole, a sextupole, and two dipoles. This complex system facilitates a high-precision measurement of particle momentum and angle inside a relatively large acceptance of up to 28 msr. Spectrometer B consists of only a single dipole in a clamshell configuration, leading to a slim design with higher spatial resolution but smaller acceptance (5.6 msr); for out-of-plane measurements, spectrometer B can be tilted by up to 10$^\circ$.

Each of the three spectrometers is equipped with similar detector systems consisting of two scintillator planes, two packets of two vertical drift chamber layers (VDC), and a gas-\v Cerenkov detector. The scintillators are used for triggering, particle identification, and for a time reference. The drift chambers are used for the reconstruction of the particle trajectory. The \v Cerenkov detector distinguishes between muons (and heavier particles) and electrons.

\subsection{Target system}
\label{sectarget}

The target system is enclosed in a vacuum scattering chamber located on the rotation axis of the spectrometers and directly connected to the beam vacuum tube. A target ladder holds several interchangeable solid state materials like graphite, polyethylene, HAVAR foil, copper, etc., of varying thicknesses. Additionally, a luminescent screen (an $\mathrm{Al_2O_3}$ plate with a cross-hair printed on it) is mounted; it is used for beam position calibration. The target ladder has a vertical translation degree of freedom that is actuated by an electric motor to select the target material. \\
The normal lid of the barrel-shaped scattering chamber can be exchanged for two different target constructions: A high-pressure gas target and a cryogenic target. The present experiment used the latter filled with liquid hydrogen as a proton target. 

The cryogenic target system is composed of two loops.  An inner loop (the ``Basel loop'') is filled with the target gas, which is liquefied before the beginning of the beam time. The completely liquefied material is continuously recirculated by a fan. The loop contains an interchangeable target cell; two types were used in this experiment: a 5-cm-long, cigar-shaped cell with its axis in the beam direction and a cylindrical cell with a diameter of 2~cm and the axis perpendicular to the scattering plane. A heat exchanger couples the inner loop to the outer loop, which is coupled to a Philips compressor. The outer loop is also filled with hydrogen and works like a heat pipe: Hydrogen is liquefied at the Philips compressor. It flows down to the target, cooling down the target heat exchanger. The warmed up hydrogen then evaporates and returns to the Philips compressor.\\
The hydrogen inside the inner loop is sub cooled to ensure that the beam load does not substantially change the density of the hydrogen by local heating above the boiling point. Nevertheless, for higher currents the beam is rastered in the transverse directions to reduce the effective power density (in both directions +-1mm for currents above $1~\mathrm{\mu A}$ and +-2mm for currents above $5~\mathrm{\mu A}$).

\subsection{Additions to the standard experimental setup}
\subsubsection{pA-meter}
\label{pameterdesc}
In a typical experiment at the three-spectrometer facility, the beam current is measured with a F\"orster probe located in a part of RTM 3 where all recirculations of the beam pass through. Accordingly, the accuracy of the measurement is best with the highest number of recirculations, i.e., for a beam energy of 855~MeV.  For 180~MeV, the beam does, however, not pass the probe.\\
Therefore, for measurements at small energies, a pA meter was installed at a collimator right before the linear accelerator segment. When the beam is deflected on the collimator, a current proportional to the beam current can be measured precisely. The pA-meter has been calibrated versus the F\"orster probe with a range of beam currents using a beam energy of 855~MeV, where the F\"orster probe is most sensitive.  Since the pA meter is situated in a section of the accelerator where the beam energy is constant, the calibration is valid for all beam energies.
The pA-meter system also has a better precision for low beam currents, which are needed for the measurements at small scattering angles.  

\subsubsection{Beam position stabilization}
\label{chptbeamstab}
A shift of the beam position on the target results in a drift of the measured cross section \cite{Jcb04}. The beam is normally stabilized by the circulation in RTM 3, which dampens beam position changes introduced in the earlier stages of the accelerator. This self-stabilization is less effective with lower recirculation number, i.e., lower energies, and is absent in the case of an incident beam energy of 180~MeV when the beam bypasses RTM 3. To eliminate beam position drifts, a beam-position control system has been installed by the MAMI group \cite{dehn}: The beam position is measured with two cavities in front of the target. Their signal is digitized and a correction current for the beam steering dipoles in the beam line is generated. The cavities need high beam currents for adequate sensitivity. Therefore, the beam has to be switched to a diagnostic mode where the beam is modulated as a train of high current pulses with a low duty cycle. These periods have to be excluded from the cross section measurements. During the data taking, the A1 computer system periodically disables the data acquisition and generates a signal to the MAMI control system to start the adjustment process. When the correction has been performed, MAMI signals back to the A1 system and data acquisition is resumed. The analysis codes have been modified to account for these pauses in the data acquisition.\\ 
The system was installed in the beginning of the second measurement period (see Table \ref{tabbeamtimes}) and was used for all later measurements.  After the installation of the system, there was no beam position drift detectable. For the first period, we observed beam position shifts of less than 0.3~mm, leading to a change in the cross section of less than 0.1\%. Because this change is matched by the luminosity measurement, the error in the ratio of the two is negligible.
 
\subsection{Kinematic coverage}
 \label{subsecprog}
\begin{figure*}
\begin{center}
 \includegraphics{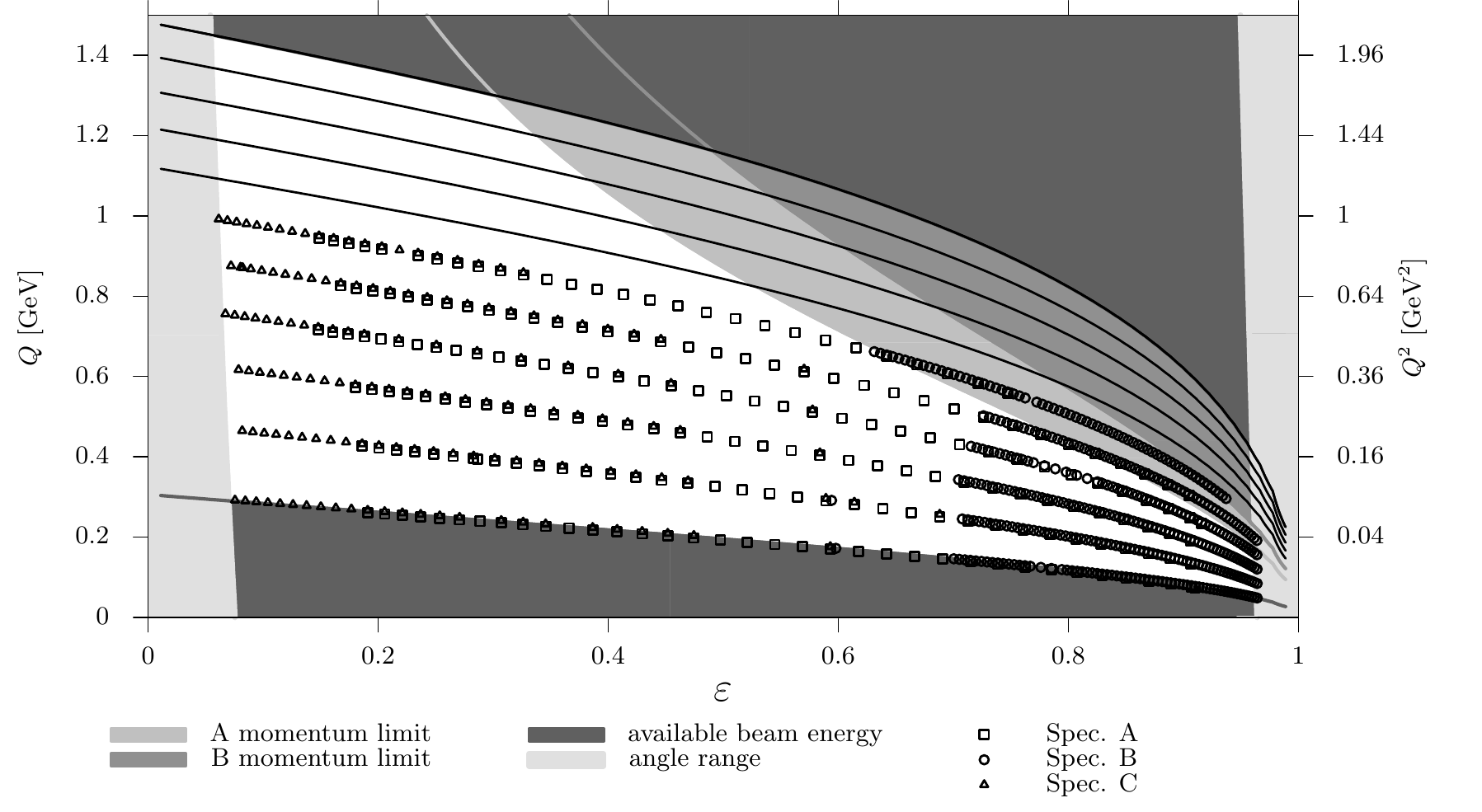}
 \end{center}
\caption[Accessible kinematical region and measured setups]{\label{figkine}The accessible kinematical region in the $\varepsilon$-$Q^2$-plane. Shaded regions are excluded because of minimum or maximum beam energy (darkest gray), maximum detectable momentum of spectrometer A and B (medium grays), and minimum angle of B and maximum angle of C (lightest gray). The centers of the acceptances of the different kinematical settings are denoted with symbols (squares, circles, and triangles correspond to spectrometer A, B, and C). The symbols are slightly shifted vertically to distinguish overlapping measurements. The upper black lines represent a possible future extension using MAMI C energies. To stretch the low-$Q^2$ part, the (left) $y$ axis is presented linear in $Q=\sqrt{Q^2}$.}
\end{figure*}

At a given (negative) four-momentum transfer squared,
\begin{equation}
Q^2=4EE^\prime\sin^2\frac{\theta}{2},
\end{equation}
where $E\ (E^\prime)$ is the energy of the incoming (outgoing) electron and $\theta$ is the electron scattering angle,
the relative contributions of $G_E$ and $G_M$ to the cross sections depend on the polarization of the virtual photon 
\begin{equation}
\varepsilon=\left(1+2\left(1+\frac{Q^2}{4m_p^2}\right)\tan^2\frac{\theta}{2}\right)^{-1}.
\end{equation}
In order to extract the form factors from the measured cross sections using the traditional Rosenbluth technique, it is mandatory to measure at different $\varepsilon$ for a given, constant $Q^2$ value. While this constraint does not have to be fulfilled when performing the mentioned global fit, which is in focus here, the range and number of different $\varepsilon$ values in a given $Q^2$ range determine the accuracy of the separation. \\
Figure \ref{figkine} displays the part of the $\varepsilon$-$Q^2$-plane accessible by the accelerator and detector setup in this experiment. To help the readability, only the centers of the overlapping acceptances are marked. The shaded areas are excluded because of the various experimental limitations (see the figure caption). 

In order to vary $\varepsilon$ at constant $Q^2$, both the scattering angle and the incident beam energy have to be changed. A beam energy change takes about 6 h at MAMI, which is quick in comparison to other accelerator facilities but still too costly to be done frequently. Thus, the measuring program was organized to minimize beam energy changes. Since the energy gain in RTM 1 and RTM 2 is fixed, the minimum beam energy of MAMI is achieved when the beam passes RTM 3 without further acceleration, resulting in a 180-MeV beam. The beam time allocation permitted to measure at six energies, spread out evenly between 180 and 855~MeV in 135-MeV steps. In this experiment, no use was made yet of the 1.6-GeV stage of MAMI. While the HDSM stage (MAMI C) was already commissioned for productive use, there was no experience with the quality of the beam, and it was not yet possible to extract the beam at energies between 0.855 and 1.6~GeV.

At each beam energy, the measured angle range was maximized. The geometric designs of target and spectrometers allow each spectrometer to cover different but overlapping angular ranges. To maximize the angular range covered by the data set and the internal redundancy in the data, all three spectrometers were used in parallel. Spectrometers A and C were used alternately as the ``production spectrometer'' (changing angle from run to run) and ``luminosity monitor'' (at a fixed angle). The spectrometer angle is changed each time only by one-fourth of the acceptance, i.e., by $2.5^\circ$ for A and C and by $0.5^\circ$ for B, so that at each angle the cross section is measured four times with the same spectrometer but with different parts of the spectrometer acceptance. While the quality of the reconstruction is not good enough to split up the acceptance of a single measurement in smaller bins with the aimed-for precision, this fourfold oversampling of the scattering angle allows us to recover the $Q^2$ dependence inside the acceptance of a single measurement. 
Spectrometer A is situated on the opposite side of the beam line from spectrometers B and C. Hence, a comparison in the overlap region between spectrometer A and the others is testing the beam, target, and rotation-axis alignment. Spectrometer A can be used in the range between $25^\circ$ and $110^\circ$. Due to the construction of the target, the angle of A had to be limited to $90^\circ$ for the long target cell and to $110^\circ$ for the short target cell. Spectrometer C extends the angle range to over $130^\circ$. The slim construction of spectrometer B allows it to reach scattering angles down to $15.5^\circ$.

When the field of a spectrometer is changed, the magnetic fields change due to eddy currents and have to stabilize. Since this takes some time, the momentum was adjusted only every second angle change in order to keep the elastically scattered electrons at roughly a constant position on the focal plane of the spectrometers.\\
For most of the individual points, the required time to achieve the envisaged statistics of below 0.2\% was around 30 min. Most settings were divided into two 5- to 20-min-long submeasurements. This reduces the statistical accuracy per submeasurement (which is compensated by the higher number of measurements) but increases the accuracy of the luminosity determined by the pA-meter, which measures the beam current before and after each submeasurement and facilitates the search and elimination of time dependent effects.

 The measurements were performed during three beam time periods, summarized in Table \ref{tabbeamtimes}.

\begin{table}
\begin{center}
\begin{ruledtabular}
\begin{tabular}{l | c|c| c}
& 08/2006 & 11/2006 & 05/2007\\
\hline
Duration& 10 days & 11 days & 17 days\\
Setup / calibration & 2 days & 2 days & 6 days\\
Beam energies [MeV] & 585, 855 & 180, 720& 315, 450, 720\\
Target cell used & long & short & long \\ 
No.\ of setup changes & 152 & 173 & 217\\
No.\ of measurements & 358 & 490 & 574 \\
Beam currents & 585: 11~nA & 180: 2.8~nA& 315: 28~nA\\
& to 5.5~$\mu$A & to 360nA & to 1.6~$\mu$A\\
& 855: 0.8~$\mu$A& 720: 90~nA & 450: 30~nA\\
& to 10~$\mu$A & to 14~$\mu$A & to -4.8~$\mu$A\\
&&&720: 8-10~$\mu$A\\
\end{tabular}
\end{ruledtabular}
\end{center}

\caption[Overview of beam times] {Overview of the beam times. Setup changes are changes of  momentum and/or angle of at least one spectrometer.}
\label{tabbeamtimes}
\end{table}
 
\section{Theory for elastic electron proton scattering}
 \subsection{Cross section in first Born approximation}
\label{chpttfecs}
\begin{figure}[h]
\begin{center}
 \includegraphics{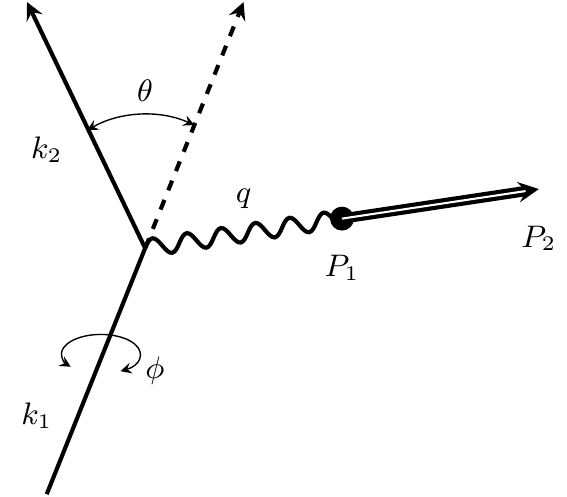}
 \end{center}
\caption[Kinematic parameter in the laboratory frame]{\label{figlabkin}The kinematic parameters for the elastic scattering of an electron on a target initially at rest.}
\end{figure}

The kinematic parameters of the elastic scattering of an electron on a
target at rest is depicted in Fig.\ \ref{figlabkin}.

The incident electron has a four-momentum $k_1=(E_1=E,\vec{p_1})$. It is scattered in the direction $\Omega=(\theta,\phi)$ with four-momentum $k_2=(E_2=E',\vec{p_2})$. In the scattering process, the four-momentum $q=k_1-k_2$ is transferred to the target via the exchange of a virtual photon. The target of mass M is initially at rest, $P_1=(M,\vec{0})$.\\
The unpolarized cross section is independent of the azi\-muthal
angle $\phi$. Therefore, there are only two degrees of
freedom and the cross section can be expressed in terms of the
energy $E$ of the incoming electron and the scattering angle $\theta$
or, equivalently, by the (negative) four-momentum transfer squared $Q^2$ and
the photon polarization $\varepsilon$.

In the one-photon-exchange approximation, the unpolarized cross
section for the elastic scattering of an electron on a proton with
internal structure is given as follows:
\begin{align}
&\left(\frac{\mathrm{d}\sigma}{\mathrm{d}\Omega}\right)_{0}=\nonumber\\
&\left(\frac{\mathrm{d}\sigma}{\mathrm{d}\Omega}\right)_\mathrm{Mott}\left[\left(F_1^2 +\tau\left(\kappa F_2\right)^2\right) + 2\tau\left(F_1+\kappa F_2\right)^2\tan^2\frac{\theta}{2}\right],
\end{align}
with the dimensionless quantity $\tau=Q^2/(4m_P^2)$ and where 
\begin{equation}
\label{eqmott}
\left(\frac{\mathrm{d}\sigma}{\mathrm{d}\Omega}\right)_\mathrm{Mott}=\frac{4\alpha^2E'^2}{Q^4}\frac{E'}{E}\left(1-\beta^2\sin^2\left(\frac{\theta}{2}\right)\right)
\end{equation}
is the recoil-corrected Mott cross section, which is the cross section
for the scattering of a point-like spin-$\frac{1}{2}$-particle on a
scalar point-like target.
The internal structure is expressed here in terms of the Dirac and
Pauli form factors,  $F_1$ and $F_2$, respectively.

The relations
\begin{align}
G_E&=F_1-\tau \kappa F_2\nonumber,\\
G_M&=F_1+\kappa F_2
\end{align}
translate the Dirac $F_1$ and Pauli $F_2$ form factors into the Sachs
form factors $G_E$ and $G_M$. They were first proposed by Yennie {\em et al.}\ \cite{yennie57}. Sachs {\em et al.}\ \cite{Ernst, Sachs} proposed that this choice provides a more physical insight than $F_1$ and $F_2$ since in the Breit frame, defined as $P^B_1+P^B_2=(2E_B,\vec{0})$, the transition current reduces to
\begin{equation}
J_B=e\chi^T_{p_2}\left(2M\cdot G_E,i\vec{\sigma}\times\vec{q}\cdot G_M\right)\chi_{p_1}.
\end{equation}
In this frame, $G_E$ and $G_M$ are the Fourier transforms of the
spatial charge and magnetization distributions. A frequent criticism
of this idea is that the Breit frame is equivalent to an infinitely
heavy proton or alternatively, a proton affixed to the coordinate
origin and, therefore, the charge and magnetization distributions are
not ``real,'' i.e., frame dependent. A more detailed discussion of
this point can be found in Ref.\ \cite{Vanderhaeghen:2010nd}.

 With the Sachs form factors, the cross section is given by
\begin{align}
\left(\frac{\mathrm{d}\sigma}{\mathrm{d}\Omega}\right)_{0}&=\left(\frac{\mathrm{d}\sigma}{\mathrm{d}\Omega}\right)_\mathrm{Mott}\times\nonumber\\
&\left[\frac{G_E^2\left(Q^2\right)+\tau G_M^2\left(Q^2\right)}{1+\tau}+2\tau G_M^2\left(Q^2\right)\tan^2\frac{\theta}{2}\right] \nonumber\\
&=\left(\frac{\mathrm{d}\sigma}{\mathrm{d}\Omega}\right)_\mathrm{Mott}\frac{\varepsilon G_E^2+\tau G_M^2}{\varepsilon\left(1+\tau\right)}
\label{eqrosen}.
\end{align}
The choice of the Sachs form factors eliminates the mixed term in the cross section, which now depends on the squares of $G_E$ and $G_M$ only. 

In the static limit $Q^2=0$, the form factors normalize to the charge and magnetic moment of the proton in units of the electron charge and of the nuclear magneton $\mu_K$, $G_E(0)=1$ and $G_M(0)=\mu_p$.
\\
The standard method to extract the form factors from measured cross
sections is the Rosenbluth separation \cite{Rosenbluth}. It exploits
the linear structure in $\varepsilon$ of Eq.\ (\ref{eqrosen}) and separates
the form factors by measurements at constant $Q^2$ but different 
$\varepsilon$ values.

The somewhat unfamiliar method used in this paper consists in
inserting many distinctly different form-factor models into Eq.\ (\ref{eqrosen}) and fitting their parameters directly to the measured cross sections. This will be discussed in detail in Sec.\ \ref{chptmodels} and following subsections.

As mentioned above, in the Breit frame the Sachs form factors are the
Fourier transforms of the charge and magnetization
distributions. Expanding the kernel of the Fourier integrals in terms
of $Q^2$ yields
\begin{equation}
G\left(Q^2\right)/G\left(0\right)=1-\frac{1}{6}\left<r^2\right>Q^2+ \frac{1}{120}\left<r^4\right>Q^4 -\dots,
\end{equation}
where $<r^n>$  is the $n$-th moment of the electric or magnetic distribution.
Therefore, the second moments can be determined by
\begin{equation}
\label{eqradius}
\left<r^2\right>=-\frac{6}{G\left(0\right)}\left.\frac{\mathrm{d}G\left(Q^2\right)}{\mathrm{d}Q^2}\right|_{Q^2=0},
\end{equation}
i.e., from the slope of the form factors at $Q^2=0$.

\subsection{Radiative corrections}
\label{chapterbrems}
\begin{figure*}

\begin{center}

(b)
 \includegraphics{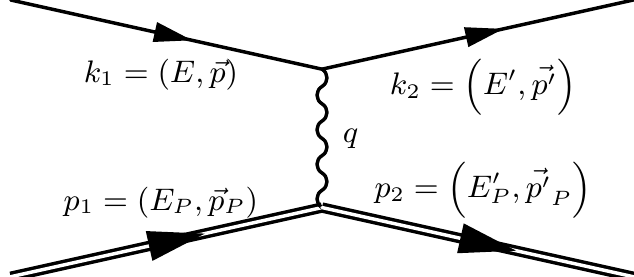}
 \hspace{3em}
(v1)
 \includegraphics{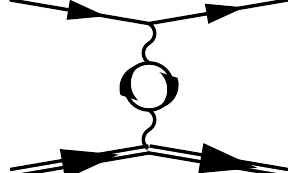}
 \\
\vspace{4ex}
(v2)
 \includegraphics{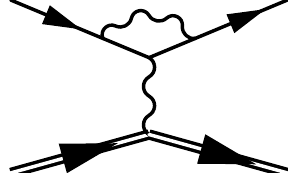}
 \hspace{2em}
(v3)
 \includegraphics{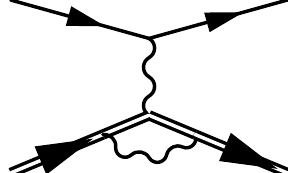}
 \hspace{2em}
(v4)
 \includegraphics{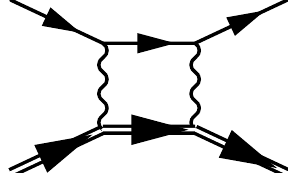}
 \hspace{2em}
(v5)
 \includegraphics{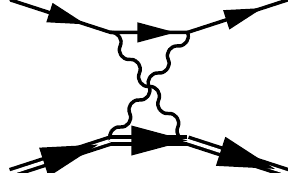}
 \\
\vspace{4ex}
(r1)
 \includegraphics{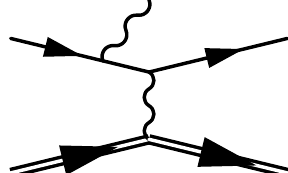}
 \hspace{2em}
(r2)
 \includegraphics{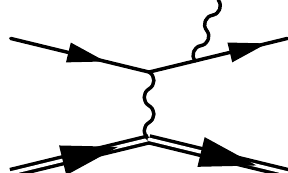}
 \hspace{2em}
(r3)
 \includegraphics{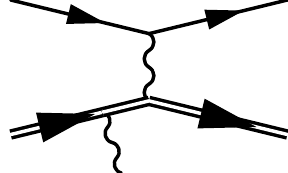}
 \hspace{2em}
(r4)
 \includegraphics{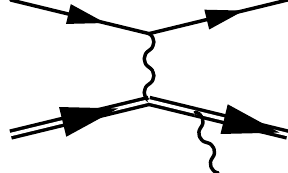}
 \end{center}

\caption[Elastic-scattering Feynman graphs] {\label{figfeyn} Feynman
  graphs of leading and next to leading order for elastic
  scattering. (b) Leading order, (v1--v5) next to leading order with
  an additional virtual photon, and (r1--r4) leading-order graphs with a radiated real photon.}

\end{figure*}
It is not possible to measure the lowest for-order cross section
directly since higher-order diagrams, as depicted in Fig.\ \ref{figfeyn}, 
always contribute to the elastic scattering process. It is
common practice to divide these contributions into groups with an
additional virtual (v1--v5 in Fig.\ \ref{figfeyn}) or real photon
(r1--r4). However, this grouping is problematic: Divergences in one
group cancel against divergences in the other group, hence all graphs
have to be evaluated at once. This leads to a correction $\delta$
to the one-photon-exchange calculation
\begin{equation}
\left(\frac{\mathrm{d}\sigma}{\mathrm{d}\Omega}\right)_{1}=\left(\frac{\mathrm{d}\sigma}{\mathrm{d}\Omega}\right)_0\left(1+\delta\right).
\end{equation}
Here $\left(\frac{\mathrm{d}\sigma}{\mathrm{d}\Omega}\right)_0$ is
the cross section for one-photon exchange alone [Fig.\ \ref{figfeyn}(b)] as given by Eq.\ (\ref{eqrosen}), while $\left(\frac{\mathrm{d}\sigma}{\mathrm{d}\Omega}\right)_{1}$ is the cross section when next-to-leading-order contributions are taken into account (graphs v1--v5 and r1--r4 in Fig.\ \ref{figfeyn}).\\
Conversely, the non-radiative cross section $\left(\frac{\mathrm{d}\sigma}{\mathrm{d}\Omega}\right)_0$ can be determined in a first-order approximation by identifying the experimental cross section with $\left(\frac{\mathrm{d}\sigma}{\mathrm{d}\Omega}\right)_{1}$ and dividing it by $(1+\delta)$.

The integrals over the internal four-momenta of the graphs v1--v3 are
logarithmically divergent for large momenta. This can be treated
theoretically by charge and mass renormalization. Details can be found
in Ref.\ \cite{Maximon2000,Vanderhaeghen2000}. Graph v2 leads to an infrared divergence, but it can be shown \cite{Bloch1937,Jauch1954} that
this cancels with corresponding divergences of the graphs r1 and r2. 

In the following, the formulas for the contributions from different
groups of diagrams used in this work will be presented. For details of
the calculation see Refs.\ \cite{Maximon2000,Vanderhaeghen2000}.\\
The vacuum polarization (v1) gives rise to the term
\begin{align}
\delta_\mathrm{vac}&=\frac{\alpha}{\pi}\frac{2}{3}\left\{\left(v^2-\frac{8}{3}\right)+v\frac{3-v^2}{2}\ln\left(\frac{v+1}{v-1}\right)\right\},\label{eqtruevac}\\
&\stackrel{Q^2\gg m_l^2}{\longrightarrow}\frac{\alpha}{\pi}\frac{2}{3}\left\{-\frac{5}{3}+\ln\left(\frac{Q^2}{m_l^2}\right)\right\},\label{eqapproxvac}
\end{align}
with $v^2=1+\frac{4m_l^2}{Q^2}$, where $m_l$ is the mass of the
particle in the loop. The approximation (\ref{eqapproxvac}) is valid
for loop electrons. However, at the energy scales of this experiment and within the envisaged accuracy, the vacuum polarization via muon and tau loops has to be accounted for and must be evaluated with Eq.\ (\ref{eqtruevac}).\\
The finite part of the electron vertex correction (v2, the infinite part cancels later on) is given in the ultrarelativistic limit by
\begin{equation}
\delta_\mathrm{vertex}=\frac{\alpha}{\pi}\left\{\frac{3}{2}\ln\left(\frac{Q^2}{m^2}\right)-2-\frac{1}{2}\ln^2\left(\frac{Q^2}{m^2}\right)+\frac{\pi^2}{6}\right\}.
\end{equation}\\
In the same limit, the contribution from real photon emission by the
electron (r1, r2) yields the following:
\begin{align}
\delta_R=&\frac{\alpha}{\pi}\left\{\ln\left(\frac{\left(\Delta E_s\right)^2}{E\cdot E'}\right)\left[\left(\frac{Q^2}{m^2}\right)-1\right]-\frac{1}{2}\ln^2\eta\right.\nonumber\\
&\left.+\frac{1}{2}\ln^2\left(\frac{Q^2}{m^2}\right)-\frac{\pi^2}{3}+\mathrm{Sp}\left(\cos^2\frac{\theta_e}{2}\right)\right\},
\end{align}
where $\eta=E/E'$, $\Delta E_s=\eta\cdot \Delta E'$. $E'$ is the
energy of an electron scattered elastically through an angle $\theta$
when no photon is emitted. An electron which radiates a photon has a
lower energy than $E'$. $\Delta E'$ is the maximum difference to $E'$
allowed by the radiative tail cut-off; it is called the cut-off
energy. Details about the Spence function $\mathrm{Sp}\left(x\right)$
can be found in Appendix B of Ref.\ \cite{friedjm00}. 

The terms where the proton contributes (v3--v5, r3, and r4) are complicated
and an exact calculation requires the knowledge of the internal
structure of the proton. Maximon and Tjon \cite{Maximon2000} divide
the correction in three parts, one proportional to the charge $Z$ ($\delta_1$), one to $Z^2$ ($\delta_2$), and a third part in which they include all of the structure dependence ($\delta^{(1)}_{el}$).
The last part is believed to be small for the kinematics of this work
and is therefore neglected. The other two correction terms (the $Z$-dependence is divided out) are given by
\begin{align}
\delta_1=&\frac{2\alpha}{\pi}\left\{\ln\left(\frac{4\left(\Delta
        E_s\right)^2}{Q^2x}\right)\ln\eta+\mathrm{Sp}\left(1-\frac{\eta}{x}\right)\right.\nonumber\\
&\hspace{5em}\left.-\mathrm{Sp}\left(1-\frac{1}{\eta x}\right)\right\},\\
\delta_2=&\frac{\alpha}{\pi}\left\{\ln\left(\frac{4\left(\Delta E_s\right)^2}{m^2_p}\right)\left(\frac{E'_P}{\left|\vec{p'}_P\right|}\ln x-1\right)+1\right.\nonumber\\
&\left. +\frac{E'_P}{\left|\vec{p'}_P\right|}\left(-\frac{1}{2}\ln^2x-\ln x\ln\left(\frac{\rho^2}{m^2_P}\right)+\ln x\right.\right.\nonumber\\
&\left.\left. -\mathrm{Sp}\left(1-\frac{1}{x^2}\right)+2\mathrm{Sp}\left(-\frac{1}{x}\right)+\frac{\pi^2}{6}\right)\right\},
\end{align}
with
\begin{equation}
x=\frac{\left(Q+\rho\right)^2}{4m^2_P},\hspace{1cm}\rho^2=Q^2+4m^2_P.
\end{equation}

For vanishing cut-off energies and, hence, $\Delta E_s$, the
corrections $\delta_R$, $\delta_1$, and $\delta_2$ get infinitely
large. In this case, however, more photons than just one are emitted
in each scattering event. It has been shown in Refs.\ \cite{Bloch1937,Yennie1961} that this can be approximately taken into account by exponentiation of the corresponding correction terms as well as for the vertex correction. For the vacuum polarization
contribution, Vanderhaeghen {\em et al.}\ \cite{Vanderhaeghen2000} iterate the first order contribution to all orders, which does not lead to an exponentiation. In total, they find 
\begin{equation}
\label{oldcorr}
\left(\frac{\mathrm{d}\sigma}{\mathrm{d}\Omega}\right)_\mathrm{exp}=\left(\frac{\mathrm{d}\sigma}{\mathrm{d}\Omega}\right)_0\frac{e^{\delta_\mathrm{vertex}+\delta_R+Z\delta_1+Z^2\delta_2}}{\left(1-\delta_\mathrm{vac}/2\right)^2},
\end{equation}
which, for the kinematics used in the present work, differs marginally
(below 0.05\%) from the fully exponentiated form, which will
therefore be used in the analysis of the measured cross sections as
follows ($Z$ has been set to 1):
\begin{equation}
\left(\frac{\mathrm{d}\sigma}{\mathrm{d}\Omega}\right)_\mathrm{exp}\hspace{-1em}\left(\Delta E'\right)=\left(\frac{\mathrm{d}\sigma}{\mathrm{d}\Omega}\right)_0e^{\delta_\mathrm{vac}+\delta_\mathrm{vertex}+[\delta_R+\delta_1+\delta_2]\left(\Delta E'\right)} .
\end{equation}

It will be described in Sec.\ \ref{chptradtail} how these higher-order contributions are accounted for in the determination of the first-order cross section from the measured data.

\subsection{Coulomb distortion and two-photon exchange}
\label{chpttpe}
The Coulomb distortion, i.e., the scattering process via the exchange
of many soft photons, and the related two-photon exchange, where both
photons have a sizable momentum, is not fully included in the
radiative corrections. There is yet some theoretical uncertainty in
the modeling of these two effects. For the two photon effect, the
off-shell nucleon and its excited states have to be modeled. For a
discussion of the Coulomb distortion, see Ref.\ \cite{friedjm00}. A complete treatment of these effects is not the topic of this paper.\\
Nevertheless, the Coulomb distortion can not be ignored completely,
especially for the determination of the radius, as has been shown by
Rosenfelder \cite{Rosenfelder99}. He finds that the extracted radius
is enlarged by about 0.018 fm when Coulomb distortion is accounted
for. For $Q^2=0$, the correction is in agreement with the simple additional
correction factor $(1+\delta_F)$, the so-called Feshbach correction
\cite{McKinley:1948zz,tsai61}, by McKinley and Feshbach as follows:
\begin{equation}
\delta_F=Z\alpha\pi\frac{\sin\frac{\theta}{2}-\sin^2\frac{\theta}{2}}{\cos^2\frac{\theta}{2}}.
\label{feshbach}
\end{equation}
This correction has been applied to the measured cross sections and is
maximal for $180^\circ$ scattering yielding a downward correction of 1.2\%.\\

The TPE becomes more important at larger $Q^2$
and may explain the difference between polarized and unpolarized
measurements at large $Q^2$ \cite{Guichon03}. Therefore, a lot of
theoretical work focuses on the energy scales above 1~GeV. In 2007,
Arrington {\em et al.}\ \cite{Arrington07} have reanalyzed the world data
set with a model for two-photon-exchange corrections and made two
fits, one with the corrections applied and one without. The ratio of
these fits represents an estimate of the two-photon effect on the form-factor ratio in the higher $Q^2$ region of the present
experiment, which can be used to compare the form-factor ratio from fits to
Rosenbluth data with previous polarized measurements. However, one has to keep in mind that these fits rely on just one model \cite{Arrington:2011kv,Bernauer:2011zz} and a generalization is uncertain. Therefore, we have chosen a different phenomenological approach (Sec.\ \ref{fwaextdata} and \ref{fwextdata}).

\section{Determination of the cross sections}
\subsection{Overview}
\label{dcslum}
In order to calculate the cross section from counting rates one has to
know the luminosity and the acceptance of the detector. However, the
acceptance is not just a fixed number given purely by the collimator
geometry; it also depends on the target length and position and on the
spectrometer angle and the momentum of the particles. The only
feasible way to determine the cross section from the measured number
of scattering events is by comparing this number to the result of a
full simulation of the experiment, $\sigma_\mathrm{sim}$, including
all aspects of the detector response, external energy loss of the
electrons in the target material and all radiative corrections.
The measured cross section is then found as
\begin{equation}
\label{eqsigma}
\sigma_\mathrm{rel,exp}=\frac{A-B}{\sigma_\mathrm{sim}\mathcal{L}}.
\end{equation}
Here, $A$ is the number of counts in the peak region integrated to the cut-off energy $\Delta E^\prime $, $B$ is the estimated background in this region, $\sigma_\mathrm{sim}$ is the simulated cross section including radiative corrections integrated over the acceptance of the spectrometers, 
and $\mathcal{L}=\int\mathcal{L}_\mathrm{eff}\mathrm{d}t$ is the
time-integrated effective, i.e., prescaling and dead-time corrected,
luminosity. For the calculation of $\sigma_\mathrm{sim}$, one has to
make use of an assumed cross section which should be sufficiently
close to the true one. As a result, Eq.\ (\ref{eqsigma}) yields the
measured cross section relative to the assumed one.

We define the data taken at one energy and angle over some time as a
``run.''
The runs are grouped into a ``set of runs,'' where the relative
normalizations of the runs to each other in the same set are determined by the luminosity-monitor
measurements (see Sec.\ \ref{lumi}).
With the  setup of this experiment, as well as with any other, the
absolute normalization of each run, i.e., the luminosity, which
comprises the absolute know\-ledge of the target length, the absolute
current calibration and the absolute detector efficiencies, can be
determined only to the few-percent level. Therefore, the relative
normalization between different sets of runs will be left floating in the
fits of the final analysis. The absolute normalization, finally, is fixed by the known values of the form factors at $Q=0$, as already mentioned.

\subsection{Cross section simulation}
\label{chptsim}

Instead of correcting the measured number of events for efficiencies, acceptance problems, and the radiative processes, as was the case in the classical electron-scattering experiments, these ingredients are better incorporated into the simulated cross section as discussed in the following.

\subsubsection{Internal radiative corrections}
\label{chptradtail}
As described above, the extraction of the first-order Born cross section $\left(\frac{d\sigma}{d\Omega}\right)_0$ of the process $ep\longrightarrow e^\prime p^\prime$ also requires the calculation of a (radiation) correction factor $f_\mathrm{corr}$. This factor depends on the kinematics and cut-off energy $\Delta E^\prime $ in the spectrum of the scattered electrons,
\begin{equation}
\left(\frac{\mathrm{d}\sigma}{\mathrm{d}\Omega}\right)_\mathrm{exp}=\left(\frac{\mathrm{d}\sigma}{\mathrm{d}\Omega}\right)_0 \cdot f_\mathrm{corr}\left(\Delta E^\prime, E, \theta\right),
\end{equation}
or, differential in the cut-off energy,
\begin{equation}
\left(\frac{\mathrm{d}\sigma}{\mathrm{d}\Omega\mathrm{d}\Delta E^\prime }\right)_\mathrm{exp}=\left(\frac{\mathrm{d}\sigma}{\mathrm{d}\Omega}\right)_0\cdot\frac{\mathrm{d}\,f_\mathrm{corr}}{\mathrm{d}\Delta E^\prime },
\end{equation}
where $\frac{\mathrm{d}\,f_\mathrm{corr}}{\mathrm{d}\Delta E^\prime }$ is the radiative tail. 
A generator algorithm for a Monte Carlo calculation has to produce events replicating this cross section in an efficient manner, i.e., without too great computational costs and with a choice of the kinematic quantities so the variance of the weights of the individual events is minimized. The generator is based on a generator for events for a virtual Compton-scattering experiment described in Ref.\ \cite{Jovas03}. In its original form, it was limited to a description of the shape of the tail without correct global normalization. In the course of the present work, it was extended to also describe accurately the peak region and to have the correct normalization.
The generator first generates a vertex position, a scattering angle, and an azi\-muthal angle from pseudorandom number sequences. Then, it follows the principles laid out in Vanderhaeghen {\em et al.}\ \cite{Vanderhaeghen2000} in calculating the energy of a radiated photon.
In the next step, a direction of this photon has to be generated. Since the cross section depends strongly on the photon angle, peaking close to the directions of the electrons, but vanishing at the exact directions, it is important to generate the photon direction with importance sampling, that is, generating more events where the cross section is higher so the weights of the events are nearly constant. To this end, one generates the directions using a suitable approximation of the Bethe-Heitler part of the cross section. Such an approximation is given by the sum of individual cross sections for radiation off the incoming or outgoing electron, neglecting the interference,
\begin {align}
\left(\frac{\mathrm{d}\sigma}{\mathrm{d}\Omega}\right)_\mathrm{approx.}&=\left(\frac{1}{2}\left(\frac{\mathrm{d}\sigma}{\mathrm{d}\Omega}\right)_e+\frac{1}{2}\left(\frac{\mathrm{d}\sigma}{\mathrm{d}\Omega}\right)_{e^\prime }\right)\nonumber\\
\left(\frac{\mathrm{d}\sigma}{\mathrm{d}\Omega}\right)_{e}&=\frac{1}{\mathrm{N}\left(E,\vec{p}\right)}\cdot \frac{1-\cos^2\theta_{e,\gamma}}{\left(\frac{E}{\left|\vec{p}\right|}-\cos\theta_{e,\gamma}\right)^2}\label{eqBHapprox}\\
\mathrm{N}\left(E,\vec{p}\right)&=-4-2\frac{E}{\left|\vec{p}\right|}\cdot\ln\left(\left(\frac{E}{\left|\vec{p}\right|}-1\right) \left/ \left(\frac{E}{\left|\vec{p}\right|}+1\right)\right) \right.\nonumber
\end{align}
and the same expressions for $E^\prime $, $p^\prime $, and $\theta_{e^\prime \gamma}$. $E\ (E^\prime )$ is the incoming (outgoing) electron energy, $\vec{p}\ (\vec{p^\prime })$ the corresponding momentum, and $\theta_{e\gamma}\ (\theta_{e^\prime \gamma})$ the angle between the incoming (outgoing) electron and the photon. 
The generator selects with equal probability whether the photon is radiated from the incoming or outgoing electron.\\ 
Then the transformation method is used to generate random values with a distribution according to Eq.\ (\ref{eqBHapprox}). The cumulative distribution is given by
\begin{align}
F\left(\theta_{e,\gamma}\right)&=\frac{\int_{-1}^{\cos(\theta_{e,\gamma})}\left(\frac{\mathrm{d}\sigma}{\mathrm{d}\Omega}\right)_{e}\mathrm{d}\cos\theta_{e\gamma}}{\int_{-1}^1\left(\frac{\mathrm{d}\sigma}{\mathrm{d}\Omega}\right)_e \mathrm{d}\cos\theta_{e\gamma}}\nonumber\\
&=\frac{1}{\mathrm{N}\left(E,\vec{p}\right)}\left(\frac{1-\left(\frac{E}{\left|\vec{p}\right|}\right)^2}{\frac{E}{\left|\vec{p}\right|}-\cos\theta_{e,\gamma}}-\cos\theta_{e,\gamma}\right. \nonumber\\
&\left.-2\frac{E}{\left|\vec{p}\right|}\ln\frac{\frac{E}{\left|\vec{p}\right|}-\cos\theta_{e,\gamma}}{\frac{E}{\left|\vec{p}\right|}+1}-2+\frac{E}{\left|\vec{p}\right|}\right).
\end{align}
A uniformly distributed number $r$ between 0 and 1 is now transformed by solving $r=F(\theta_{e,\gamma})$ to the new random variable $\theta_{e\gamma}$ with the correct distribution (see Ref.\ \cite{NR}, Sec.\ 7.2). The required inversion of $F$ is realized numerically via a bisection method.

The innermost part of the generator calculates the Feynman graphs of the lowest order describing the Bethe-Heitler (radiation from the electron) + Born (radiation from the proton) processes for the now-fixed kinematics. Here a Jacobian for the transformation $\mathrm{d}\Omega_k^{Lab}$ to $\mathrm{d}\Omega_k^{c.m.}$ has to be taken into account. It is calculated numerically using finite differences.\\
The cross section calculated with these graphs is infrared divergent. This is accounted for by a modification of the propagators. Their denominators are
\begin{eqnarray*}
\mathrm{Bethe\textrm{-}Heitler:}&2k_1\cdot q^\mathrm{rad} ,\,&-2k_2 \cdot q^\mathrm{rad} \\
\mathrm{Born:}&-2p\cdot q^\mathrm{rad} ,\,&p^\prime \cdot q^\mathrm{rad} ,
\end{eqnarray*}
where $k_1 (k_2 )$ is again the four-vector of the incoming (outgoing) electron and $q^\mathrm{rad} $ the four-vector of the radiated photon evaluated in the center-of-mass (c.m.) system. Here, $q^\mathrm{rad} $ is replaced with $q^\mathrm{rad} _{mod}=q^\mathrm{rad} /\left|\vec{q}^\mathrm{rad}\right|$. Hence, the calculation yields the correct cross section multiplied with a factor $K^2=\left|\vec{q}^\mathrm{rad}\right|^2$ since the matrix element enters quadratically into the cross section. One order of $K$ is then divided out at the cross section level, and the remaining order has to be accounted for later when the different parts of the generator are combined.

\subsubsection{External radiation}
In an extended target, the material in the path of the particles before and after the scattering inflict an energy loss in addition to the internal processes. 
When the cryogenic target is used, the incoming beam has to pass through different layers of matter until the scattering process occurs. This includes the walls of the target and the liquid hydrogen inside the target. Additionally, the cold target acts as a cold trap.  ``Snow,'' i.e., frozen water and nitrogen from the residual gas inside the vacuum chamber, can build up on the target surface and is easily identified as additional peaks in the scattering spectrum. Switching to high current melts the snow where the beam enters and exits the target. However, snow remains on the sides of the cell. The outgoing electron has to pass part of the hydrogen, the wall of the target, possibly snow, and then the windows between the spectrometer and vacuum chamber and a short distance of air between them. In all these layers, the electron loses energy by external bremsstrahlung and ionization of the atoms. These processes have to be folded with the internal bremsstrahlung spectrum; the simulation does this numerically.

\subsubsection{Resolution}
The resolution of the drift chambers, the characteristics of the electronics, and the knowledge of the transport inside the magnetic system give rise to specific error distributions for the extracted kinematical variables of the detected particles at the target. In the simulation, we employ a simple Gaussian to model the error-distribution in the reaction vertex and the sum of two Gaussians with different weights and width to model the error distributions in the in-plane and out-of-plane angles and in the momentum. With two Gaussians, the longer tails in the distributions of the errors in these variables are better reproduced than by a single Gaussian.  

\subsubsection{Test of the description of the radiative tail}
\begin{figure}
\begin{center}
 \includegraphics{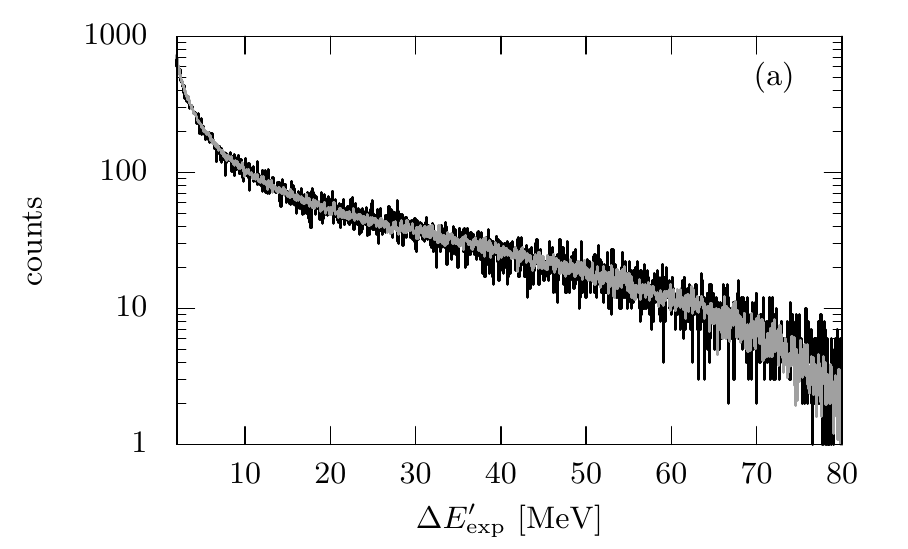}
  \includegraphics{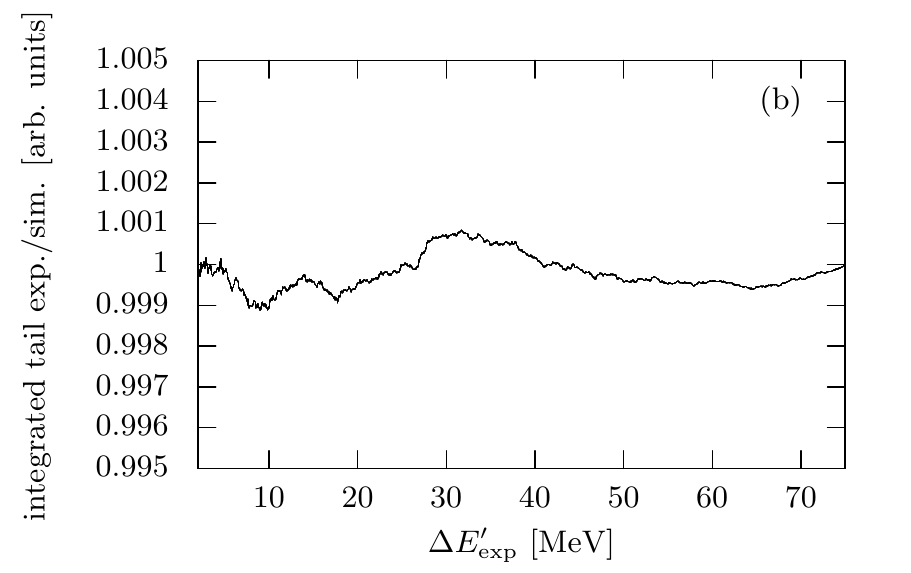}
 \end{center}
\caption[Comparison of elastic tail] {\label{figtail1} (a) Comparison of experimental (black) and simulated (gray) $\Delta E^\prime $ histogram in the region of the tail. A cut on the vertex position was applied to suppress background from reactions off the walls. (b) Ratio of the integral of the experimental data to the integral of the simulation integrated up to the cut-off energy $\Delta E^\prime $. The ratio varies by less than 0.1\% for cut-off energies up to 75\ MeV.  The ratio is scaled to start at 1. Data: Spectrometer A, $53^\circ$, and 855-MeV incident beam energy.  }
\end{figure}

In the classical electron scattering experiments the radiative correction was calculated
from a cut-off energy by Eq.\ (\ref{oldcorr}). By contrast, we calculate in the simulation
the full radiative tail, accounting for the convolution over the complicated acceptance
of the spectrometers, for the convolution of internal and external energy loss, and for
the dependence of the cross section on the energy.
With the classical radiative correction one meets a dilemma: A large cut-off
is favored by the then-small correction; however, there comes in the unaccounted dependence
of the cross section on the energy. The latter disadvantage is minimized by a small
cut-off, but then one has fewer events, the correction is large, and there enters
an uncertainty from the resolution of the spectrometer. 
Compared to the evaluation of the old measurements, the specific choice of the cut-off energy is considerably less important in our method, where the measured cross section is given by a comparison of the simulated and measured spectrum, giving full account to the acceptance and to the energy dependence of the cross section.  We demonstrate the quality of the description of the radiation tail in Fig.\ \ref{figtail1}, which shows a comparison of the measured $\Delta E^\prime_\mathrm{exp}$ spectrum and the simulated tail.
For this comparison, the background in the measured spectrum is suppressed by a vertex cut.
The upper panel compares the two spectra directly in a logarithmic scale. The downward bend
at large $\Delta E^\prime_\mathrm{exp}$ is due to the finite momentum acceptance of the spectrometer.
The lower panel shows the ratio of the integrals over the two distributions which, finally,
gives the experimental cross section: Over the wide range of cut-off energies the result depends
on $\Delta E^\prime_\mathrm{exp}$ by less than $10^{-3}$.

\subsection{Determination of the experimental cross sections}
\label{chptana}
In this section we discuss the determination of the number of events $A$, the background $B$, and of the luminosity $\cal{L}$ for Eq.\ (\ref{eqsigma}) and of some further corrections and anomalies.
\subsubsection{Data selection}
Electrons scattered elastically without the emission of a photon have an energy 
\begin{equation}
E^\prime\left(\theta\right)=\frac{E}{1+\frac{E}{m_p}\left(1-\cos\theta\right)}.
\end{equation}
Internal and external bremsstrahlung as well as ionization reduce the energy of the detected electron. In order to identify the elastic scattering, one defines
\begin{equation}
\Delta E^\prime_\mathrm{exp}=E^\prime\left(\theta_\mathrm{exp}\right)-E^\prime_\mathrm{exp},
\end{equation}
the difference of the detected energy $E^\prime_\mathrm{exp}$ to the energy calculated from the detected scattering angle, $E^\prime(\theta_\mathrm{exp})$. 

\begin{figure}[bt!]
\begin{center}
 \includegraphics{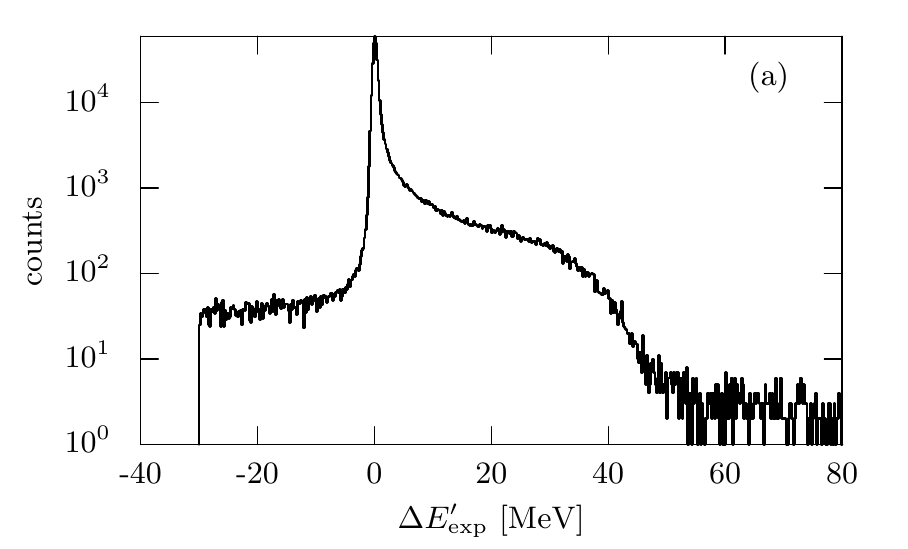}
  \includegraphics{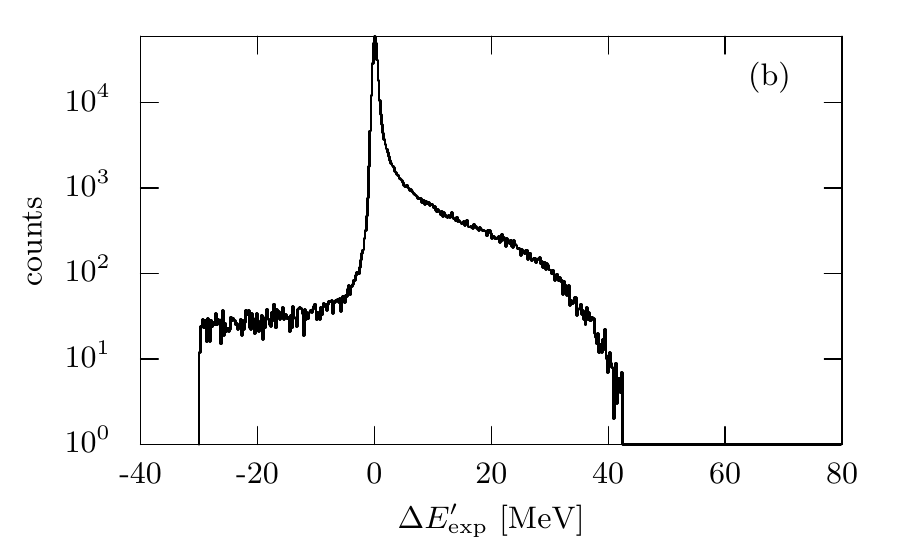}
  \includegraphics{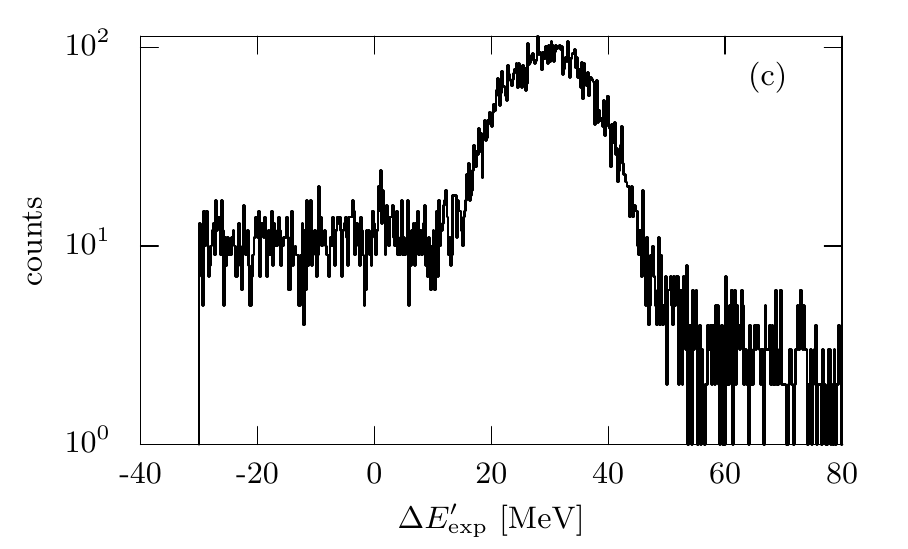}
 \end{center}
\caption[Effect of cuts on $\Delta E^\prime $]{\label{figcuts} Typical $\Delta E_\mathrm{exp}^\prime$ spectrum for a measurement with spectrometer A at $90.5^\circ$ at an incident energy of $E=585$~$\mathrm{MeV}$. (a) Spectrum without any cut. (b) After all cuts except a cut in $\Delta E_\mathrm{exp}^\prime $. (c) Events that are rejected by the cuts. Random events give rise to the nearly constant level between $-$30 and 15~MeV. The bump around 30\ MeV originates from events detected near the edges of the detector plane.}
\end{figure}

The angular acceptance of the spectrometer is given by the collimator. In order to reduce the background we apply cuts on the reconstructed in-plane ($\Phi_0$) and out-of-plane ($\Theta_0$) angles relative to the central trajectory. The cuts are chosen to be  outside of the nominal acceptance so the acceptance is still defined by the geometry. However, these cuts suppress badly reconstructed tracks and noise hits.
The momentum acceptance is not well defined by the physical construction of the spectrometer. We therefore apply cuts on $\Delta p_c$, i.e., the deviation of the particle momentum from the spectrometer's central momentum, to have a well defined momentum acceptance.  Figure \ref{figcuts} shows a typical measured spectrum of  $\Delta E^\prime_\mathrm{exp} $ and the effect of these cuts, which are summarized in Table \ref{tblcuts}.
To select the elastic reaction for the extraction of the cross section, we apply a cut in $\Delta E^\prime_\mathrm{exp}$  accepting events only in the elastic peak region up to a certain cut of $\Delta E^\prime_\mathrm{max}$. 

\begin{table}
\begin{ruledtabular}
\begin{tabular}{l||c|c|c}
Cut & spec. A & spec. B & spec. C\\
\hline
$\Delta p_c$ & $-10\%<\Delta p_c<9\%$ & $\left|\Delta p_c\right|<7.3\%$ & $\left|\Delta p_c\right|<12.3\%$ \\
$\Phi_0$ &  $\left|\Phi_0\right|<6.5^\circ$ &  $\left|\Phi_0\right|<3^\circ$ &  $\left|\Phi_0\right|<6.7^\circ$ \\
$\Theta_0$ &  $\left|\Theta_0\right|<5^\circ$ &  $\left|\Theta_0\right|<3^\circ$ &  $\left|\Theta_0\right|<6.5^\circ$ \\
$y_\mathrm{snout}$ & -- & $\left|y_\mathrm{snout}\right| < 30\ \mathrm{mm}$ & --\\
\end{tabular}
\end{ruledtabular}
\caption[Overview of analysis cuts]{\label{tblcuts}Overview of the cuts used in the analysis. We apply a cut in $\Delta p_c$ to define our momentum acceptance. Additionally, events with reconstructed in-plane ($\Phi_0$) and out-of-plane ($\Theta_0$) angles outside of the acceptance are cut to suppress background. For spectrometer B, an additional cut removes trajectories which stem from scattering off the inner surfaces of the spectrometer snout.}
\end{table}

\begin{figure}
\begin{center}
 \includegraphics{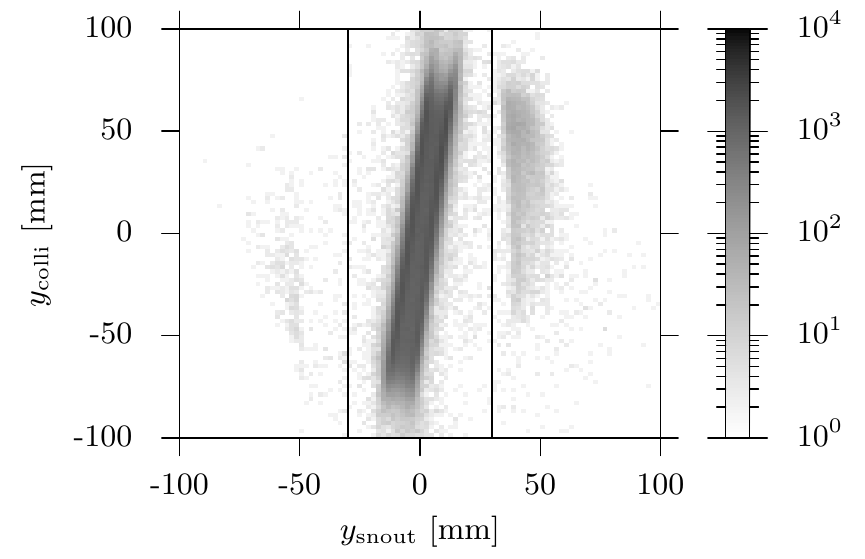}
 \end{center}
\caption[Snout-scattering] {\label{figsnoutcolli} Histogram of $y_\mathrm{colli}$ versus $y_\mathrm{snout}$ of the measurement with spectrometer B at $32.5^\circ$ at an incident energy $E=180\ \mathrm{MeV}$. The events around $y_\mathrm{snout}=0$ correspond to good events, and the sidebands result from back scattering from the snout walls. The black vertical lines indicate the cut used in the analysis. The gray scale is logarithmic to emphasize the events in the left and right sideband. The diagonal ridges seen in the good events are caused by scattering off the target walls.}
\end{figure}

\begin{figure}
\begin{center}
 \includegraphics{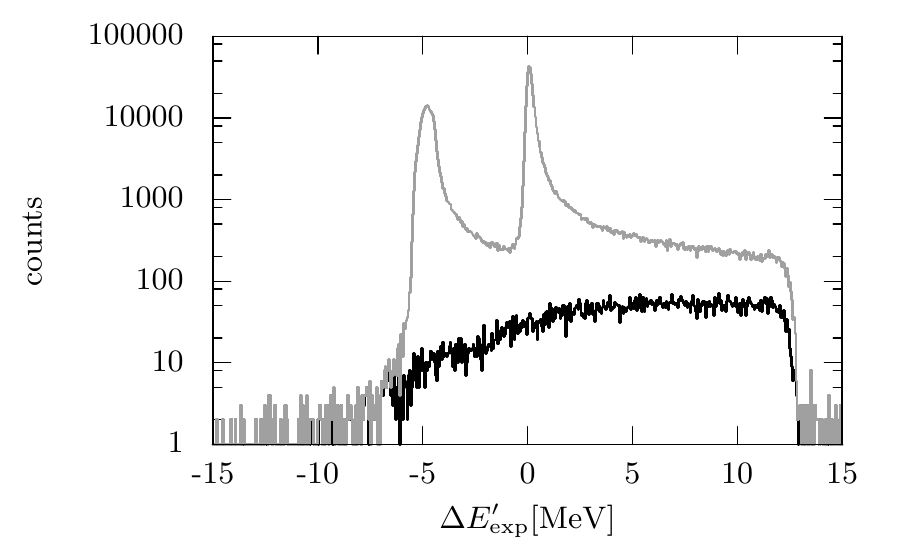}
 \end{center}
\caption[Effects of snout-scattering in $\Delta E^\prime_\mathrm{exp}$] {\label{figsnoutcut} $\Delta E^\prime_\mathrm{exp}$ histogram of the same measurement as shown in Fig.\ \ref{figsnoutcolli}. The events cut away (black) contribute over the complete acceptance with an increase toward smaller energies. The peak at $-5\ \mathrm{MeV}$ (gray curve) originates from elastic scattering off the entrance and exit walls of the cryogenic target cell.}
\end{figure}

For spectrometer B, an additional cut has to be applied. With this spectrometer particles can be detected whose initial trajectory between target and spectrometer lies outside the acceptance defined by the collimator. These particles hit the long snout in front of the collimator and may be scattered back into the acceptance and arrive at the focal plane. They are identified by the horizontal coordinate at the entrance of the snout, $y_\mathrm{snout}$. Figure \ref{figsnoutcolli} shows a two-dimensional histogram of the horizontal coordinate at the collimator, $y_\mathrm{colli}$, versus $y_\mathrm{snout}$. The events around $y_\mathrm{snout}=0$ correspond to good events. On the other hand, one identifies a shadow on the right and a dimmer shadow on the left side belonging to events from snout scattering. In the final analysis, a cut with $\left|y_\mathrm{snout}\right|<30\ \mathrm{mm}$ was applied. As one can see in Fig.\ \ref{figsnoutcut} showing the distribution of the events that are suppressed by this cut, the electrons scattered by the entrance snout are located in the region of the radiative tail.

\subsubsection{Determination of resolution and central momentum}
\label{chptexpsimmatch}
In the simulation, the accuracy of the determination of the particle coordinates is parameterized in the target reference frame, i.e., the simulation contains parametrizations for the resolution in the vertex position, in the momentum and in the in-plane and out-of-plane angles. These parametrizations include parameters for the width of the distributions, which depend on the kinematics and which are determined for each run individually. 

It is particularly important that the peak position in the $\Delta E^\prime_\mathrm{exp}$ histogram matches between experiment and simulation, otherwise the cut in $\Delta E^\prime_\mathrm{exp} $ would fail to select the equivalent part of the peak and tail region in both the experiment and the simulation. This is governed by the knowledge of the central momentum of the spectrometer, determined by the strength of the magnetic field. 

The determination of the central momenta and resolutions is done in a two-step process.
The first step is to find the vertex resolution and, together with this, a possible target offset. For this, a standard nonlinear least-squares optimization is performed: For each variation of the parameters, the simulation is run, and the vertex histograms of experiment and simulation are compared. Since the reconstruction of the measured data is not dependent on these values, only the simulation has to be updated at each fit iteration.

In the second step, the central momentum and the remaining resolutions are optimized. To this end, the spectra of $\Delta E^\prime_\mathrm{exp} $, of the angles, and of $\Delta p_c$ (the momentum relative to the central momentum) are compared. Since the central momentum value affects the reconstruction of the electron kinematics from the measured data, both data analysis and simulation have to be updated at each step.

\subsubsection{Background subtraction}
\label{chptbackgroundsubtract}
As mentioned in Sec.\ \ref{secexperiment}, the liquid hydrogen is contained in a cryo cell. The electron beam has to pass through the walls of this cell, a $10$-$\mathrm{\mu m}$-thin foil made of HAVAR, an alloy of several metals, and also possibly pass through a layer of snow. Scattering off the wall or snow nuclei produces background in the energy region of the elastic peak of the hydrogen, both from the radiative tail from elastic scattering and from quasi-elastic scattering.

The shapes of the background from both elastic and quasi-elastic scattering are taken from simulations based on a physical model, while their amplitude is fitted to the data. As verified by empty target measurements, the inelastic peaks are either so small that they can be ignored ($\leq$ 0.035\% of the elastic hydrogen peak in the worst case) or they are outside the region around the elastic hydrogen peak accepted for the cross-section determination.  In the latter case, the region of the inelastic peak is excluded from the fit of the background amplitudes.

\begin{figure}[t!]
\begin{center}
 \includegraphics{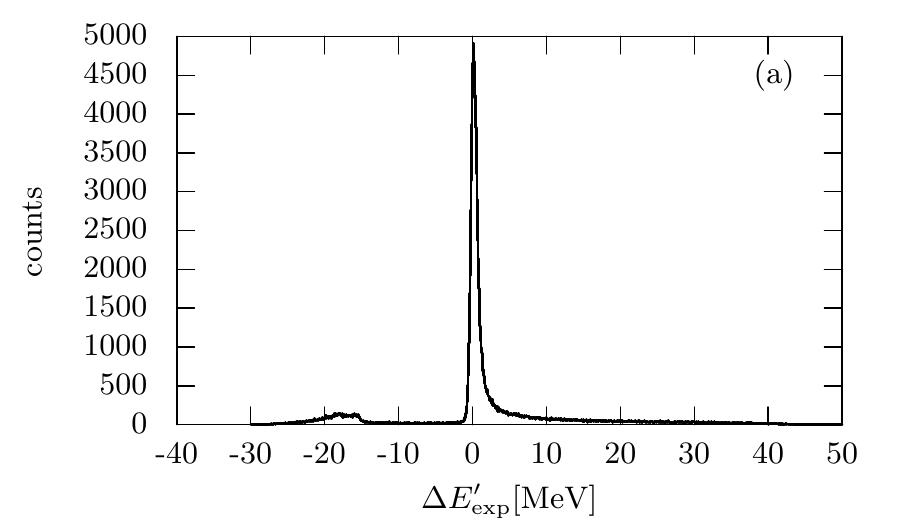}
  \includegraphics{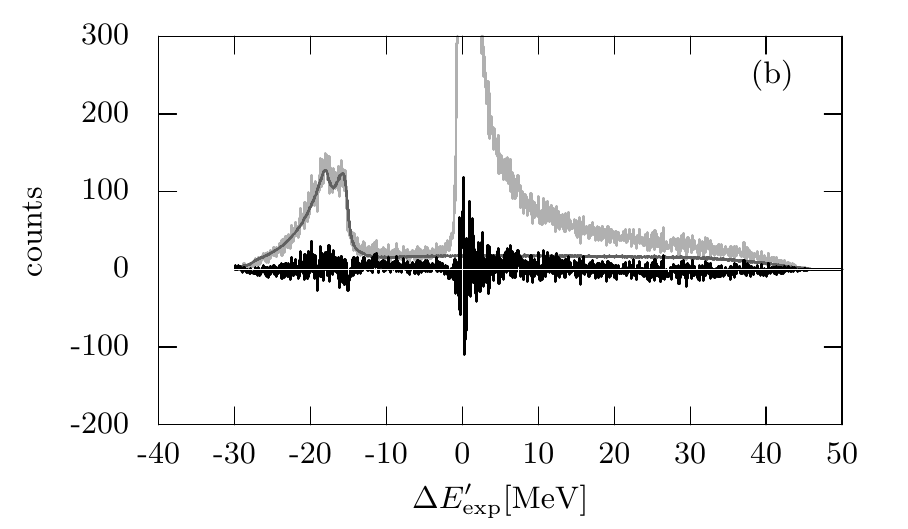}
  \includegraphics{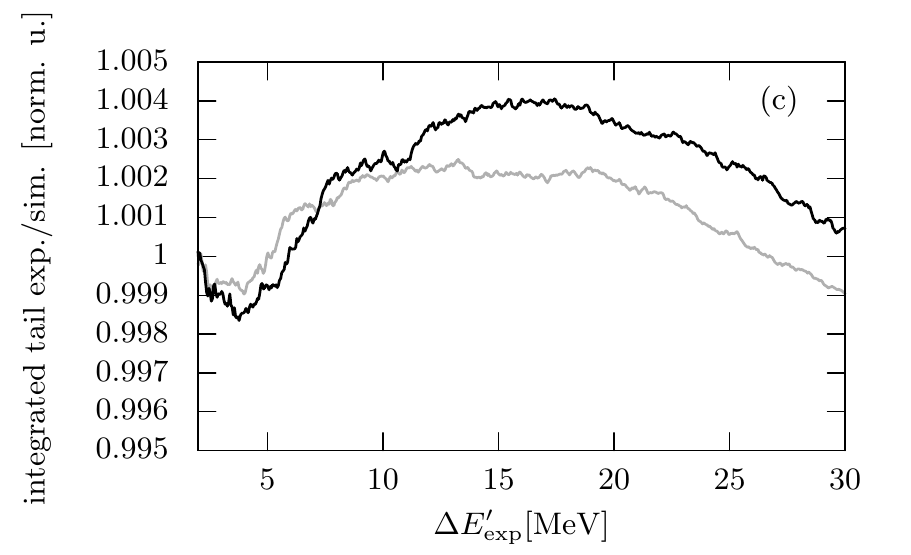}
 
\end{center}
\caption[Background subtraction] {\label{figbackpeak} (a) $\Delta E^\prime_\mathrm{exp}$ spectrum measured with spectrometer A at $28^\circ$ and 450-MeV incident beam energy. (b) Same spectrum (light gray),  background estimate (dark gray), and difference of the data to the sum of simulated hydrogen peak and background (black). (c) The ratio of the integration of the peak in data and simulation as a function of the right limit, i.e., of the cut-off energy. Black: First of two measurements, same data as the upper two panels. Gray: Second measurement. The curves are scaled to start at 1. } 
\end{figure}

Figure \ref{figbackpeak} displays a measured spectrum and the difference spectrum, i.e., the data histogram minus the three simulated and scaled spectra, elastic off hydrogen, elastic off wall or snow atoms, and quasi-elastic off wall or snow atoms. One sees the excellent agreement in the region of the radiative tail from elastic scattering off the proton, while there are slight imperfections at $\Delta E^\prime_\mathrm{exp}=0$ around the steep fall offs of the hydrogen peak, which, however, level out to zero in the integral (cf.\ Fig.\ \ref{figbackpeak} bottom). The first measurement at this kinematic, shown in the upper plots and in black in the bottom plot, has about half of the statistics of the second measurement, which was taken shortly after. The depression close to the peak is caused by a minuscule offset in the peak position---the second measurement is better in this regard. It has to be noted that for both measurements, the integral ratio varies less than $\pm0.15\%$ between 7 and 30~MeV. (N.B.: In Ref.\ \cite{Bernauer:2010wm}, the same data were erroneously labeled differently.)

\subsubsection{Luminosity}
\label{lumi}
In order to determine the integrated luminosity, the beam current is continuously measured during data taking with a fluxgate magnetometer. At the start and end of each run, the beam current measurement is compounded with the pA-meter measurement on the collimator as described above. This reduces the uncertainty for small currents and short runs.

The target density is calculated from continuous pressure and temperature measurements. In order to avoid local overheating of the liquid hydrogen due to the heat load of the passing electron beam, at high currents the beam is rastered over the (curved) frontal face of the target. The small change in the effective target thickness due to this rasterization is accounted for by the simulation.
Still, the absolute length of the cooled-down cryo cell is hard to determine to better than 1\%. Uncertainties in its determination, however, enter as a constant factor in the normalization which will be taken as a fit parameter as already mentioned. 

For all measurements, one of the three spectrometers was used as a luminosity monitor, i.e., this spectrometer stayed with the same field at the same angle, thus measuring the count rate for a fixed momentum transfer for a time where many runs at different angles were taken with the other spectrometers, i.e., for one set of runs. This spectrometer thus monitors the relative luminosity. 
In the course of the measurements at one energy, only a few changes of the monitor angle are necessary to ensure that its event rate is high enough.\\
Each measurement of the luminosity monitor is analyzed in the same way as the normal cross section measurements, that is, the normal procedure of background subtraction, dead time correction and normalization to the estimated luminosity is performed. 

From the $n$ individual luminosity results in a set of runs, the average cross section is calculated and the  cross section values $\sigma_{\mathrm{exp},i}$, measured with the other spectro\-meters, are now normalized:
\begin{eqnarray}
\sigma_{\mathrm{exp,norm},i}&=&\sigma_{\mathrm{exp},i}\cdot\frac{\sigma_\mathrm{lum,avg}}{\sigma_{\mathrm{lum},i}}.
\end{eqnarray}
Hence, the common factors in the calculation of $\sigma_\mathrm{exp,i}$ and $\sigma_{\mathrm{lum}},i$, i.e., beam current, target density, and target length, cancel out and uncertainties in their determination play no role apart from the set-to-set normalization taken as a fit factor in the final analysis.

In this procedure, the statistical error of the normalized data is enhanced by the statistical error of the luminosity measurement,
\begin{equation}
\frac{\Delta\sigma_{\mathrm{exp,norm},i}}{\sigma_{\mathrm{exp,norm},i}}=\sqrt{\left(\frac{\Delta \sigma_{\mathrm{exp},i}}{ \sigma_{\mathrm{exp},i}}\right)^2+\left(\frac{\Delta \sigma_{\mathrm{lum},i}}{\sigma_{\mathrm{lum},i}}\right)^2}.
\end{equation}

This method provides a stable and precise determination of the relative normalization. It is therefore applied to all data points with exception of the 315-MeV data set. In the analysis of this subset, an error was found in the setup of spectrometer C, which was used as the luminosity monitor for most of these data. Therefore, for this subset we rely on the luminosity provided by the pA meter, as described in Sec.\ \ref{pameterdesc}. This leads to slightly larger uncertainties (cf.\ Sec.\ \ref{ptpebcs} ). Compared to the pA-meter-based luminosity, we observe changes with a one-$\sigma$ width of about 0.1--0.2\% for energies of 450~MeV and higher. We find a similar width for the higher currents at 180~MeV. For the lowest currents, the pA-meter measurement and the beam current itself are more noisy and the corrections have a width of about 0.6\%.

As described in Sec.\ \ref{dcslum}, the absolute normalization will be determined by fitting normalization constants to groups of cross sections, determined in a first step with the measured beam currents and target thickness and corrected by the relative luminosity measurements. It is reassuring, that the normalizations of the different sets of measurements are in the expected range of $\pm$3.5\% and that they do not depend on the form factor model used in the fit in any significant way. The largest model spread of the normalization constants is 0.26\% for the flexible models discussed below and 0.74\%, including the Friedrich-Walcher model. This large difference occurs for cross sections with the highest $Q^2$ and is caused by the different high-$Q^2$ behavior of this model (see Sec.\ \ref{moddep}). The average standard deviation is 0.074\%.

\subsubsection{Further corrections and anomalies}
The statistical precision achieved in this experiment together with the conceptual design of overlapping acceptances made several anomalies apparent, which would have been missed in a traditional type of experiment. In the course of the analysis it was found that the acceptances of spectrometers A and C are not completely given by the sheer geometry of the collimators, instead, they depend to some extent on the vertex position. A corresponding correction has been developed and implemented in the simulation software. 

Furthermore it was found that the stray magnetic field of spectrometer C influences the measurement with spectrometer B when the spectrometers are close to each other. In a dedicated beam time, a correction formula has been determined which has been applied to the data.  Further details can be found in Ref.\ \cite{bernauerphd}.

\section{Data analysis}
\subsection{ World data basis}

\label{chworld}
In order to provide a coherent analysis and fit of all world data based on the methods developed by this work we included all readily available data related to the determination of the proton form factors known up to today. The inclusion  of experiments with polarized electrons \cite{Ron:2011rd,Crawford07,Gayou01,Gayou02,Puckett:2011xg,Jones00,Jones06,MacLachlan06,Meziane:2010xc,Milbrath98,Pospischil01,Puckett10,Punjabi05,Ron07,Zhan11} allows us, in particular, to extract a phenomenological two-photon effect. To this end we simply add the published form factor ratios to the database, with systematic and statistic errors added linearly. Some of the listed references are reanalysis of older measurements---we then only include the updated values.

For Rosenbluth-type measurements \cite{Christy04, Janssens65, Qattan,Sill93,Simon80,Walker,Andivahis,Borkowski74,Borkowski74-2,Goitein70,Litt69,Price71,Bosted90,Rock92,Stein75} we include the cross sections instead of extracted form factors in order to make full use of the available information without any bias. We use the quoted statistical errors and let the normalization of the different data sets float. These data sets were taken over the time period of several decades and naturally include a diverse set of radiative corrections. We divided the old corrections out and applied  the updated corrections as far as possible to a common standard, following the Maximon-Tjon prescription \cite{Maximon2000}, with the extension to muon and tau loops and exponentiation of the correction to account for higher orders, matching the treatment of the new data as close as possible.

In Refs.\ \cite{Christy04, Janssens65, Qattan,Sill93,Simon80,Walker}, the radiative corrections are based on Ref.\ \cite{MoTsai}, in different variations, and they are straight forward to update. Reference \cite{Andivahis} also belongs to this group. For this data set, we include two normalization constants; the data were taken with two different spectrometers and the two subsets show a clear normalization mismatch in the overlap.

The corrections in Refs.\ \cite{Borkowski74,Borkowski74-2,Goitein70,Litt69,Price71}  are based on Ref.\ \cite{MeisterYennie}. For the two Borkowski {\em et al.}\ data sets \cite{Borkowski74,Borkowski74-2} we assume independent normalizations and include a normalization parameter for each data set.

For Ref.\ \cite{Bosted90,Rock92,Stein75} the radiative corrections could not be updated since no details on the applied corrections could be obtained.  

\subsection{Form factor models}
 \label{chptmodels}

For a direct fit of the measured cross sections, an ansatz has to be made for the description of the form factors.
Since the true form-factor functional form is unknown, we have to rely on a subjective choice, possibly introducing a bias. However, we can reduce the impact of this by employing a wide variety of models. The model bias can then be judged in this context.

In the frequentist picture, each of these models, together with a specific choice of the parameters, constitutes a hypothesis we can test against the data. For each model we then choose the parameter set with the highest $p$-value via a least-squares fitter. From the goodness of fit of the different models, we can rule out some of them; however, this actually leaves the domain of the strict frequentist view.
In the Bayesian picture, a selection of a model constitutes a prior. From the infinite function space, we reduce our selection to those which are representable by the model, attributing zero probability to all the others. Additionally, the fit essentially assumes a flat prior for the probability distribution for the parameters.

 In the following the models used in this work will be discussed. For the magnetic form factor, $G_M$, the factor $\mu_p$ has been suppressed to improve readability. All models are normalized to 1 at $Q^2=0$. This will be used by the fit to fix the global normalization.

\subsubsection{Dipole}
The designation ``standard dipole,''
\begin{equation}
G_\mathrm{standard\ dipole}(Q^2)=\left(1+\frac{Q^2}{0.71\mathrm{GeV}^2}\right)^{-2}
\end{equation}
was coined by Hand {\em et al.}\ \cite{hand63}. For a long time it was the accepted form for the electric form factor of the proton and---scaled with $\mu_{p/n}$--- also for the magnetic form factor of both the proton (``scaling relation'') and the neutron, and it is today found in many text books (e.g., Ref.\ \cite{povh04}). While the choice of the dipole form was originally purely phenomenological, the related exponential falloff in $r$-space comes about as the probability function of a quantum mechanical particle trapped in a narrow potential well.

In the present analysis, the scaling relation is not enforced. Instead different parameters for the electric and magnetic form factor are used,
\begin{equation}
\label{singledipole}
G^{E,M}_\mathrm{dipole}(Q^2)=\left(1+\frac{Q^2}{a^{E,M}}\right)^{-2}.
\end{equation}
With only two free parameters, $a^E$ and $a^M$, this model is very rigid, and it will be seen that it is not able to describe the data of this experiment, as was the case already for earlier data (e.g., Simon {\em et al.}\ \cite{Simon80}).

\subsubsection{Double dipole}\label{ddipol}
A somewhat more flexible ansatz consists of the sum of two dipoles,
\begin{align}
\label{doubledipole}
G^{E,M}_\mathrm{double\ dipole}(Q^2)=& 
a^{E/M}_0\left(1+\frac{Q^2}{a^{E/M}_1}\right)^{-2}\nonumber\\
&+\left(1-a^{E,M}_0\right)\left(1+\frac{Q^2}{a^{E/M}_2}\right)^{-2}.
\end{align}

\subsubsection{Polynomials}
\label{chptp}
\paragraph{Simple polynomial.}

A polynomial is a simple model without theoretical idea of the nature of the form factors except some level of continuity or smoothness. The constant term is fixed to 1 by the normalization. With a polynomial of the order $n$, the form factors are parameterized as follows:
\begin{equation}
G^{E,M}_\mathrm{polynomial,n}(Q^2)=1+\sum_{i=1}^n a^{E,M}_i Q^{2\, i}.
\end{equation}
Since the form factors drop rapidly with $Q^2$, high orders are needed to describe them adequately over a larger $Q^2$ range. 

\paragraph{Polynomial $\times$ dipole.}\label{polymaldipol}
\label{chptpmd}
In order to free the polynomial from the necessity to describe the gross behavior of the form factors, the latter may be accounted for by multiplying the polynomial by the standard dipole as follows:
\begin{align}
&G^{E,M}_{\mathrm{polynomial} \times \mathrm{dipole,n}}(Q^2)=\nonumber\\
&\hspace{1em} G_\mathrm{standard\ dipole}(Q^2)\times \left(1+\sum_{i=1}^na^{E,M}_i Q^{2\, i}\right).
\end{align}
In principle, it would be possible to optimize also the parameter of the dipole. It was found, however, that this additional freedom does not improve the fits and has a high computational cost. 

\paragraph{Polynomial + dipole.}
\label{chptpad}
A variation of the aforementioned splitting-off of the gross behavior of the form factors is the sum of a polynomial and the standard dipole instead of the product,
\begin{align}
&G^{E,M}_{\mathrm{polynomial} + \mathrm{dipole,n}}(Q^2)=\nonumber\\
&\hspace{1em}G_\mathrm{standard\ dipole}(Q^2) + \left(\sum_{i=1}^na^{E,M}_i Q^{2\, i}\right).
\end{align}
While the multiplication parameterizes the relative deviation from the standard dipole, the sum parametrizes the absolute deviation.  

\paragraph{Inverse polynomial.}
\label{chptip}
A variation of the polynomial model is the inverse polynomial ansatz as in Ref.\ \cite{Arrington03},
\begin{equation}
G^{E,M}_\mathrm{inv.\ poly.,n}(Q^2)=\frac{1}{1+\sum_{i=1}^n a^{E,M}_i Q^{2\, i}}.
\end{equation}

\subsubsection{Splines}
\label{chptspline}
In all other models described in this section, the behavior of the function in different $Q^2$ regions is highly correlated. Therefore, possible shortcomings in the description of the data in one $Q^2$ region may influence negatively the description in other regions. Functions that decouple the behavior in different $Q^2$ regions to a great extent are splines.

A spline ansatz has multiple advantages. Depending on the number of knots, a spline can be very flexible. Nevertheless, the fit converges even for a large number of knots quickly since each parameter essentially affects a limited part of the curve only without long-range biases.

\paragraph{Cubic spline.}
Cubic splines are assembled from polynomials of the third order. Due to the C2 continuity constraint, a cubic spline with $k$ knots ($k-1$ polynomials with four parameters each) has only $k+2$ parameters. The spline segment between the $i$th and $(i+1)$th knot can be written in matrix notation as follows:
\begin{equation}
S_i\left(t\right)=\frac{1}{6}\left[\begin{array}{cccc}t^3&t^2&t&1\end{array}\right]
\left[\begin{array}{cccc}-1&3&-3&1\\3&-6&3&0\\-3&0&3&0\\1&4&1&0\end{array}\right]
\left[\begin{array}{c}p_{i-1}\\p_{i}\\p_{i+1}\\p_{i+2}\end{array}\right].
\end{equation}
Here $t\in\left[0,1\right]$ denotes the position between the two knots $Q^2_i$ and $Q^2_{i+1}$ as follows:
\begin{equation}
t=\frac{Q^2-Q^2_i}{Q^2_{i+1}-Q^2_i}.
\end{equation}
For the fits to our new data alone, we use uniform splines, i.e., constant knot spacing.
To impose the normalization constraint the ansatz is chosen as
\begin{equation}
G^{E,M}_\mathrm{spline}(Q^2)=1+Q^2 S^{E,M}(Q^2).
\end{equation}

\paragraph{Cubic spline $\times$ dipole.}
\label{chptsmd}
Following the same considerations as in Sec.\ \ref{polymaldipol} it is advantageous to multiply the spline ansatz with the standard dipole. This leads to the ansatz
\begin{align}
&G^{E,M}_{\mathrm{spline} \times \mathrm{dipole}}(Q^2)=\nonumber \\
&\hspace{3em} G_\mathrm{standard\ dipole}(Q^2) \times \left(1+Q^2 S^{E,M}(Q^2)\right).
\end{align}

\subsubsection{Friedrich-Walcher parametrization}
\label{chptfw}
In their analysis of the before-2003 world data of the proton and neutron form factors, Friedrich and Walcher \cite{fw03} used an ansatz that is composed of a smooth part and a ``bump.''
The smooth part is identical to the double dipole ansatz as follows:
\begin{align}
&G_S\left(Q^2,a_0,a_1,a_2\right)=\nonumber \\
&\hspace{3em}a_0\left(1+\frac{Q^2}{a_1}\right)^{-2}+\left(1-a_0\right)\left(1+\frac{Q^2}{a_2}\right)^{-2}.
\end{align}
The bump contribution consists of a Gaussian in $Q^2$ with an amplitude $a_b$, position $Q_b$  and a width $\sigma_b$. To suppress odd powers in $Q$ in the Taylor expansion of the Gaussian for $Q_b\neq0$, another Gaussian is added which is mirrored at $Q^2=0$, as has been done by Sick \cite{Sick74} for a model independent analysis of nuclear charge distributions in r-space. The bump contribution is hence described by the following:
\begin{equation}
G_b\left(Q^2,Q_b,\sigma_b\right)=e^{-\frac{1}{2}\left(\frac{Q-Q_b}{\sigma_b}\right)^2}+e^{-\frac{1}{2}\left(\frac{Q+Q_b}{\sigma_b}\right)^2}
\end{equation}
To attribute the full normalization to the smooth part, the bump contribution is multiplied by $Q^2$. The complete model is, therefore,
\begin{align}
&G^{E,M}_\mathrm{Friedrich-Walcher}(Q^2)=\nonumber \\
&G_S\left(Q^2,a^{E/M}_{0,1,2}\right)+a^{E/M}_b\cdot Q^2 G_b\left(Q^2,Q^{E/ M}_b,\sigma^{E/M}_b\right).
\end{align}

\subsubsection{Continued fraction}
A popular model introduced by Sick \cite{Sick03} is the continued fraction ansatz. However, it turns out that this model produces slowly converging fits and the results are difficult to control due to poles in the denominator. While this was studied in some detail, it was not included into the final analysis.

\subsubsection{Extended Gari-Kr\"umpelmann model}
\label{chptgk}
While all previous models are just mathematical procedures with no physical meaning for the description of the data, the extended Gari-Kr\"umpelmann model \cite{Gari92,Lomon01,Lomon02,Lomon06}---actually a group of models which differ in their details---is based on physical considerations. In this work, the version called DR-GK`(1) \cite{Lomon01} (respectively, GKex(01) \cite{Lomon02,Lomon06}) is selected, since it had the best results in Ref.\ \cite{Lomon06} for existing proton form factor data when the normalization of the data sets is not varied.\\
Under the assumption that QCD is the fundamental theory of the strong interaction, the $Q^2$ dependence of the electromagnetic form factors can be calculated in perturbative QCD (pQCD) for very high momentum transfers. For small momentum transfers, the confinement property of QCD leads to an effective hadronic description with vector meson dominance (VMD), the coupling of a photon to a vector meson which itself couples to the nucleon.\\
Earlier models that were based solely on VMD introduced multiple, phenomenological poles of higher mass in addition to the $\rho$, $\omega$, and $\phi$-poles. Gari and Kr\"umpelmann limit the VMD contributions to these three poles but enforce the asymptotic $Q^2$ behavior dictated by the scaling behavior of pQCD by additional terms.\\ 
In the model used here, the dispersion integral approximation of the $\rho$-meson contribution is replaced by an analytical form. The model was extended to include the $\rho'(1450)$ pole (for details see Refs.\ \cite{Gari92,Lomon01}).\\
As mentioned in Sec.\ \ref{chpttfecs}, the Sachs form factors can be rewritten in terms of the Dirac and Pauli form factors $F_1$ and $F_2$, which are preferred by VMD models. Those can be divided into an isoscalar and an isovector component as follows:
\begin{equation}
2F^p_{1,2}=F^{is}_{1,2}+F^{iv}_{1,2},\ \ \ \ \ \ 2F^n_{1,2}=F^{is}_{1,2}-F^{iv}_{1,2}.
\end{equation}
The model GKex(01) is formulated in terms of these  four form factors with the poles for  $\rho$, $\rho'$, $\omega$, $\omega'$, and $\phi$ mesons,
\allowdisplaybreaks
\begin{align}
F_1^{iv}\left(Q^2\right)=&\frac{N}{2}\frac{1.0317+0.0875\left(1+\frac{Q^2}{0.3176~\mathrm{GeV}^2}\right)^{-2}}{\left(1+\frac{Q^2}{0.5496~\mathrm{GeV}^2}\right)}F_1^\rho\left(Q^2\right)\nonumber\\
&+\frac{g_{\rho'}}{f_{\rho'}}\frac{m_{\rho'}^2}{m_{\rho'}^2+Q^2}F_1^\rho\left(Q^2\right)\nonumber\\
&+\left(1-1.1192\frac{N}{2}\frac{g_{\rho'}}{f_{\rho'}}\right)F_1^D\left(Q^2\right),\nonumber\\
F_2^{iv}\left(Q^2\right)=&\frac{N}{2}\frac{5.7824+0.3907\left(1+\frac{Q^2}{0.1422~\mathrm{GeV}^2}\right)^{-1}}{\left(1+\frac{Q^2}{0.5362~\mathrm{GeV}^2}\right)}F_2^\rho\left(Q^2\right) \nonumber\\
&+\kappa_{\rho'}\frac{g_{\rho'}}{f_{\rho'}}\frac{m_{\rho'}^2}{m_{\rho'}^2+Q^2}F_2^\rho\left(Q^2\right)\nonumber\\
&+\left(\kappa_\nu-6.1731\frac{N}{2}-\kappa_{\rho'}\frac{g_{\rho'}}{f_{\rho'}}\right)F_2^D\left(Q^2\right),\nonumber\\
F_1^{is}\left(Q^2\right)=&\sum_{\alpha=\omega,\omega',\phi}\frac{g_{\alpha}}{f_{\alpha}}\frac{m_{\alpha}^2}{m_{\alpha}^2+Q^2}F_1^\alpha\left(Q^2\right)\nonumber\\
&+\left(1-\frac{g_{\omega}}{f_{\omega}}-\frac{g_{\omega'}}{f_{\omega'}}\right)F_1^D\left(Q^2\right),\nonumber\\
\label{eqgekff}
F_2^{is}\left(Q^2\right)=&\sum_{\alpha=\omega,\omega',\phi}\kappa_\alpha\frac{g_{\alpha}}{f_{\alpha}}\frac{m_{\alpha}^2}{m_{\alpha}^2+Q^2}F_2^\alpha\left(Q^2\right)\nonumber\\
&+\left(\kappa_s-\kappa_\omega\frac{g_{\omega}}{f_{\omega}}-\kappa_\omega'\frac{g_{\omega'}}{f_{\omega'}}-\kappa_\phi\frac{g_\phi}{f_\phi}\right)F_2^D\left(Q^2\right).
\end{align}

In this model, the form factors  $F_i^\alpha$ ($\alpha=\rho$, $\omega$, $\omega'$, $\phi$, meson-nucleon) and  $F_i^D$ (quark-nucleon) are parameterized as follows:
\begin{align}
F_1^{\alpha,D}\left(Q^2\right)&=\frac{\lambda_{1,D}^2}{\lambda_{1,D}^2+\tilde{Q}^2}\frac{\lambda_2^2}{\lambda_2^2+\tilde{Q}^2},\nonumber\\
F_2^{\alpha,D}\left(Q^2\right)&=\frac{\lambda_{1,D}^2}{\lambda_{1,D}^2+\tilde{Q}^2}\left(\frac{\lambda_2^2}{\lambda_2^2+\tilde{Q}^2}\right)^2,\nonumber\\
F_1^\phi\left(Q^2\right)&=F_1^\alpha\left(Q^2\right)\cdot\left(\frac{Q^2}{\lambda_1^2+Q^2}\right)^{\frac{3}{2}},\nonumber\\
F_2^\phi\left(Q^2\right)&=F_2^\alpha\left(Q^2\right)\cdot\left(\frac{\lambda_1^2}{\mu_\phi^2}\frac{Q^2+\mu_\phi^2}{\lambda_1^2+Q^2}\right)^{\frac{3}{2}},
\end{align}
with
\begin{equation}
\tilde{Q}^2=Q^2\frac{\ln\left[\left(\lambda_D^2+Q^2\right)/\lambda^2_\mathrm{QCD}\right]}{\ln\left(\lambda_D^2/\lambda_\mathrm{QCD}^2\right)}.
\end{equation}
The parametrization fulfills the normalization constraint for $Q^2=0$.
The constants $\kappa_\nu$ and $\kappa_s$ and the masses $m_\rho$, $m_\omega$, $m_\phi$, $m_{\rho'}$, and  $m_{\omega'}$ are taken as $\kappa_\nu=3.706$, $\kappa_s=-0.12$, $m_\rho=0.776\ \mathrm{GeV}$, $m_\omega=0.784\ \mathrm{GeV}$, $m_\phi=1.019\ \mathrm{GeV}$, $m_{\rho'}=1.45\ \mathrm{GeV}$ and $m_{\omega'}=1.419\ \mathrm{GeV}$. \\
There remain at most 14 free parameters: eight couplings (four $g_\alpha/f_\alpha$, four $\kappa$); four cut-off masses ($\lambda_1$, $\lambda_2$, $\lambda_D$, and $\mu_\phi$); the mass $\lambda_\mathrm{QCD}$, which gives the size of the logarithmic $Q^2$ behavior; and the normalization parameter $N$ for the dispersion relation part of the $\rho$ meson.\\
In Ref.\ \cite{Lomon02}, at most 12 of these parameters were varied, since either the $\omega'$-meson contribution was neglected or  $N$ and $\lambda_\mathrm{QCD}$ were fixed to $N=1$  and $\lambda_\mathrm{QCD}=0.150\ \mathrm{GeV}$, the physical value. The latter constraints are also used in the present work. Still, the fit shows slow convergence and a high time complexity because of the type of mathematical operations used and the mathematical properties of the formulas.

\subsubsection{Other models not described in this paper}
The new high-precision Mainz data were used in several fits not described in this paper, including simultaneous fits to both proton and neutron form factors.
Bauer {\em et al.} \cite{Bauer:2012pv} calculate the electromagnetic form factors of the nucleon to third chiral order in manifestly Lorentz-invariant effective field theory and fit to the Mainz data for the proton and the world data for the neutron. Lorenz {\em et al.} \cite{Lorenz:2012tm}  use a dispersion relation approach to analyze the Mainz data.

\subsection{Fits to cross sections and polarization measurements}
 \label{chptmodsel}
\subsubsection{Fit strategy}
The experimentally determined cross sections are analyzed performing a direct fit of the different models for the form factors. 

For the new Mainz data, the fit minimizes the sum of the weighted deviations squared,
\begin{equation}
M^2=\sum_i\left(\Pi_i\cdot r_i-\frac{\int_{A_i}\left(\frac{\mathrm{d}\sigma}{\mathrm{d}\Omega}\right)_\mathrm{model}\mathrm{d}\Omega}{\int_{A_i}\left(\frac{\mathrm{d}\sigma}{\mathrm{d}\Omega}\right)_\mathrm{std.\ dipole}\mathrm{d}\Omega}\right)^2/(\Pi_i\cdot\Delta r_i)^2.
\label{eq:MM}
\end{equation}
By dividing the measured count rate by the simulated count rate, we have extracted the cross sections from the measurement as ratios $r_i$ to the cross sections for the standard dipole. This is compared to the ratio of  $\left(\frac{\mathrm{d}\sigma}{\mathrm{d}\Omega}\right)_\mathrm{model}$ and $\left(\frac{\mathrm{d}\sigma}{\mathrm{d}\Omega}\right)_\mathrm{std.\ dipole}$, the cross sections calculated 
with the fit model and with the standard dipole, respectively, individually integrated over the acceptance $A_i$ of run $i$. $\Delta r_i$ is the uncertainty of $r_i$ and $\Pi_i$ the product of normalization parameters to be discussed further below. 
As discussed, it is not possible to determine the normalization of the different measurements, i.e., runs, to much better than 2\%, an order of magnitude larger than the statistical errors. However, the overall normalization can be determined from the measured data themselves due to the knowledge of the cross section at $Q^2=0$. Furthermore, the relative normalization of sets of runs, grouped by the relative luminosity determination method described above, are well constrained through the kinematical overlaps of the different sets and can be easily determined as free parameters in the fit. 
Overall, 31 normalization constants $n_j$ are used as free parameters in addition to the model parameters in the fit to a total of 1422 measurements from this experiment (not yet including the external world data). 
These parameters model the overall normalization difference between spectrometers and between different sets of runs.  Each individual measurement by a spectrometer has an individual combination, i.e., a product $\Pi_i$, of a subset of these normalization constants. For example, to a measurement of spectrometer B might belong the product of the overall efficiency of spectrometer B compared to A, and a constant for the normalization of the set this measurement belongs to. All normalization parameters are used in different combinations for a large number of individual measurements.

$M^2$ can be identified with $\chi^2$ only when we compare with a known true theory curve and when the deviations from the true theory curve are Gaussian distributed with a true variance matching our estimated errors $\Pi_i\cdot\Delta r_i$, which one cannot prove. For the problem at hand the true theoretical curve is not known and it
has to be estimated by the best fit curve which is nonlinear in the
fit parameters by construction. As is so often the case in
experimental physics, the frequentist interpretation of $M^2$ is
problematic and the identification $\chi^2 = M^2$ can be only approximate. It is, however, customary to use $\chi^2 \approx M_\mathrm{min}^2$ or, better, the reduced $\chi^2$, $\chi^2_\mathrm{red}=\chi^2/(N_\mathrm{data\ points}-N_\mathrm{parameters})$, as an approximate measure of the quality of the different fit models.

The external cross-section data are included in a similar way. In addition, for each data set, we add a term $[(1-n_{\mathrm{ext},j})/\Delta n_{\mathrm{ext},j}]^2$ with the normalization fit parameter $n_{\mathrm{ext},j}$ and the normalization uncertainty quoted in the corresponding paper, $\Delta n_{\mathrm{ext},j}$. 
For the form factor ratio results from experiments using polarization techniques, we use terms of the form
$[(G_E/G_M)_\mathrm{model}(Q^2_i)-R_i)/\Delta R_i]^2$, where $R_i$ is one of the externally measured ratios and $\Delta R_i$ the quoted error, with systematic and statistic errors added linearly.

\subsubsection{Point-to-point errors beyond counting statistics}
\label{ptpebcs}
Besides the errors from counting statistics, different additional effects contribute to the point-to-point error of the cross sections. These include the dead-time estimation, the uncertainty of the current measurement for the 315-MeV data (see Sec.\ \ref{lumi}), the uncertainty of the  background estimation, and undetected slight variations of the detector and accelerator performance. At this level of precision these effects are hard to quantify, even with direct measurements. In order to estimate these effects we group the data by incident beam energy and by the spectrometer with which the data sre measured. We then inspect for each group the distribution of the deviation of the data points from the fit divided by the error from the counting statistics. These distributions follow a bell-shape curve and it is therefore safe to assume that the point-to-point errors are, to a large extent, also normally distributed. We therefore scale the statistical errors of the different data groups by the width of the bell curve to account for the additional point-to-point uncertainty. This effectively assumes that the additional error contribution scales with the statistics of the individual measurement, which should be true for the dominant source, the background estimation, and also for the dead-time estimation. Iterating the fit with updated scaling factors leads to a meta-stable situation, in which the scaling factors oscillate between two solutions, without significant changes in the fit function. We suppress these oscillations by hand and find a set of factors which, when iterated, are stable but result in a $\chi^2$ slightly larger than 1.  In order to achieve a $\chi_\mathrm{red}^2$ of unity for the best models the statistical errors would need to be increased further by less than 7\%. In view of the smallness of this change, we choose not to apply such a further modification.
The factors are determined using the spline fit, which gives the best $\chi_\mathrm{red}^2$ without any scaling, and the same set of factors is used for all models. 

The scaling factors of the point-to-point errors lie in the range between 1.07 and 1.8 with exception of the 315-MeV data. For these the luminosity could only be determined by the less precise measurement of the beam current and target density, hence the errors had to be scaled by 1.7 to 2.3.

It has to be noted that an overall scaling of the errors by a common factor does not change the best fit. However, since the factors differ for each group, we observe that cross-section values calculated from the fits with and without these error renormalizations change at most by 0.3\%.  In any case, it is possible to rank the fits by the $\chi^2_\mathrm{red}$ when the same set of fixed scaling factors are used for all fits.  This ranking is almost independent of the chosen set of scaling factors---only widely varying scaling factors from data set to data set can change the rank of a fit. Such a large variation in the point-to-point uncertainty can be ruled out from the data.

\subsubsection{Fits without external data}
The spline and polynomial models allow for a varying number of parameters. For the determination of the optimal number, one encounters the basic fact that it is not possible to determine simultaneously which model describes the data and how statistically pure a data sample is. In the extreme case, a model goes through all data points, i.e., it interpolates the data. The choice of the number of parameters is therefore a trade-off: With too few parameters, the model cannot describe the gross behavior of the data and the deduced quantities cannot be trusted; on the other hand, a fit with too many parameters starts to follow local random deviations instead of averaging over fluctuations.

$\chi^2_\mathrm{red}$ as a function of the parameter number reaches a  plateau at around 10 parameters per form-factor model, signaling that the underlying functional shape has been approached and any further reduction in $\chi^2$ is from fitting statistical fluctuations. In each group, the model with the standard dipole multiplied in reaches the plateau with one to two fewer parameters. Interestingly, for the spline models, $\chi^2_\mathrm{red}$ starts to drop again when the parameter number reaches 12; the fits then start to show oscillations at a $Q^2$ above 0.4~$\mathrm{GeV}^2$.

The number of parameters were selected as the lowest number where the plateau was surely reached. While not directly visible in the $\chi^2$ value, the polynomial $\times$ dipole model starts to oscillate at higher $Q^2$ for orders above nine, so an order of eight has been chosen. The inverse polynomial reaches a plateau already with seven parameters. Table \ref{tabparam} summarizes the used parameter numbers.
\begin{table}[h]
\begin{center}
\begin{ruledtabular}
\begin{tabular}{c|c|c||c||c|c}
Poly.&Poly.+dip.&Poly.$\times$dip.&Inv. poly. &Spline&S.$\times$dip.\\
\hline
10&10&8&7&8&7
\end{tabular}
\end{ruledtabular}
\end{center}
\caption[Selected orders for polynomial and spline models]{\label{tabparam}Chosen orders for polynomial and spline models.}
\end{table}

The flexible spline and polynomial models reach $\chi^2$ values below 1600 (for 1422 data points). This is the baseline against which the other models have to be judged. Table \ref{tabchi} lists the achieved $\chi^2$ value and the number of parameters of the different models.
\begin{table}
\begin{center}
\begin{ruledtabular}
\begin{tabular}{c|c|c|c}
Model & $\chi^2$ & Number of param. & $\chi^2_\mathrm{red}$\\
\hline
Single dipole & 3422 & $2\times 1 + 31$ & 2.4635 \\
Double dipole & 1786 & $2\times 3 + 31$ & 1.2893 \\
Polynomial & 1563  & $2\times 10 + 31$ & 1.1399 \\
Poly. + std. dipole & 1563  & $2\times 10 + 31$ & 1.1400 \\
Poly. $\times$ std. dipole & 1572 & $2\times 8 + 31$ & 1.1436 \\
Inv. poly. & 1571 & $2\times 7 + 31$ & 1.1406 \\
Spline & 1565 & $2\times 8 + 31$ & 1.1385 \\
Spline $\times$ std. dipole  & 1570 & $2\times 7 + 31$ & 1.1403 \\
Friedrich-Walcher & 1598 & $2\times 7 + 31$ & 1.1588 \\
ext. Gari-Kr\"umpelmann & 1759 & $14+31$ & 1.2777 \\
\end{tabular}
\end{ruledtabular}
\end{center}
\caption[$\chi^2$ of the models]{\label{tabchi}The achieved total $\chi^2$, the number of parameters (factor 2: two form factors; 31: number of normalization parameters) and $\chi^2_\mathrm{red}$ for the different models. The degree of freedom is given by 1422 minus the number of parameters.}
\end{table}
The single-dipole fit results in a $\chi^2$ of more than 3400 and is therefore excluded.
The double dipole achieves a $\chi^2$ of 1786, which is much closer to the results with the flexible models. Nevertheless, the model dependency analysis (see Sec.\ \ref{moddep}) shows that the extraction of the radius by the double dipole is not reliable and, depending on the exact shape of the true form factor, the deviations of the double-dipole fit from the true value can be large.

The Friedrich-Walcher model reaches a $\chi^2$ that is less than 2.5\% larger than the best flexible model, well below the width of the $\chi^2$ distribution ($\sigma_{\chi^2}\approx 58$); it is therefore included in the analysis.

The extended Gari-Kr\"umpelmann model achieves a $\chi^2$ of 1759, which is only slightly better than the double dipole. This fit is rather unstable and it seems that there are ambiguities in the solutions. Since the calculation and convergence is slow due to the large number of logarithms and the numerical properties of the model, it was not possible to perform a full study of this model. Such a study would require varying the starting conditions and constraints.  For a reliable fit of this model, it may be necessary to fix the 31 normalization parameters obtained with one of the flexible models beforehand. Subsequently, the data could be used for a fit with this model. This has not been pursued further in this work.

\begin{figure}
\begin{center}
 \includegraphics{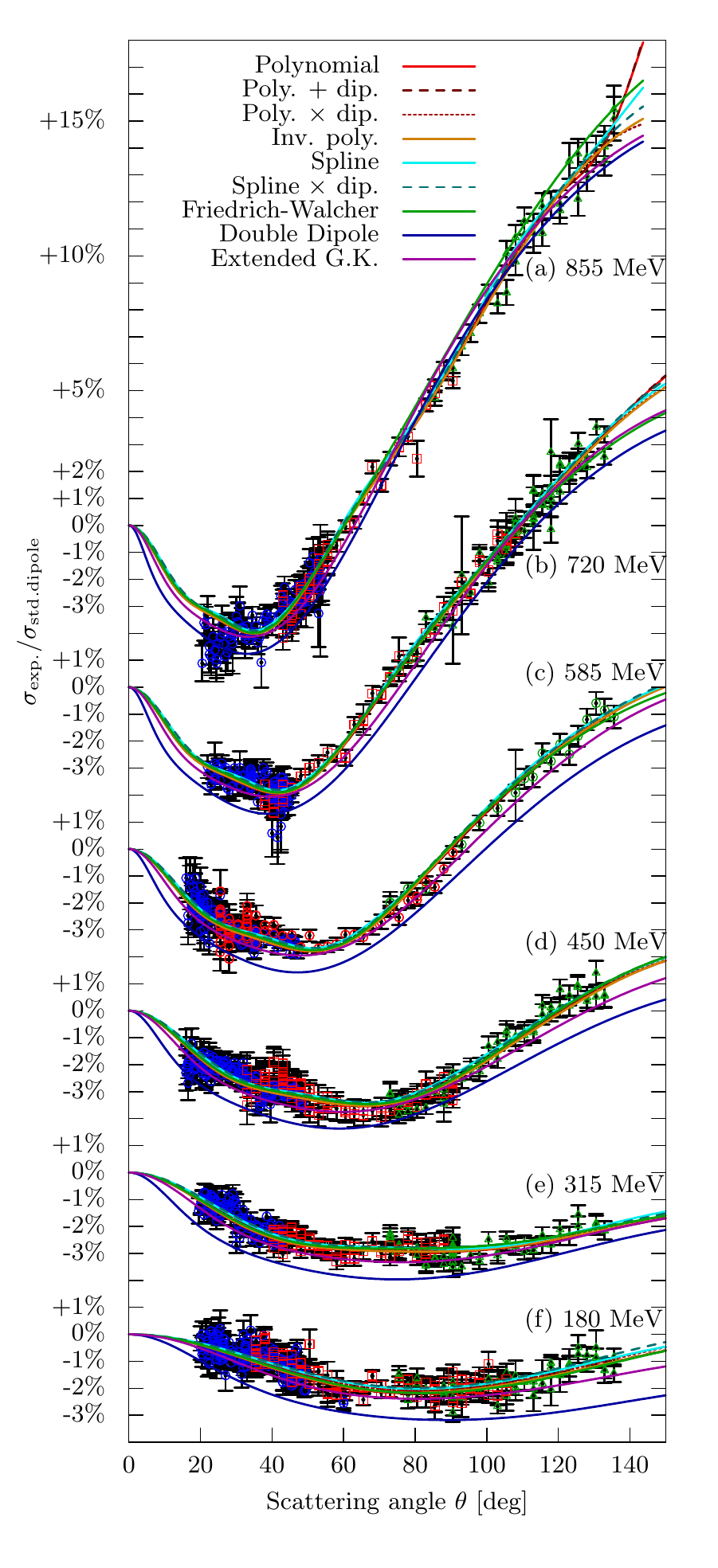}
 \vskip-1cm
\end{center}
\caption{\label{figcs180} (Color) The cross sections and the fits for 855, 720, 585, 450, 315 and 180\ MeV [(a)-(f)] incident beam energy divided by the cross section calculated with the standard dipole, as functions of the scattering angle (red: measured with spectrometer A; blue: spectrometer B; green: spectrometer C). The normalization parameters $n_j$ applied to the measured cross section data are taken from the spline fit. The cross sections of the fits that achieve a good $\chi^2 < 1600$ differ by at most 0.7\%. The normalization parameters $n_j$ from the double-dipole fit would shift the data down by 1.6\% at most. Accordingly, its curve lies below the data with the normalizations from the spline fit.}
\end{figure}

Figure \ref{figcs180} shows the measured cross section divided by the cross section calculated with the standard dipole and with the scaled statistical errors, compared to the different fits. The measured cross sections are normalized with the normalization parameters extracted with the spline fit. The precision is better than 0.4\% (average) per data point. It is noted that the normalization parameters depend slightly on the model.  Therefore, for a comparison of the data to a fitted model, the normalization found in the fit of that model should be used. However, the models that achieve a small $\chi^2$ yield very similar normalizations, so it is reasonable to present the data normalized only to the spline model which has the smallest $\chi^2_\mathrm{red}$. For the flexible models the maximum difference in a normalization parameter is 0.26\% and the average standard deviation is 0.073\%. The largest difference for the ``good'' models occurs for the 855~MeV data, where the (less flexible) Friedrich-Walcher model shifts the data slightly upwards by 0.7\% for the data measured with spectrometer C. 

For the models that do not yield a $\chi^2 < 1600$, i.e., the double-dipole and the extended Gari-Kr\"umpelmann model, the differences in the normalizations are larger (up to 1.6\% in the case of the double-dipole fit).  Both models would shift the cross sections down, therefore both fit curves are below the data with normalizations from the spline fit (see Fig.\ \ref{figcs180}).  

The analyses with the ``good'' models yield cross sections which differ by less than 1\% for almost all of the $Q^2$ range of the data. In the high-$Q^2$ range the fits start to diverge. Above 0.55\ $\mathrm{GeV}^2$ only data from 720 and 855~MeV contribute. Therefore, the separation into $G_E$ and $G_M$ is not well determined. In the $Q^2$ region covered only by 855-MeV data, the allocation of the cross-section strength to the electric or magnetic part is undetermined, giving rise to the larger spread of the models.

\subsubsection{Fits including the world cross section and polarization data; possible two-photon exchange effect}
\label{fwextdata}
The addition of the world data extends the range in $Q^2$ considerably, with data points reaching above $30\ \mathrm{GeV}^2$. However, their uncertainty and density vary widely. Fits with high-order polynomials are therefore problematic and spline fits with a constant  knot spacing small enough to accommodate the low-$Q^2$ data are impossible. We therefore extended the spline $\times$ dipole model to nonconstant knot spacing and placed knots roughly according to the data point density at 0, 0.25, 0.5, 0.75, 1, 1.5, 3, 5, 10, and 40\ $\mathrm{GeV}^2$.  

A fit including only external Rosenbluth data in addition to the new Mainz data results in $\chi_\mathrm{red}^2=2074.64/1810=1.146$, well comparable to the numbers above. 

Including all available data, i.e., also polarization data on the form factor ratio, raises this to
$\chi_\mathrm{red}^2=2282.24/1868=1.222$, a rather large increase in $\chi^2$ of 207.6 for only 58 additional data points. This demonstrates that the difference between the Rosenbluth and polarization methods seen at higher $Q^2$ does not vanish with our floating normalization of the cross-section data. 

The most likely explanation for this discrepancy is the effect of hard two-photon exchange which is believed to have a larger effect on the Rosenbluth separation than on the ratio determined from polarization measurements \cite{Guichon03}. The data basis is not broad enough to disentangle the contributing amplitudes over the whole $Q^2$ range as has been done in Refs.\ \cite{Guttmann:2010au,Borisyuk:2010ep} for a single $Q^2$ point. In fact, the experimental information is just enough to constrain rather simple models. Therefore, we assume a simple linear dependence on $\varepsilon$ which vanishes at $\varepsilon=1$ and a logarithmic dependence on $Q^2$, similarly to Ref.\ \cite{Blunden:2003sp}, as an additional multiplicative term $(1+\delta_\mathrm{TPE} )$ on top of Feshbach's Coulomb correction Eq.\ (\ref{feshbach}), 
\begin{equation}
\delta_\mathrm{TPE}=-(1-\varepsilon) \, a\, \ln(b\,Q^2+1),
\label{eq:TPE}
\end{equation}
where $a$ and $b$ are fit parameters. The global fit of Alberico {\em et
al.}\ \cite{alberico09} uses a
similar approach, with a two-parameter model introduced by Chen {\em et
al.}\ \cite{Chen07}. In contrast to our approach, their model assumes a given $Q^2$
dependence, but gives more freedom in the epsilon dependence.
The fit including our TPE parametrization to all data gives $\chi_\mathrm{red}^2=2151.72/1866=1.153$ now as good as the ``good'' fits above.

We also performed fits of all data with a low order Pad\'e model,
\begin{equation}
G^{E,M}_\mathrm{Pad\acute{e}}(Q^2)=\frac{1+\sum_{i=1}^3 a^{E,M}_i\, Q^{2\cdot i}}{1+\sum_{j=1}^5 b^{E,M}_j\, Q^{2\,j}},
\label{eq:Pade}
\end{equation} 
i.e., the same  parametrization as in Ref.\ \cite{Arrington07}. This model has only 8 instead of 11 parameters per form factor and therefore has much lower flexibility and the achieved $\chi^2$s are considerably higher. Table \ref{wdchitable} gives an overview over the achieved $\chi_{\mathrm{red}}^2$s.

\begin{table}
\begin{center}
\begin{tabular}{c|c|c}
Data base  & Model & $\chi^2_\mathrm{red}$\\
\hline
cross sections& Spline with var. knots &$\frac{2074.64}{1810}=1.146$\\
c.s.+pol. & Spline with var. knots & $\frac{2282.76}{1868}=1.222$\\
c.s.+pol. & '' + TPE& $\frac{2151.72}{1866}=1.153$\\
cross sections& Pad\'e & $\frac{2289.14}{1816}=1.261$\\
c.s.+pol. & Pad\'e & $\frac{ 2465.51}{1874}=1.316$\\
c.s.+pol. & Pad\'e + TPE & $\frac{2360.81}{1872}=1.261$\\
\end{tabular}
\end{center}
\caption{\label{wdchitable} The $\chi^2_\mathrm{red}$ for several fits to 1) all cross section data, i.e., to the new data of this paper and previous cross section world data, 2) all cross section and polarization results without any TPE model, 3) same as 2) with the simple TPE model. All fits where performed for the variable spline model and the Pad\'e model. }
\end{table}
 \subsection{Statistical significance and model dependence \label{significance}}
 Our model for uncertainties divides up all sources into three parts: the first part is the global component, i.e., uncertainties which affect all points in a normalization group the same way. These are automatically corrected for by the floating normalization. The next component affects each measurement individually---we treat those via the scaling of the counting statistics errors described above and in the following subsection. The third, and most critical, component covers effects which change the cross sections in a systematic way. We describe this in Sec.\ \ref{npperrors}. A given source of uncertainty does not necessarily fall into only one of these categories. For example, the detector efficiencies bulk contribution is a global effect, but we also treat this source in the third category. While point-to-point efficiency differences are less probable, they are taken care of by the error scaling.

A source of systematic error is the selection of models used in the fit which needs special attention. We describe our approach in Sec.\ \ref{moddep}.

\subsubsection{Point-to-point uncertainties}
\label{mcbands}
To express the uncertainty in the result due to uncertainties in the data and extraction procedure we must construct a confidence region around the best fit. However, the meaning of this region is a delicate point. 

As described above the uncertainties from counting statistic are scaled to account for point-to-point uncertainties from other sources like the luminosity determination and detector efficiency fluctuations. Classically, such point-to-point  uncertainties are treated by standard error propagation, which assumes a linear approximation. In order to circumvent this limitation, we calculate the confidence bands using a Monte Carlo technique: From the best fit, we generate a large number of pseudo data sets, with data varied according to the individual uncertainties. We additionally vary the normalization of the data groups with a $\pm5$\% uncertainty. Each of these data sets is now fitted, and from the resulting fit ensemble one can construct envelopes as the confidence bands to a selected confidence level.

For the interpretation of this band, it has to be noted that it is dependent on the model function. One uses an implicit prior assumption, namely, that the true curve can actually be expressed by the model function. From this, it is clear that a less flexible model will have a smaller uncertainty; the assumption is then a stronger statement and therefore reduces the uncertainty. It is also clear that it is impossible to define a model-independent band: Without any prior, that is, allowing any function (or distribution), the uncertainty at positions off the exact points of the input data is infinite. 

Per se, standard error propagation and the Monte Carlo method construct the pointwise confidence bands, that is, one expects the value derived from the experiment to be inside the confidence band around the true value at a given $Q^2$ with the specified confidence level without any limitation on the behavior at a different $Q^2$. In the linear approximation, this is commonly reformulated in the not-quite-precise inversion: One expects that the unknown true value is inside the confidence band around the experimentally determined value. In a non-linear model, this inversion is even less accurate, a fact which can not be cured without the introduction of further prior assumptions.

However, with our method, it is possible to overcome the other limitation, i.e., the reduction to single $Q^2$ points. We can construct simultaneous confidence bands, i.e., the statement is extended to express that, with the chosen confidence level, the true function is inside the band for a chosen $Q^2$ range and not just at a single point. These bands are therefore strictly wider than the point-to-point bands of the same confidence level. Assuming the same shape of the bands, the Monte Carlo technique allows us to find scaling factors: for the models at hand, a 68\% simultaneous band is about 2.3 times wider than the pointwise band with the same confidence level. To achieve 95\%, one has to scale the 68\% pointwise bands by a factor of around 3.

In this paper, we present pointwise bands with a confidence level of 68\%, the usual ``one-$\sigma$ errors.''

\subsubsection{Non point-to-point uncertainties}
\label{npperrors}
Besides statistical errors, one has to take into account uncertainties which affect several measurements in a systematic way. 
Most of these are irrelevant as they affect either all points or all points of a set in the same way. For example, an error in the target density or thickness will shift all points up or down. Due to the way the fits are constructed these shifts are subsumed in the fitted normalization constants. A $\pm 5\%$ uncertainty of this normalization scaling is included in the simulation and therefore also in the confidence bands as described above. 
What remains are slow drifts over time or scattering angle which may affect the outcome of the fits. We identified several experimental sources:
\begin{itemize}
\item Energy cut in the elastic tail. This error can be estimated by varying the cut-off energy. It changes the form-factor results by at most 0.2\% for high $Q^2$ and by less than 0.1\% for $Q^2<0.55\ \mathrm{GeV}^2$.
\item Drift of the normalization. This error might occur due to unaccounted dead-time effects in the detectors or electronics when the event rate changes. From the long-term experience with the detector setup, this error on the cross sections is estimated to be below 0.05\%.
\item Efficiency change due to different positions of the elastic peak on the focal plane. The detector efficiency is position dependent because of different wire tension, missing wires or quality of the scintillators. By adjusting the central momentum, the position of the electron trajectories in the focal plane was almost constant. This effect on the cross sections is estimated to be at most 0.05\%.
\item The vertex-dependent acceptance correction for spectrometers A and C. A comparison of the 720-MeV data, measured with the long and short target cells, leads to a cross-section uncertainty below 0.1\%.
\item The influence of spectrometer C on the measurement with spectrometer B. We split this uncertainty in a part which is effectively point to point, reflected by the error scaling, and a part which behaves systematically as a function of the angle. The latter is estimated to be below 0.1\%.
\item The background estimation. Depending on the size of the background below the elastic hydrogen peak this error is estimated to be between 0.1\% and 0.5\%.
\end{itemize}
While the first point can be tested directly by fitting data with varied cut-off energy, the other uncertainties have to be treated by hand. To this end the cross sections are grouped by the energy and by the spectrometer with which they are measured. For each group, we define a linear function $c(\theta)=a\cdot (\theta-\theta_\mathrm{min})$ interpolating from 0 for the smallest scattering angle to the full estimated uncertainty at the maximum angle of the group. The cross sections are then multiplied by $1+c(\theta)$.  The sign of $a$ was kept constant for all energies. The so modified cross sections were then refitted with the form factor models. In order to determine an upper and a lower bound the fits were repeated with negated $a$. The uncertainties found in this way are added quadratically to the uncertainties from the radiative tail cut-off. The choice of a linear function in $\theta$ is certainly arbitrary, but we checked several different reasonable functional dependencies on $\theta$ and $Q^2$ , e.g., imitating the effect of a spectrometer angle offset or target position offset. They all produced similar results. The so-determined uncertainties are reflected by the experimental systematic confidence bands presented in this paper.

A possible source of uncertainty not from data but from theory are the radiative corrections. The absolute value of the radiative corrections should already be correct to better than 1\% and a constant error in the correction will be  absorbed in the normalization. Any slope introduced as a function of $\theta$ or $Q^2$  by the radiation correction will be contained in the slope-uncertainty discussed above up to a negligible residual; it is therefore not considered.

\begin{figure} 
 \includegraphics{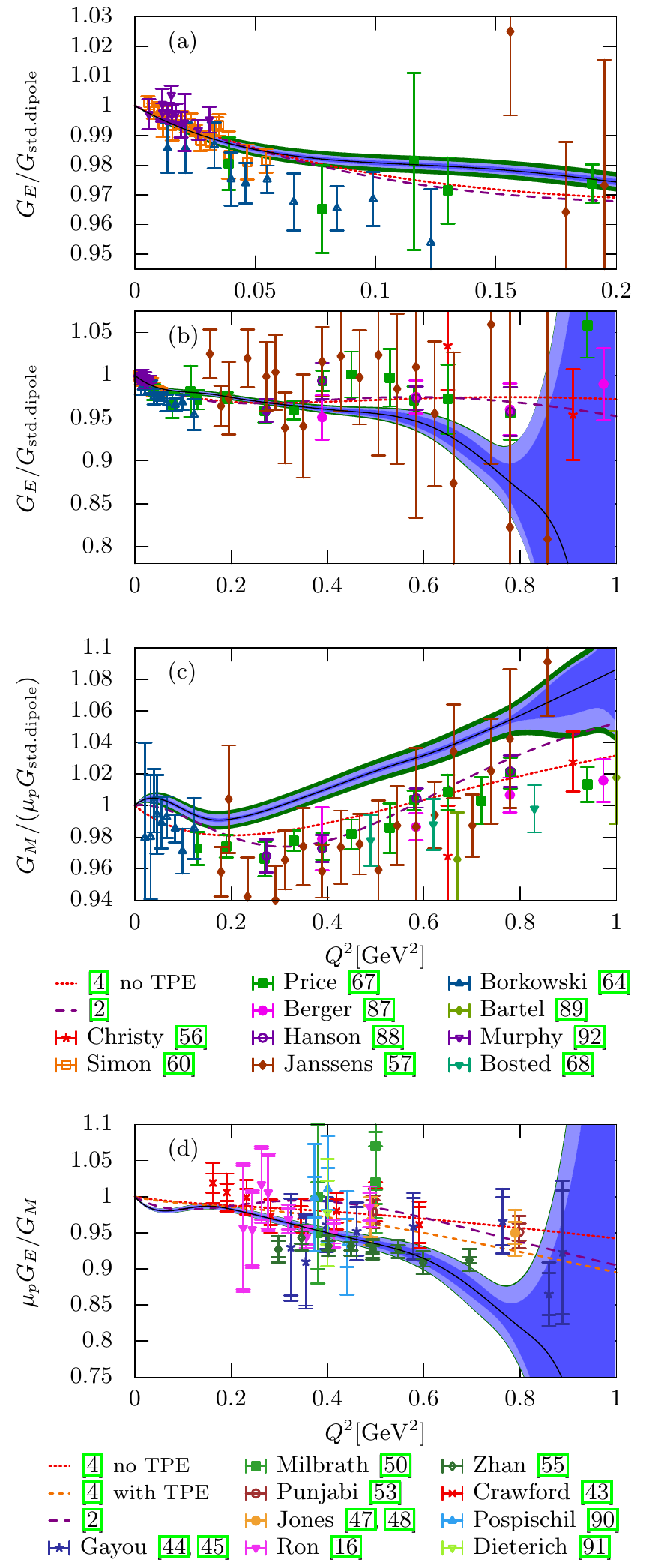}
 \caption{\label{fig_emem} (Color) The form factors $G_E$ and $G_M$, normalized 
  to the standard dipole, and $G_E/G_M$ as a function of $Q^2$. 
  Black line: Best fit to the new Mainz data, blue area: statistical  
  68\%  pointwise confidence band, light blue area: experimental systematic error,  
  green outer band: variation of the Coulomb correction by $\pm 50\%$. 
  The different data points depict the previous measurements \cite{Arrington07,fw03,Christy04,Simon80,Price71,Berger71,Hanson73,Janssens65,Bartel73,Bosted90,Gayou01,Gayou02,Milbrath98,Punjabi05,Jones00,Jones06,Zhan11,Crawford07,Pospischil:2000pu,Dieterich01}
 as in Refs.\ \cite{fw03,Arrington07} with the data points of Refs.\
  \cite{Borkowski74-2,Murphy74,Ron:2011rd} added. }
\end{figure}

In order to evaluate the influence of the applied Coulomb correction, the amplitude of the correction was varied by $\pm 50\%$. The so-modified cross sections are refitted with the different models. The differences of the extracted form factors to the results for the data with the unmodified correction are shown as a band in Fig.\ \ref{fig_emem}.

Except for the phenomenological TPE model included in the fit to the full data set, we do not include any theoretical correction of the hard two-photon exchange to the cross sections in our analysis but apply Feshbach's Coulomb correction.  Published Rosenbluth data normally do not include a Coulomb correction. This has to be considered for comparisons of our fits with old Rosenbluth separations.

\subsubsection{Model dependence}
\label{moddep}
An important issue is the question of whether the form-factor functions are sufficiently flexible to be a suitable estimator for the unknown true curve or whether they introduce any bias, especially in the extraction of the radius. We have studied this problem in two ways:

First, we used a Monte Carlo technique similar to the method described in Sec.\ \ref{mcbands}. We analyzed Monte Carlo data sets produced at the kinematics of the data of the present experiment with a series of published form factors: the standard dipole, the Pad\'e and polynomial descriptions of Refs.\ \cite{Arrington03, Arrington07} and the Friedrich-Walcher parametrization \cite{fw03}. For each model, we produced roughly 50\,000 data sets.
By construction, we know the ``truth'' we have to compare with and can evaluate the suitability of our fits. All flexible models and the Friedrich-Walcher model were able to reproduce the input form factors from these simulated data to a high precision without any notable bias. This is also reflected in the average $\chi^2_\mathrm{red}$ values obtained, presented in Table \ref{tablchi}, which deviate minimally from 1. The inflexible single-dipole model failed as expected for any input model except the standard dipole itself. The double-dipole model reproduces the general shape for most models surprisingly well; however, one cannot extract the radii reliably as can be seen in the Tables \ref{tabradiffge} and \ref{tabradiffgm} listing the bias of the radius extraction. All flexible models exhibit only a small bias here except for the spline for a single input parametrization. These tables also list the $1\sigma$ width of the distributions, i.e., these values are not the error of the bias, but describe what kind of precision one can expect from the model for a single experiment. In that sense, the spline models are more efficient than the polynomial models.

\begin{table}
\begin{ruledtabular}
\begin{tabular}{c|r|r|r|r|r}
Fit model & \multicolumn{5}{c}{Input parametrization}\\
&  \multicolumn{1}{c|}{Std.\ dip.} & \multicolumn{1}{c|}{Arr.03P} & \multicolumn{1}{c|}{Arr.03R} & \multicolumn{1}{c|}{Arr.07} & \multicolumn{1}{c}{F.-W.}\\
\hline
Single dipole          & $1.000$& $2.193$& $2.227$ & $2.230$& $3.216$ \\
Double dipole        & $1.002$& $1.033$& $1.001$& $1.003$& $1.162$ \\
Polynomial             & $1.000$& $1.000$& $1.000$& $1.000$& $1.000$\\
Poly. + dipole         & $1.000$& $1.000$& $1.000$& $1.000$& $1.000$\\
Poly. $\times$ dipole  & $1.000$& $1.000$& $1.000$& $1.000$& $1.000$\\
Inv. poly.             & $1.000$& $1.000$& $1.000$& $1.000$& $1.000$\\
Spline                 & $1.000$& $1.000$& $1.002$& $1.002$& $1.000$\\
Spline $\times$ dipole & $1.000$& $1.000$& $1.000$& $1.000$& $1.000$\\
Friedrich-Walcher      & $1.005$& $1.004$ & $1.004$& $1.004$& $1.002$\\
\end{tabular}
\end{ruledtabular}
\caption[Achieved red. $\chi^2$ in model dependency analysis]{\label{tablchi}The average achieved $\chi^2_\mathrm{red}$  of the different model combinations. Columns: Input parametrizations. Rows: Models used in the fit.}
\end{table}

\begin{table}
\begin{ruledtabular}
\begin{tabular}{c|r@{$\pm$}l|r@{$\pm$}l|r@{$\pm$}l|r@{$\pm$}l|r@{$\pm$}l}
Fit model   & \multicolumn{10}{c}{Input parametrization}\\
& \multicolumn{2}{c|}{Std. dip.} & \multicolumn{2}{c|}{Arr.03P} & \multicolumn{2}{c|}{Arr.03R} & \multicolumn{2}{c|}{Arr.07} & \multicolumn{2}{c}{F.-W.}\\
& \multicolumn{2}{c|}{811} & \multicolumn{2}{c|}{829} &\multicolumn{2}{c|}{868}&\multicolumn{2}{c|}{878}&\multicolumn{2}{c}{860}\\
\hline
Single Dipole          & $0$&$0.7$ & $29$&$1$ & $-6$&$1$ & $-15$&$1$ & $-2$&$1$ \\
Double Dipole          & $0$&$1$ & $10$&$1$ & $0$&$2$ & $3$&$3$ & $81$&$27$ \\
Polynomial             & $0$&$7$ & $0$&$7$ & $0$&$6$ & $0$&$6$ & $0$&$6$ \\
Poly. + dipole         & $0$&$7$ & $-1$&$7$ & $0$&$6$ & $-1$&$6$ & $0$&$6$\\
Poly. $\times$ dipole  & $0$&$5$ & $0$&$5$ & $0$&$4$ & $0$&$4$ & $0$&$5$\\
Inv. poly.             & $-1$&$5$ & $-1$&$5$ & $0$&$5$ & $-1$&$5$ & $0$&$5$\\
Spline                 & $-1$&$3$ & $-1$&$3$ & $-3$&$3$ & $-5$&$3$ & $0$&$3$\\
Spline $\times$ dipole & $0$&$3$ & $1$&$3$ & $-1$&$3$ & $-2$&$3$ & $1$&$3$\\
Friedrich-Walcher      & $0$&$1$ & $3$&$2$ & $-1$&$2$ & $+2$&$3$ & $-1$&$3$ \\
\end{tabular}
\end{ruledtabular}
\caption{\label{tabradiffge} Bias of the different models for the charge radius extraction and the width of the radius distribution. Positive values correspond to an extracted radius larger than the input radius. Values are in atm. }
\end{table}

\begin{table}
\begin{ruledtabular}
\begin{tabular}{c|r@{$\pm$}l|r@{$\pm$}l|r@{$\pm$}l|r@{$\pm$}l|r@{$\pm$}l}
Fit model   & \multicolumn{10}{c}{Input parametrization}\\
& \multicolumn{2}{c|}{Std. dip.} & \multicolumn{2}{c|}{Arr.03P} & \multicolumn{2}{c|}{Arr.03R} & \multicolumn{2}{c|}{Arr.07} & \multicolumn{2}{c}{F.-W.}\\
& \multicolumn{2}{c|}{811} & \multicolumn{2}{c|}{837} &\multicolumn{2}{c|}{863}&\multicolumn{2}{c|}{858}&\multicolumn{2}{c}{805}\\
\hline
Single dipole          & $0$&$0.3$ & $-32$&$0.4$ & $-50$&$0.4$ & $-53$&$0.4$ & $5$&$0.4$ \\
Double dip.          & $0$&$1$ & $12$&$2$ & $2$&$3$ & $3$&$4$ & $49$&$2$ \\
Polynomial             & $-1$&$18$ & $-1$&$17$ & $-1$&$17$ & $-2$&$17$ & $-2$&$17$ \\
Poly. + dip.         & $0$&$15$ & $-1$&$15$ & $-1$&$14$ & $-1$&$12$ & $-1$&$15$\\
Poly. $\times$ dip.  & $-1$&$14$ & $-1$&$14$ & $-1$&$13$ & $-2$&$14$ & $-2$&$14$\\
Inv. poly.             & $0$&$13$ & $0$&$13$  & $0$&$13$  & $0$&$12$ & $0$&$13$\\
Spline                 & $1$&$7$ & $1$&$7$   & $1$&$6$   & $-1$&$7$ & $0$&$7$\\
Spline $\times$ dip. & $0$&$6$ & $0$&$6$   & $-1$&$6$  & $-2$&$6$ & $-1$&$6$\\
F.-W.     & $0$&$2$ & $1$&$5$ & $0$&$6$ & $2$&$5$ & $-1$&$6$ \\
\end{tabular}
\end{ruledtabular}
\caption{\label{tabradiffgm} As Tab. \ref{tabradiffge} but for the magnetic radius.}
\end{table}

\begin{figure}
\begin{center}
 \includegraphics{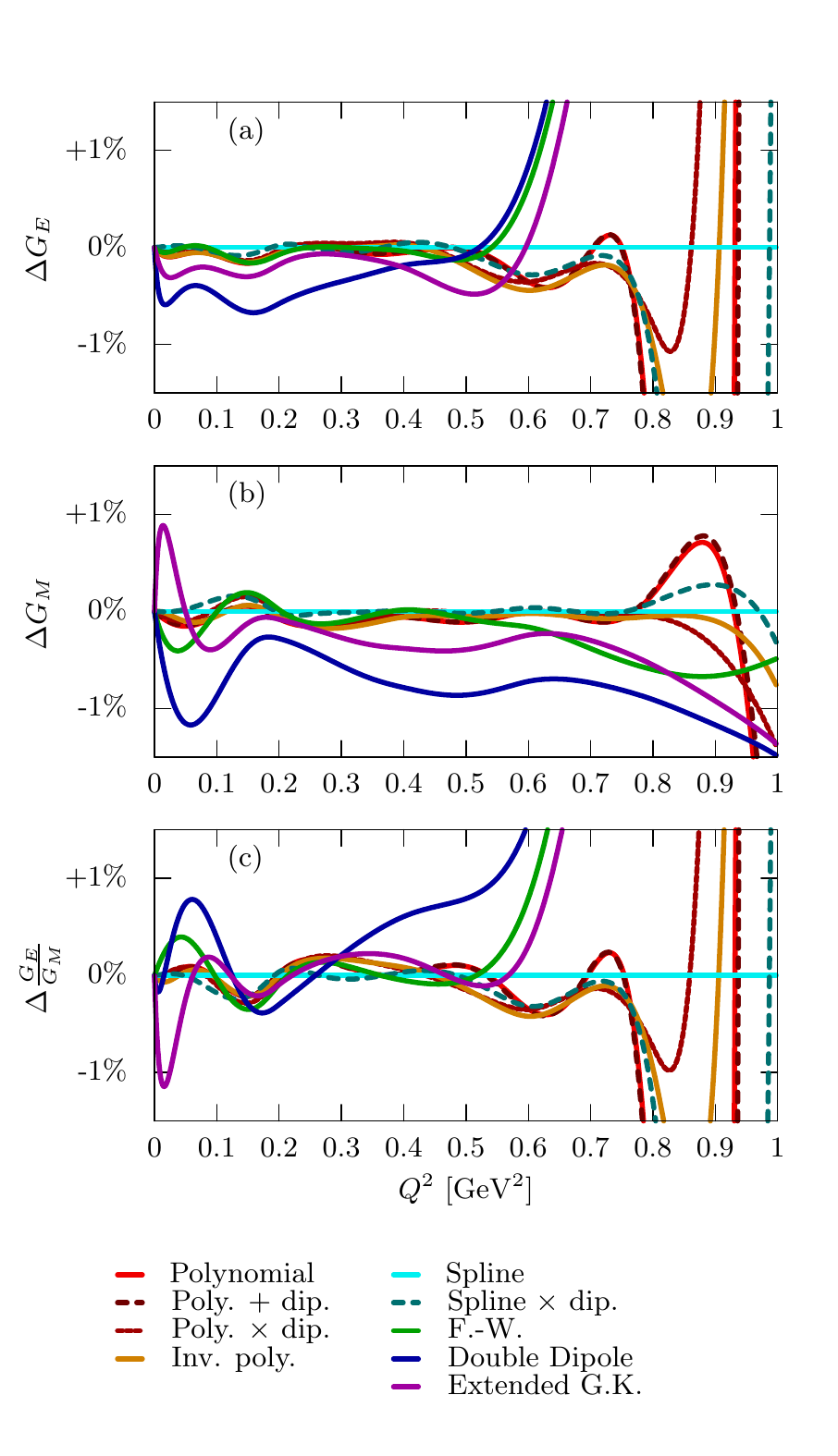}
 \end{center}
\caption{\label{figmodelgegm}(Color) Relative deviation of the different models to the spline fit. }
\end{figure}

Second, we compare the form factors determined with our broad set of models. Figures \ref{figmodelgegm} show the relative deviation of the different models from the spline fit. The flexible models have a very small spread between themselves, at least in the region where a reliable disentanglement of the form factors is possible. The less flexible fits exhibit larger deviations, especially above 0.5~$\mathrm{GeV}^2$. The Friedrich-Walcher parametrization has good agreement at lower $Q^2$, similarly to the flexible models, but exhibits the same bias as the other less flexible fits at higher $Q^2$.

\begin{figure}
\begin{center}
 \includegraphics{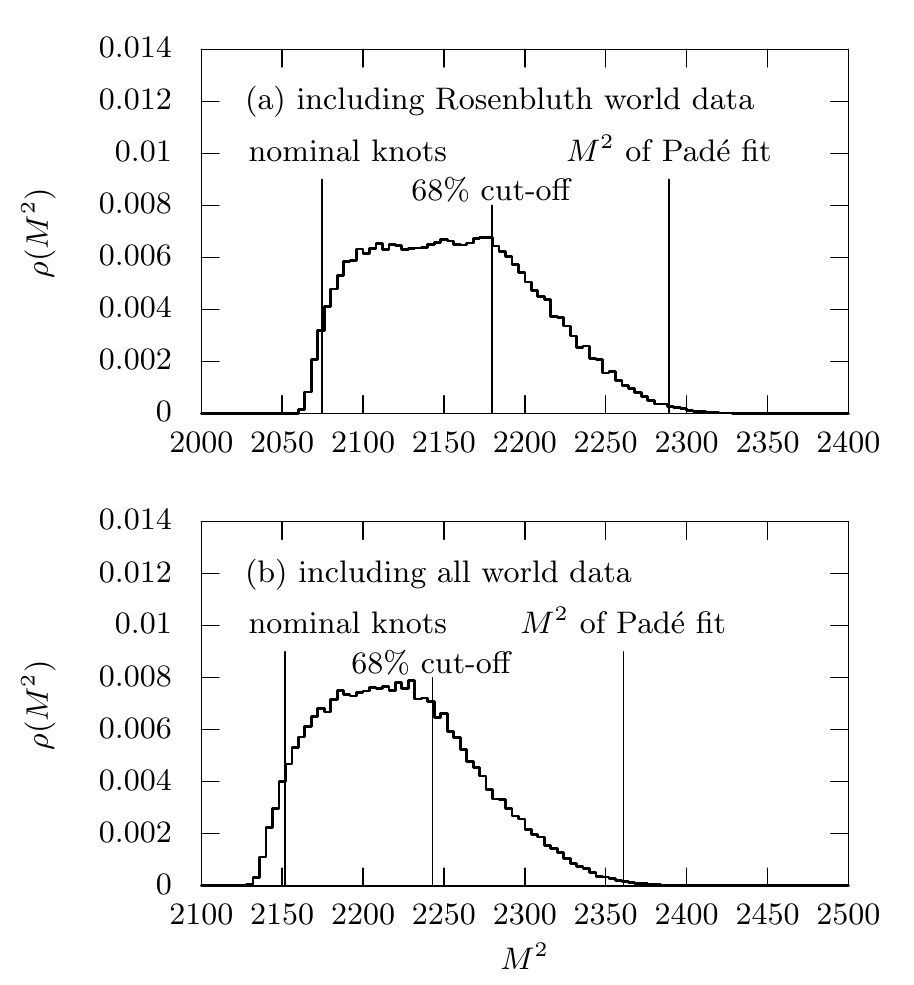}
 \end{center}
\caption{\label{figmwq} Probability distribution of $M^2$ when the knot position of the variable splines are varied as described in the text.}
\end{figure}

\begin{figure}
\begin{center}
 \includegraphics{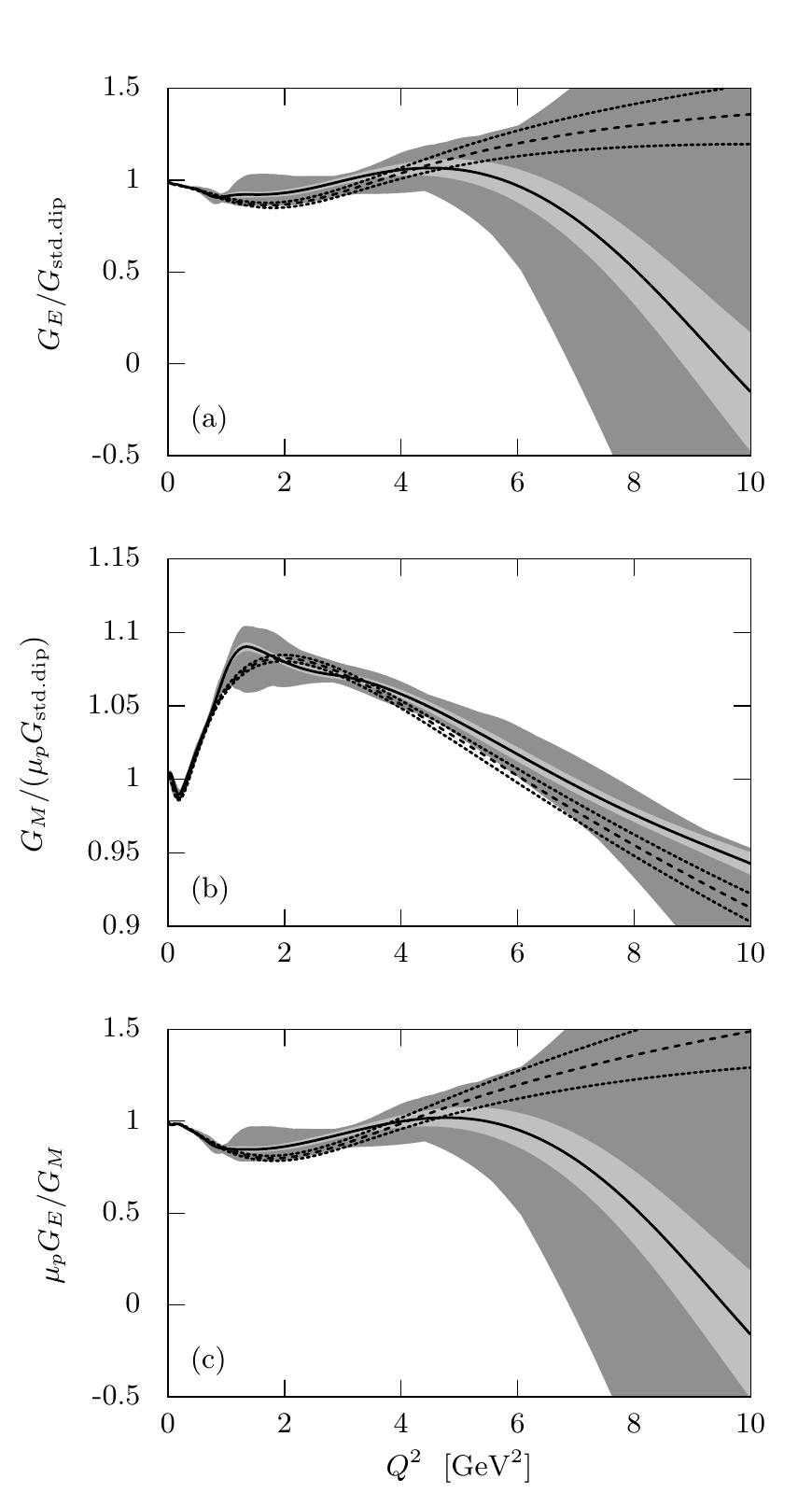}
 \end{center}
\caption{\label{figmsplinewq} Results for a global fit with the data 
  of this work together with external cross section data. Black line: Best 
  spline fit with nominal knot values. Light gray: Statistical 68\%
  pointwise confidence  band. Dark gray: model dependency from knot variation. Dashed line: 
  Pad\'e model. Dotted lines: edges of statistical confidence band for 
  Pad\'e model.}
\end{figure}

\begin{figure}
\begin{center}
 \includegraphics{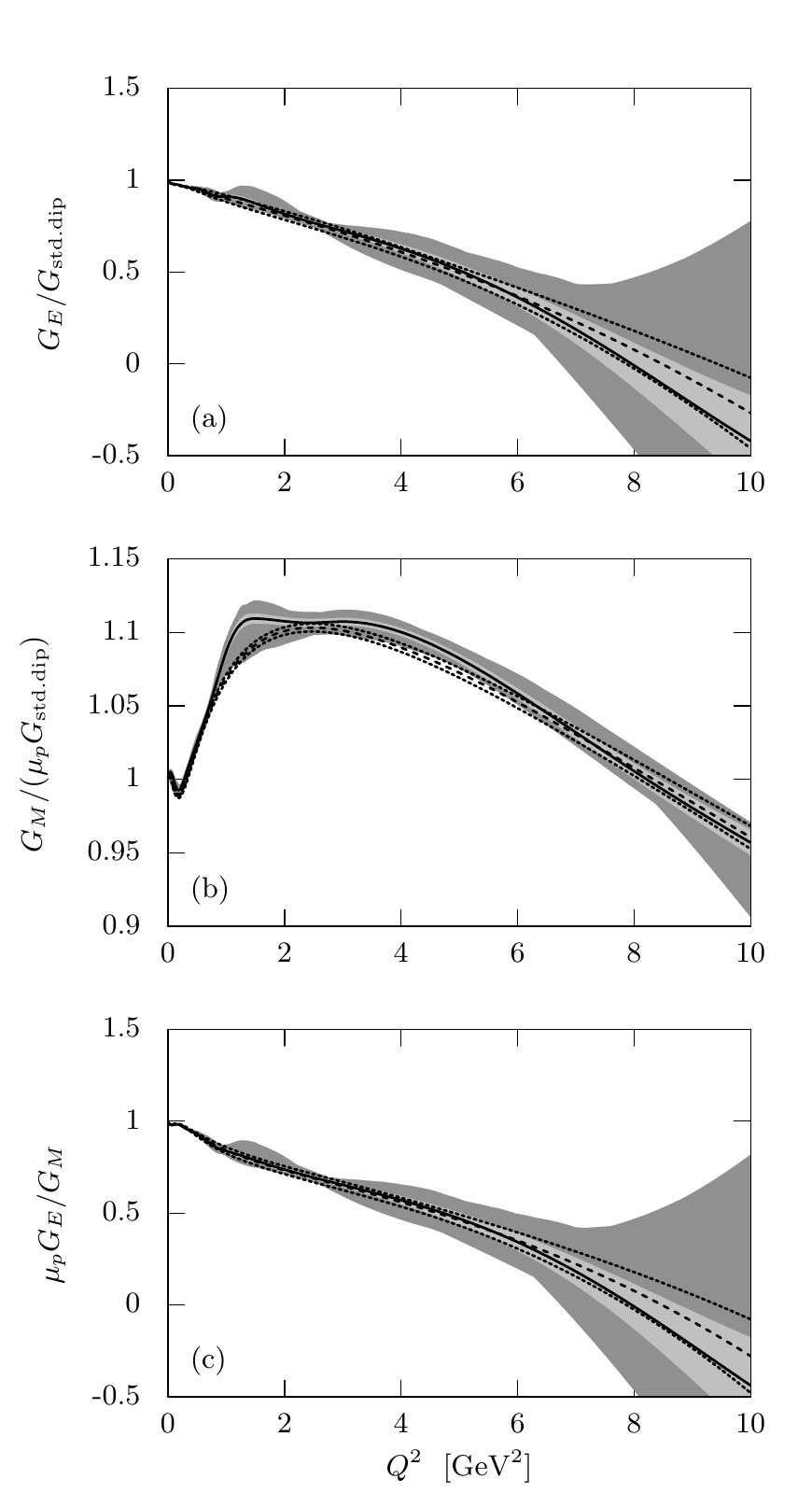}
 \end{center}
\caption{\label{figmsplinewqrt}  As in Fig.\ \ref{figmsplinewq}, but for
  fits to the
  data of this work and the external data from unpolarized and
  polarized measurements under the inclusion of the TPE parametrization Eq.\ (\ref{eq:TPE}).}
\end{figure}

For the fits which include the world data we use the mentioned variation of the spline model. The additional freedom of picking the location of the knots gives local flexibility of the model function. This variation provides an alternative handle on testing model dependencies since the flexibility of the model can be varied in an almost continuous way. For the purpose of this paper we kept the number of knots constant and only varied the positions with a Monte-Carlo approach. We select for each knot a random position between half the distance to the previous and half the distance to the next knot with a uniform distribution and refit. The distribution of $M^2$s of this ensemble of models is presented in Fig.\ \ref{figmwq}.  It has to be stressed here that, besides the points we raised earlier, this is not a $\chi^2$-like distribution---it is not a distribution of $M^2$ fitting an ensemble of repeated experiments with the same model, but the distribution of $M^2$ fitting the same data with an ensemble of models. As can be seen from Fig.\ \ref{figmwq} the original choice of the knot positions (``nominal knots'') was already close to optimal. We construct a confidence band by taking the envelope for the 68\% best models. The result is displayed in Figs. \ref{figmsplinewq} and \ref{figmsplinewqrt} together with the other contributions to the uncertainty.

\section{Results for the form factors}
 \label{ffffitssec}
\subsection{Fits to the new Mainz data alone}
In Fig.\ \ref{fig_emem} we present the results for $G_E$ and $G_M$
and of their ratio for fits of the spline model to the new data
without additional external data. In the same figure we show previous measurements and
fits to old data. It has to be noted that the previous measurements are plotted as
given in the original publication without the update to the radiative
corrections described above.  The error bars of the previous data
shown for $G_E$ and $G_M$ represent only the statistical error, the
normalization uncertainties are typically of the order of a few
percent. Since TPE corrections are not applied to any of the data the
corresponding non-TPE-corrected fit of Ref.\ \cite{Arrington07} is
shown. In the plot of the ratio the fit to the TPE-corrected data of
Ref.\ \cite{Arrington07} is also included. We show the 
Friedrich-Walcher fit from Ref.\ \cite{fw03} to the data before 2003,
 which now has to be regarded as superseded by the fit to the new data.
 
The results for $G_E$ exhibit some peculiar structure for small $Q^2$,
therefore we show this form factor also with an extended scale
[Fig.\ \ref{fig_emem} (a)].
First, $G_E$ exhibits a significant negative slope relative to the 
standard dipole at $Q^2 \approx 0$, giving rise to the larger electric
radius. 
After the slope levels out around 0.1\ $\mathrm{GeV}^2$,
there is an indication of a bump around  0.15\ $\mathrm{GeV}^2$, however, at the 
limit of significance. 
Further, the ratio to the standard dipole remains constant up to 
0.55\ $\mathrm{GeV}^2$ when the slope again  
becomes larger. In that region, however, only measurements at large  
scattering angles for only two beam energies contribute so the fit  
becomes less reliable and more sensitive to systematic errors such as  
the neglect of TPE. For even higher $Q^2$ measurements have been taken  
only at one energy and a separation of $G_E$ and $G_M$ is not possible.  

The magnetic form factor $G_M$ deviates from earlier measurements. 
We relate this to the normalization depending on the extrapolation
with an assumed analytical form which---in previous analyses---does not include the wiggle first seen by this experiment. The specifics of the maximum and the minimum 
of the wiggle structure depend, of course, on the functional form one
divides by---in our case, the standard dipole. 

The structure at small $Q^2$ seen in both form factors corresponds to 
the  scale of the pion of about  
$Q^2 \approx m_{\pi}^2 \approx 0.02\ \mathrm{GeV}^2$  
and may be indicative of the influence of the pion cloud \cite{Vanderhaeghen:2010nd}. 
 
While the deviation of $G_M$ from previous measurements seems surprising 
at first glance, it reconciles the form factor ratios from experiments  
with unpolarized electrons, like this one, with those found with polarized  
electrons, especially with the high-precision measurements in  
Refs.\ \cite{Ron07,Zhan11,Ron:2011rd}.

Due to the deviation of the results of $G_M$ from most of the previous
determinations the geometric reliability of the spectrometer motion has
been questioned by some experimenters after the publication of Ref.\
\cite{Bernauer:2010wm}. The verification of the rotational axes of the
spectrometers in 2013 found them to be within the assumed limits and
is far of from explaining the change in $G_M$ from previous
measurements.

A possible general concern with fits is the question of
convergence. In the time-like region, the form factors have poles which
limit the convergence of a polynomial expansion around 0 to $\left|Q^2\right|<4m_\pi^2$.
To test this, we modified the spline model; we add to the spline model
a  calculation of the non-analytic terms
\cite{Ledwig:2011cx}. In effect, the splines will then only have
to fit the remaining, analytical part. The result is almost
indistinguishable from the spline fit without this addition with a
relative change of below  $6\times10^{-4}$.

\subsection{Form factors via Rosenbluth separation}
\label{chptextrosen}
The classical way of determining $G_E$ and $G_M$ is the Rosenbluth
separation of cross sections measured at fixed $Q^2$ for different
polarization $\varepsilon$. Rewriting the cross section [Eq.\ (\ref{eqrosen})] as
\begin{eqnarray}
\label{eqrosenred}
\sigma_\mathrm{red}&=&\varepsilon \left(1+\tau
\right)\left(\frac{\mathrm{d}\sigma}{\mathrm{d}\Omega}\right)_{0}/\left(\frac{\mathrm{d}\sigma}{\mathrm{d}\Omega}\right)_\mathrm{Mott}\nonumber\\
&=&\left(\varepsilon G^2_E\left(Q^2\right)+\tau
  G^2_M\left(Q^2\right)\right)
\end{eqnarray}
makes it obvious that, for constant $Q^2$, the reduced cross section
$\sigma_\mathrm{red}$ depends linearly on $\varepsilon$ with
$G_E^2(Q^2)$ as slope and $\tau G_M^2(Q^2)$ as ordinate intercept.
Hence, a linear fit can be used to extract $G_E$ and $G_M$. We have
discussed in Sec.\ \ref{sec:intro} the advantages of extracting the form factors
through a global fit to the cross sections. Nevertheless, we also
perform a classic Rosenbluth separation of our data in order to
reconcile our analysis with the expectation the community might have.

One of the problems with a direct Rosenbluth separation of measured
cross sections is a coherent inclusion of a normalization of the
data. We handle this by first fixing the cross section with the
normalizations extracted by the spline fit.

Another problem is the necessity of several data points at constant
$Q^2$ but varied $\varepsilon$. Due to the large number of
measurements with overlapping acceptances, it is possible to find a set of 77 narrowly spaced $Q_i^2$ with measurements
at at least three different $\varepsilon$, so the linearity can be
tested. Obviously, not all of the measured data are being used in this
case---especially unfortunate is the loss of information on the lower end
of $Q^2$: The lowest point is 4 times larger than what is
available in the data.
To project the cross section, which has been averaged over a
finite-size $Q^2$ range given by the spectrometer acceptance, to the nearest $Q_i^2$ point, we divide by
the numerically integrated result of the Monte Carlo simulation with
the standard dipole for $G_E$ and $G_M$ and multiply by
the differential cross section evaluated with the same form factors at the given $Q_i^2$ point.

This procedure implies an error which, to first order, depends on the 
difference of the curvature of the true cross section and the one used
in the Monte Carlo calculation, multiplied by the square of the acceptance \cite{Hanson:1951zz}, when the
cross section is attributed to the central $Q^2_\mathrm{central}$; this
error is found to be negligible for our measurements. Attributing
the resulting cross section to an off-central value, say $Q^2_i$, 
results in an additional error proportional to the difference in
the slopes of the true and the reference cross sections and on the
projection distance $Q^2_i - Q^2_\mathrm{central}$. Our measurements are so narrowly
spaced in $Q^2$ that this uncertainty is below 0.15\% for the highest
$Q_i^2$ presented here and considerably less for lower $Q_i^2$.

\begin {figure}
\begin{center}
 \includegraphics{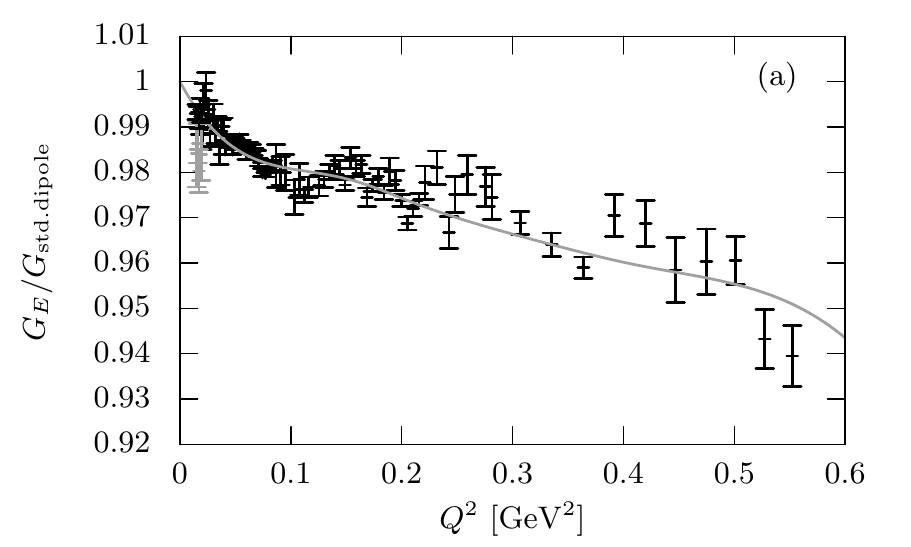}
  \includegraphics{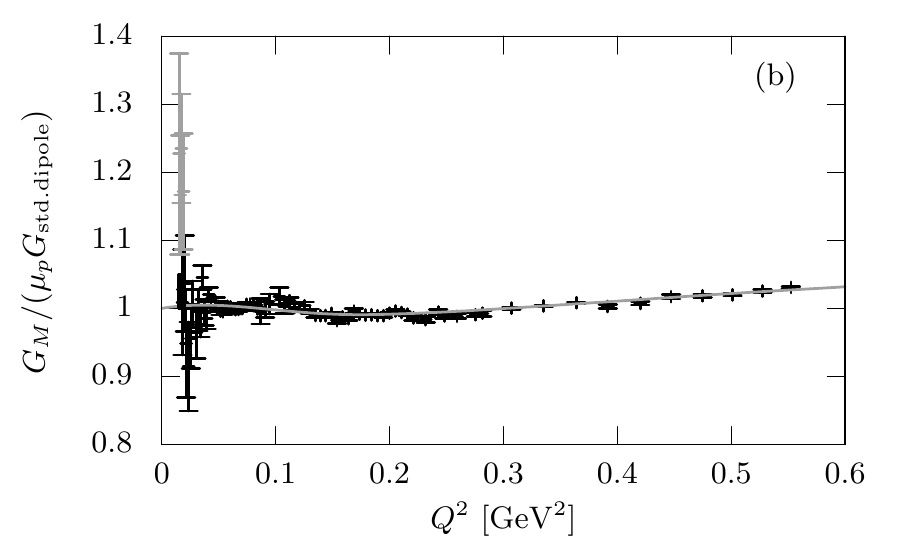}
 \end{center}
\caption[Rosenbluth separation]{\label{figrosvgl}$G_E$ and $G_M$
  determined via the Rosenbluth separation technique (black points)
  compared to the spline fit (gray curve). For the lowest points,
  $G_M/(\mu_pG_\mathrm{std.\ dip.})$ was varied between 1 and 1.05
  (the results of the unconstrained Rosenbluth fits are shown in gray). For details see text.}
\end{figure}

The Rosenbluth-separated form factors are shown  in Fig.\
\ref{figrosvgl}, together with the result of the global fit (spline
model). For the lowest $Q^2$ points, where $G_M$ is less well
determined, $G_M/(\mu_pG_\mathrm{std.\ dip.})$ was not determined by
the Rosenbluth fit but, for each point, set once to 1 and once to 1.05, as one
would expect it to be in that range and not larger. For each point,
the difference in $G_E$ of the two constraint fits and the errors of the individual
fits give the error of $G_E$ shown in Fig.\ \ref{figrosvgl}.
The points from the unconstrained fits are presented in gray for reference. The use
of the prior knowledge that the magnetic form factor cannot differ too much from the standard dipole for $Q^2 \sim 0$ helps to reduce the error
bars on $G_E$ for low $Q^2$ considerably.

The agreement of the Rosenbluth-separated form factors with those from
the global fits has been tested by calculating a reduced $\chi^2$ from
the differences of the Rosenbluth data points to the spline fit. The
rather large value of 2.2 is found, with similar numbers for a
comparison of $G_E$ or $G_M$ alone. Fits of polynomials (order 10) to
$G_E$ and $G_M$ from the Rosenbluth separation yield also
$\chi^2_\mathrm{red}$ values above 2. In order to put these numbers into
perspective, one has to note that the $\chi^2_\mathrm{red}$-distribution is 
much wider for the fit to the Rosenbluth-separated form factors, due to the 
lower number of degrees of freedom. 
In fact, interpreting the deviation of the flexible fits
from the expectancy value 1 as purely statistical, $\chi^2_\mathrm{red}$
values up to 1.7 for the fit to the Rosenbluth-separated form factors
would have the same probability as $\chi^2_\mathrm{red}$ values up to
1.14 for the global fits.

While the ``ingredients'' of the global fit and the Rosenbluth
separation are in principle similar, the explicit Rosenbluth
separation differs fundamentally from the global fit since it (a) has
to contract the large acceptance of the measurements to single
$Q^2$ points and (b) acts on subsamples of the complete data set.
In this way a large part of the dependence on the
primary kinematic variables, the scattering angle and the incident
energy, disappears and the information is
lost.  This is exacerbated by the fact that the set of $Q^2$ values
has to be the same for all energies.
The consequences of these differences have not been studied
fully, however, the robustness (see Sec.\ 8.7 of Ref.\ \cite{james}),
i.e., the insensitivity
to unaccounted non-Gaussian errors of the
input data, has been tested for both estimators, i.e., the form
factors determined via the global fit and via the Rosenbluth
separation.  To this end, statistically pure pseudo data are generated
from the spline fit and then perturbed with systematic errors. 
Unperturbed, both methods result in an average $\chi^2_\mathrm{red}$
of 1. Adding systematic shifts, however, increased the 
$\chi^2_\mathrm{red}$ for the Rosenbluth method more than for the
global fit method. 

In fact, if we shift 5\% of the data points by a systematic ``not normal distributed''
shift of 0.5\%, the fits yield a difference in the
$\chi^2$ increase comparable to the difference seen for the measured
data. We therefore conclude that the global fit is a much more robust
estimator of the form factors with respect to non-normal errors in the
measured cross sections.

\begin {figure}
\begin{center}
 \includegraphics{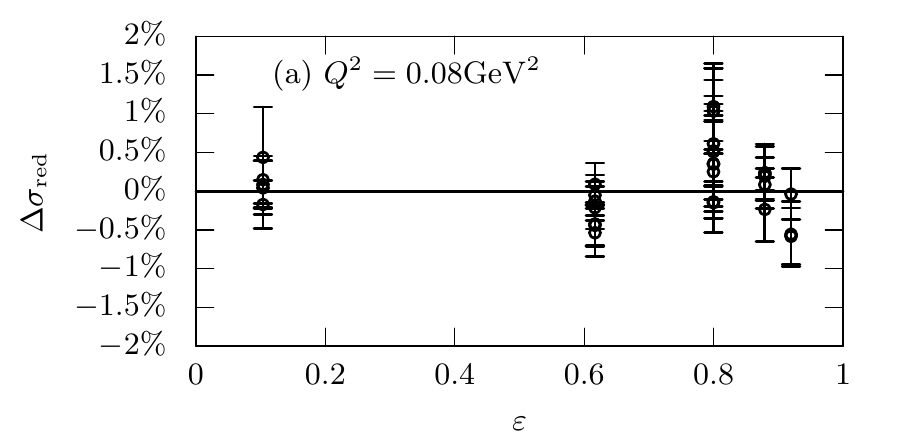}
  \includegraphics{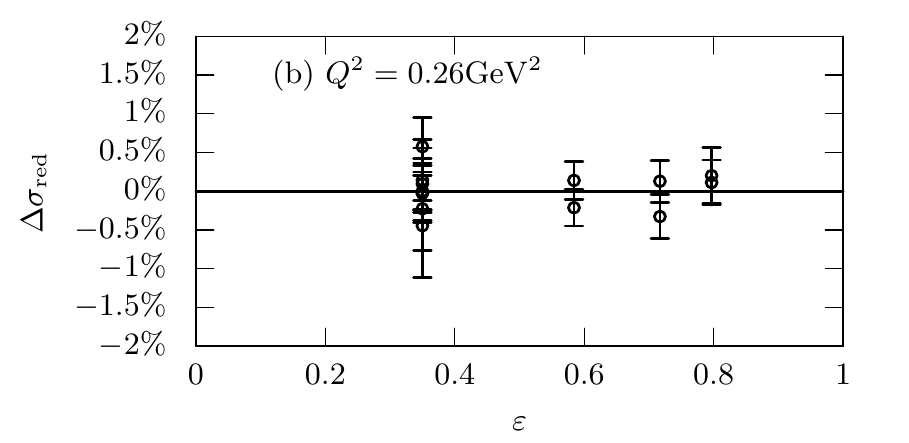}
  \includegraphics{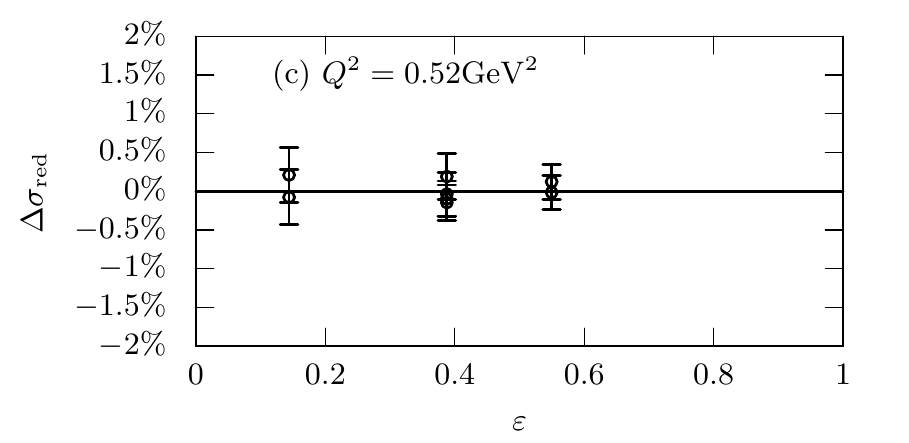}
 \end{center}
\caption[Relative deviation from the Rosenbluth
fit]{\label{figrosdev1} The relative deviation of the measured
  $\sigma_\mathrm{red}$ from the Rosenbluth straight-line
  fit for three different $Q^2$ values.  No systematic deviation from the
  linear fit is indicated.}
\end{figure}

As to other systematic insufficiencies in the measurement, the
Rosenbluth separation is also prone to errors in theoretical
corrections of the cross sections. Such insufficiencies might show up
as deviations of the reduced cross section from Eq.\ (\ref{eqrosenred}),
i.e., from a straight line in the Rosenbluth plot. Figure \ref{figrosdev1} shows the relative deviations of the measured
$\sigma_\mathrm{red}$ from the Rosenbluth fit for 3 of the 77 $Q^2$-values.  At the (quite high) level of statistical accuracy of this
experiment no systematic deviations from the straight line are
visible. This was tested further for all 77 points by
fitting polynomials of second order, where the coefficient of the
quadratic term was found to be compatible with zero.

\subsection{Fits including external data}
\label{fwaextdata}
\subsubsection{Unpolarized elastic scattering only}
For global fits with the addition of external data we have to adapt the 
models to be usable with the drastic change in data point density over 
$Q^2$ as already described. We use a flexible
spline $\times$ dipole model with variable knots described above and the less flexible
Pad\'e model given by Eq.\ (\ref{eq:Pade}). Figure \ref{figmsplinewq} shows both form factors and
the form-factor ratio from both models. 

The overall behavior of the Pad\'e model is quite similar to the spline model
up to 5 $\mathrm{GeV}^2$, without following the small-scale wiggles of
the spline fit.

Differences in the gross behavior appear in
$G_E$ for  $Q^2>5\ \mathrm{GeV}^2$, where the existing data start to
determine $G_E$ badly. Here the Pad\'e model does not exhibit the
downward bend,
however, without leaving the model confidence band of the spline model. 
For $G_M$, there is a distinct knee between 1 and 2
$\mathrm{GeV}^2$ for the spline model, which appears slightly washed out but clearly
visible also in the Pad\'e model. The
high-$Q^2$ behavior of both models is closer together in trend in
$G_M$ than
in $G_E$; however, the
confidence band of the spline fit does not overlap the Pad\'e fit over
the complete range.

The large model-dependency estimate for larger $Q^2$ illustrates a
point which is often underestimated: The standard error propagation 
used for the construction of the confidence bands gives an estimate of the 
statistical error {\it for the chosen model} only.

\subsubsection{Unpolarized and polarized elastic scattering}
Adding additional information in the form of form-factor-ratio data from
polarized experiments can help to reduce the uncertainty in the form
factor separation, especially affecting the uncertainty of $G_E$ at
large $Q^2$. 
As discussed in Sec.\ \ref{fwextdata}, we need to introduce additional parameters for a TPE parametrization given by 
Eq.\ (\ref{eq:TPE}) to reconcile the two measurement methods. Figure \ref{figmsplinewqrt} shows the results of
these fits, again for both models. While $G_E/G_\mathrm{std.dip}$ now
decreases more or less linearly, $G_M$ is shifted upward for $Q^2>1\
\mathrm{GeV}^2$. As a result, the ratio also decreases almost linearly. This behavior is similar for
 both the spline as well as the Pad\'e model. In spite of the added
 parameters, the widths of the confidence bands are reduced.

\begin{figure}
\begin{center}
 \includegraphics{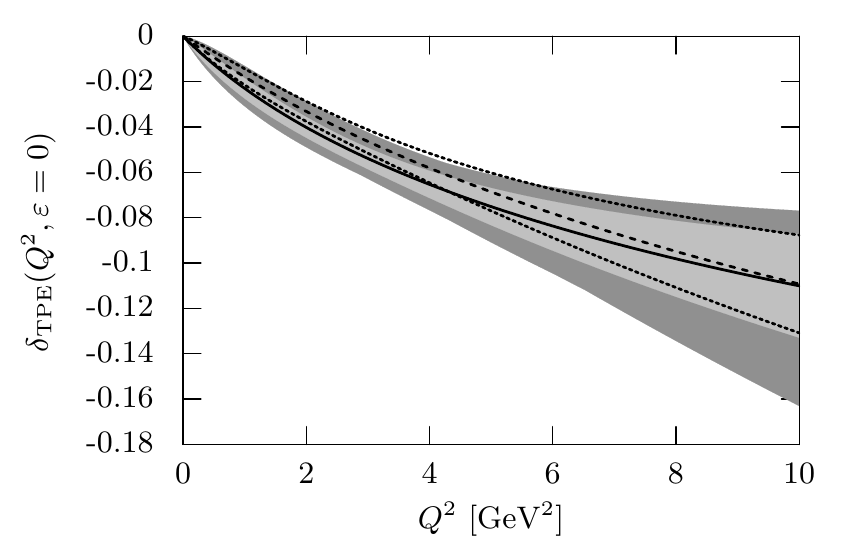}
 \end{center}
\caption{\label{figtpecontrib}The TPE contribution to the cross
  section (excluding the Feshbach-term) as determined by the
  fit to the complete world data set as a function of $Q^2$ for $\varepsilon=0$. Same
  nomenclature as in Fig.\ \ref{figmsplinewq}.}
\end{figure}

Figure \ref{figtpecontrib} shows the contribution from our
TPE parametrization as a
function of $Q^2$ at $\varepsilon=0$ where the correction is
maximal. This contribution is similar for both
models. Table \ref{tabab} lists the fit result for the parameters $a$ and $b$.

\begin{table}
\begin{center}
\begin{ruledtabular}
\begin{tabular}{c|c|c}
Model &$a$&$b$\\
\hline
Spline with var. knots&\ $0.069$\ &\ $0.394\ \mathrm{GeV}^{-2}$\ \\
Pad\'e& $0.104$& $0.188\ \mathrm{GeV}^{-2}$\\
\end{tabular}
\end{ruledtabular}
\end{center}
\caption{\label{tabab} The values of $a$ and $b$ of the TPE
  parametrization, extracted from the
  fits to all data.}
\end{table}

It is clear that the TPE model used is simple and only 
weakly motivated by theoretical considerations. However, it gives 
much better fits for the wide $Q^2$ region in which the ratio $G_E/G_M$ 
derived without accounting for TPE from the unpolarized scattering and
from the polarized measurements 
differs significantly. While the total $\chi^2$ increases by 207 when
adding 58 polarization data points without any TPE parametrization
taken into account, the increase is reduced to 77 by the inclusion of
only two free parameters (for the detailed numbers cf.\ Table \ref{wdchitable}).

\begin{figure}
\begin{center}
 \includegraphics{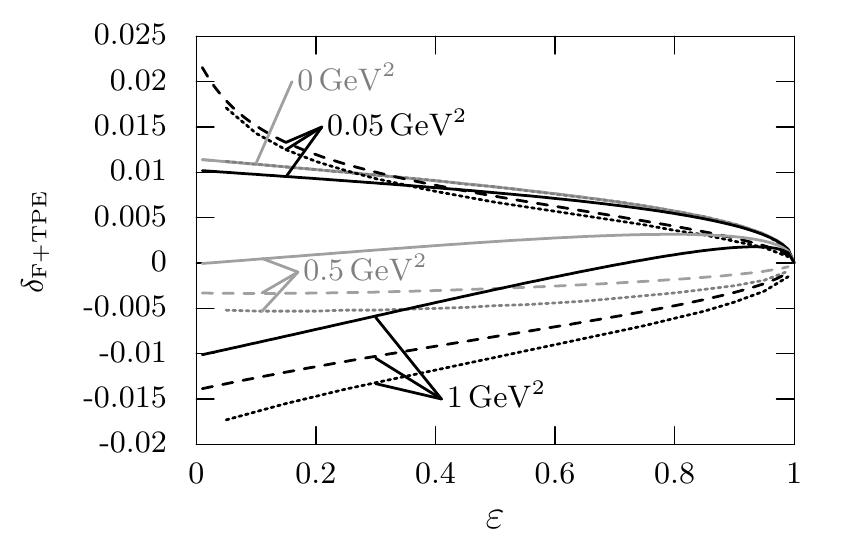}
 \end{center}
\caption{\label{figtpecomp} Comparison of the complete TPE
  correction $\delta_\mathrm{F+TPE}$
(without the soft part included in Maximon-Tjon) for four different
  $Q^2$ values. Solid: Fit to
  data, dashed: calculation with the approximation of
    Ref.\ \cite{Borisyuk:2006uq}. dotted: TPE corrections from Refs.\ \cite{Blunden05,Arrington:2011dn,blundenpc}.}
\end{figure}

Figure\ \ref{figtpecomp} shows the TPE correction as 
a function of $\varepsilon$ for selected $Q^2$ values in comparison with the 
calculation of Borisyuk {\em et al.}\ \cite{Borisyuk:2006uq}
and  the calculation by Blunden {\em et al.}\ \cite{Blunden05,Arrington:2011dn,blundenpc}.. Both
calculations are valid in the low-$Q^2$ region.  Our simple parametrization  can evidently not 
reproduce the strong curvature of these particular calculations in the 
low-$\varepsilon$ region. Most of the data however are situated in the 
mid-range of $\varepsilon$, where the calculations are almost linear
and our parametrization gives a similar slope, 
but a different overall normalization of about 0.5\%, growing slightly
with $Q^2$.  Such a correction linear in $\varepsilon$ cannot be tested for with the
Rosenbluth separation as it is indistinguishable from a change in $G_E$.
The change of  the overall normalization is mostly absorbed in our floating normalization of the
fits.  

\begin{figure}
\begin{center}
 \includegraphics{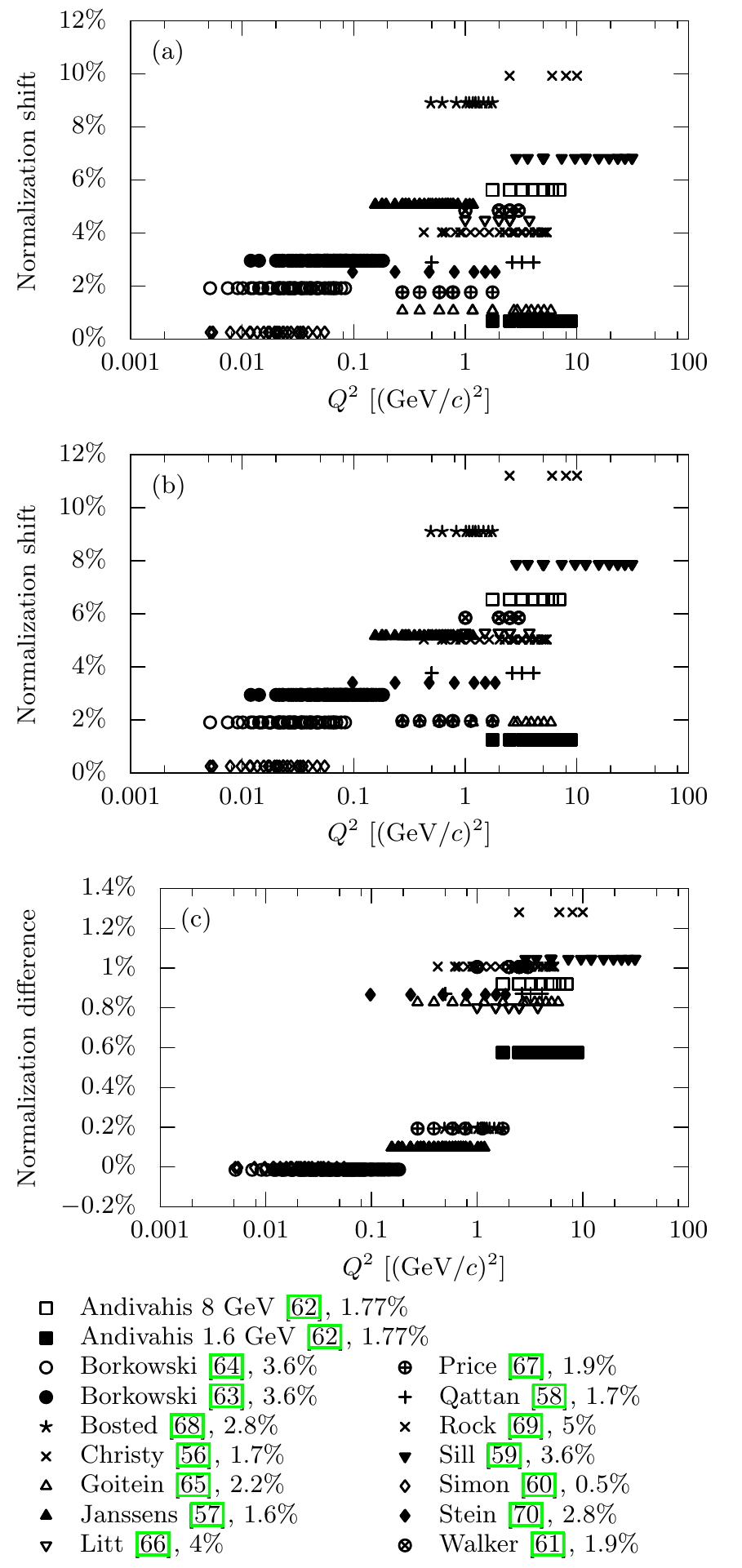}
 \end{center}
\caption{\label{fignorm} Shifts of the normalization $n_{\mathrm{ext},j}-1$ of the external
  cross section data found by the spline fits. (a) Fits to
  the data of this work and external  data from unpolarized
  scattering. (b) Data from polarized measurements are
  additionally included and the fit is extended with the TPE
  parametrization. (c) The difference of the normalization shift $[\mathrm{(b)}-
  \mathrm{(a)}]$ . The numbers in the legend indicate the normalization
  uncertainty we assumed in the fit.}
\nocite{Andivahis,Borkowski74-2,Borkowski74,Bosted90,Christy04,Goitein70,Janssens65,Litt69,Price71,Qattan,Rock92,Sill93,Simon80,Stein75,Walker}
\end{figure}

An indication for such a normalization effect can be seen by a comparison of the
normalization factors of the external data found in the fits: In
Fig.\ \ref{fignorm} the shift of the cross section normalization $n_{\mathrm{ext},j}-1$ 
of the external data is displayed. Figure \ref{fignorm}(a) shows the result when only 
the external unpolarized cross-section data are added to the data set
used for the fit. As can be seen,
all shifts are positive, i.e., the actual cross sections as reconstructed by
the fit are larger than the values quoted in the original publication.
 The spread of the shifts is quite large. However, a fit to
just the previous data sets without the new Mainz data shows a similar
spread of almost the same shifts for the low-$Q^2$ data sets and shifts
 smaller by about 4\% absolute for the large-$Q^2$ data. 
While it may look strange that all shifts are positive, the mean of
the normalization falls together with the shift of the oldest
measurement \cite{Janssens65}, for which the absolute normalization
was certainly not better than a few percent, and the other older measurements may have checked their
normalization with regard to Ref.\ \cite{Janssens65}.

Figure \ref{fignorm}(b) shows the normalization when, in
addition, the polarization data and the TPE parametrization are taken
into account. Figure \ref{fignorm}(c) shows the difference in the normalization
introduced by this. While there is virtually no change in the normalization with respect to the
analysis without the polarization data for the data
sets at $Q^2$ below 0.2~GeV$^2$, we find a shift of about 1\% for the
large-$Q^2$ data sets, similarly to the spread of the curves in Fig.\ \ref{figtpecomp}.

\begin{figure}
\begin{center}
 \includegraphics{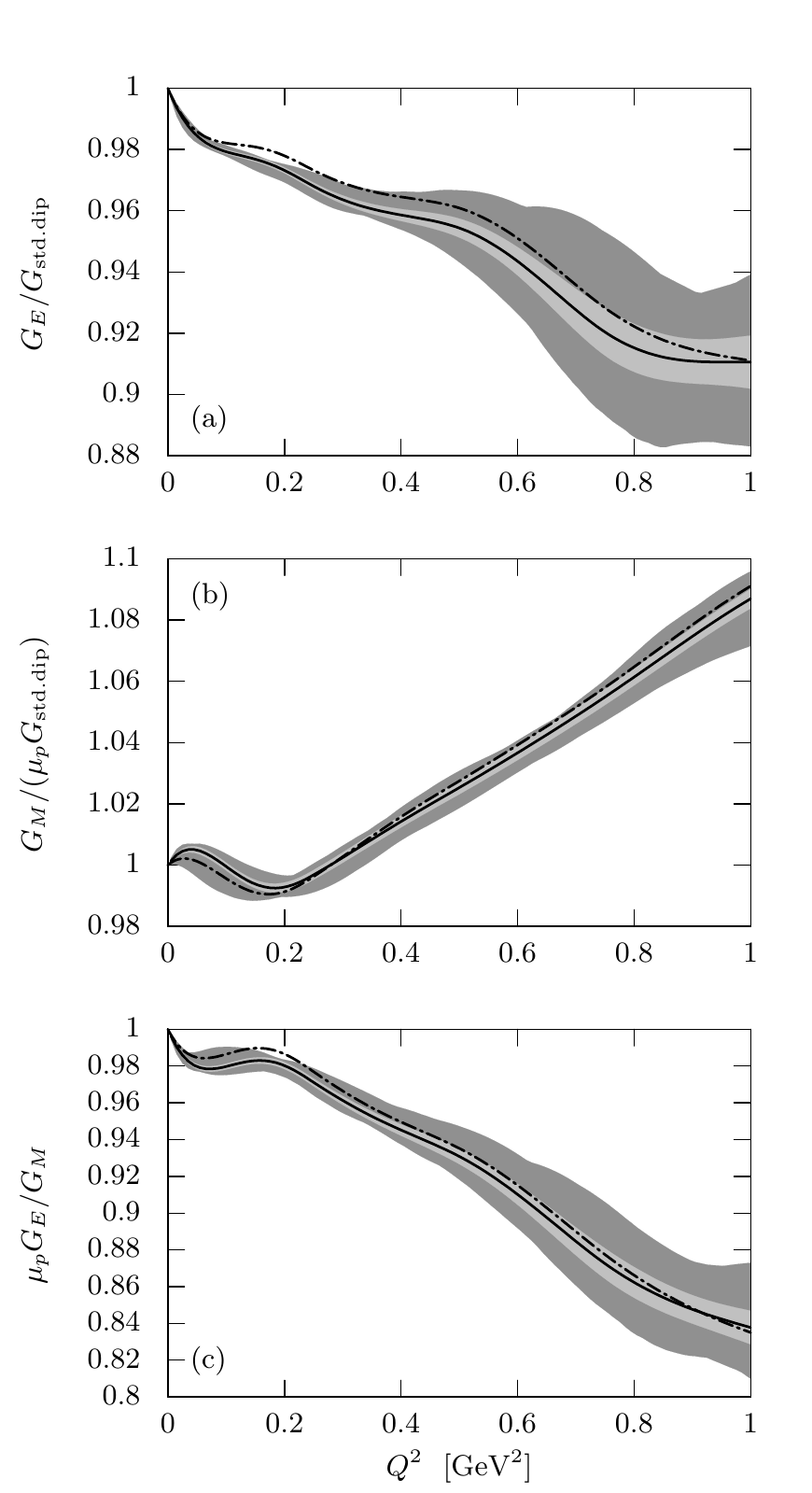}
 \end{center}
\caption{\label{figmsplinewqrtzoom}  The spline fit with variable
  knots from Fig.\ \ref{figmsplinewqrt}, zoomed in to $Q^2<1\ (\mathrm{GeV}/c)^2$. The dashed-dotted line
represents a fit of all data up to $5 (\mathrm{GeV}/c)^2$, with TPE
corrections according to Refs.\ \cite{Blunden05,Arrington:2011dn,blundenpc} applied.}
\end{figure}

Figure \ref{figmsplinewqrtzoom} shows the low-$Q^2$ region of the spline fit. We
find similar features as in the fits to the Mainz data alone which did
not include a TPE correction: Both the change in slope of $G_E$ and
the bump-dip structure in $G_M$ are visible. 

The figure also shows a fit of the data without our empirical
TPE-correction model but with the Feshbach correction replaced by a TPE
calculation by Blunden {\em et al.}\ \cite{Blunden05,Arrington:2011dn}, who
provided us with a numeric evaluation \cite{blundenpc}. Since this
calculation is valid on;y up to around $5\ (\mathrm{GeV}/c)^2$, we fit
data only up to this value. The correction reduces the disagreement between
unpolarized and polarized data: the $\chi^2$ reduces from 2232 for a
fit with the Feshbach correction to 2142 with the Blunden
correction, equal to an increase of about 1.85 for each data point from
polarization experiments. Our full fit achieves 2107, i.e.\ 1.22 for each data
point from polarization. Fits up to $3\ (\mathrm{GeV}/c)^2$ produce similar results. 

The fit for $G_E$ with the Blunden TPE correction lies higher at
around $Q^2=0.2\ (\mathrm{GeV}/c)^2$, making the change in slope even
more pronounced. On the other hand, the bump in $G_M$ is much
reduced. However, the fit still indicates that the following dip
compared to the dipole does not start at $Q^2=0$, like previous fits
extracted, but at slightly higher $Q^2$ values.

\section{Electric and magnetic radii}
 \label{secradii}
\subsection{Extraction method and model dependence}
According to Eq.\ (\ref{eqradius}), the electric and magnetic rms radii, $r_E=\sqrt{\left<r_E^2\right>}$ and $r_M=\sqrt{\left<r_M^2\right>}$, are given by the slopes of the corresponding form factors at $Q^2=0$. Therefore the accuracy with which they are determined by the measurement is given by the accuracy of the data in particular at low $Q^2$. Since the accuracy of $G_E$ is high at low $Q^2$, $r_E$ will be determined with good precision, while $G_M$ and therefore $r_M$ is less well determined  due to its small contribution to the cross section at low $Q^2$. In any case, the determination of $r_E$ and $r_M$ corresponds to an extrapolation of $G_E$ and $G_M$ to $Q^2=0$, and one has to ask the question to which extent the result depends on the ansatz for the fit model, in particular on its flexibility which depends on functional form and number of parameters, $N_p$. For too low flexibility, the resulting radii are not trustworthy since the data are not correctly reproduced. If the flexibility is too high, the fit can follow the smallest statistical fluctuations, which results in erratic determinations of the radii.
In fact, one has to compromise between these two extremes, as we did for the choice of the number of parameters $N_p$ by looking at the development of $\chi^2$ as a function of $N_p$. Here, we also inspect the resulting values of the radii and search for the range of $N_p$ in which the extracted radii are stable.

For both the electric and the magnetic radius the polynomial and the polynomial + dipole model produce a stable result for $N_p>9$. The polynomial $\times$ dipole model works comparably well for the electric radius for $N_p>8$, but shows erratic behavior for the magnetic radius already for $N_p>9$. The inverse polynomial, which has a quicker convergence to the $\chi^2$ plateau, also deteriorates faster into such erratic behavior for $r_M$. Nevertheless, these models agree quite well for both radii when one confines oneself to $N_p$ at the beginning of the plateau.

The erratic behavior of the magnetic radius stems from the less stringent determination of the magnetic form factor at low $Q^2$, where the magnetic contribution is small and where, with enough flexibility (large $N_p$), the fit follows smallest statistical deviations, resulting in  larger uncertainties. Low $N_p$ give the fit enough stability to extrapolate $G_M$ from higher $Q^2$ values, where the magnetic contribution is sizable, down to $Q^2=0$, but may be not flexible enough to capture the true behavior. It has to be noted that in previous determinations of $r_M$ only models with much less flexibility have been used and that the data had been taken at $Q^2 > 0.0053$ GeV$^2$ only and they had significantly larger errors.

The spline fits based on polynomials of third order tend to
give a smaller electric radius than the rest of the models, they
additionally exhibit a depression in the value of the radius of about
0.015~fm, that is, about 2.5 times the statistical uncertainty,  for
$N_p$ around 10. This difference between the result from the splines
and the polynomials was further investigated, but no conclusive cause
was found. The spline model's tendency to underpredict, albeit less pronounced, was
already seen in the model dependency analysis described in Sec.\ \ref{moddep}. The curvature of the spline models is limited by the order of the base polynomial. To test for a possible bias, we also use splines based on polynomials of fourth and fifth order. They produce progressively larger radii.

Focusing on the $\chi^2$ of points below $Q^2=0.06\ \mathrm{GeV}^2$ (543 data points) the spline fits yield a $\chi^2$ around 581 while the rest of the models give around 576. While this might indicate a worse fit of the low-$Q^2$ region by the spline models, the $\Delta \chi^2$ of 5 is small compared to the 1$\sigma$ width of the $\chi^2$ distribution [$\sigma_{\chi^2}(N_\mathrm{d.o.f}\approx543)\approx33$].

\begin{figure*}
  \begin{center}
     \includegraphics{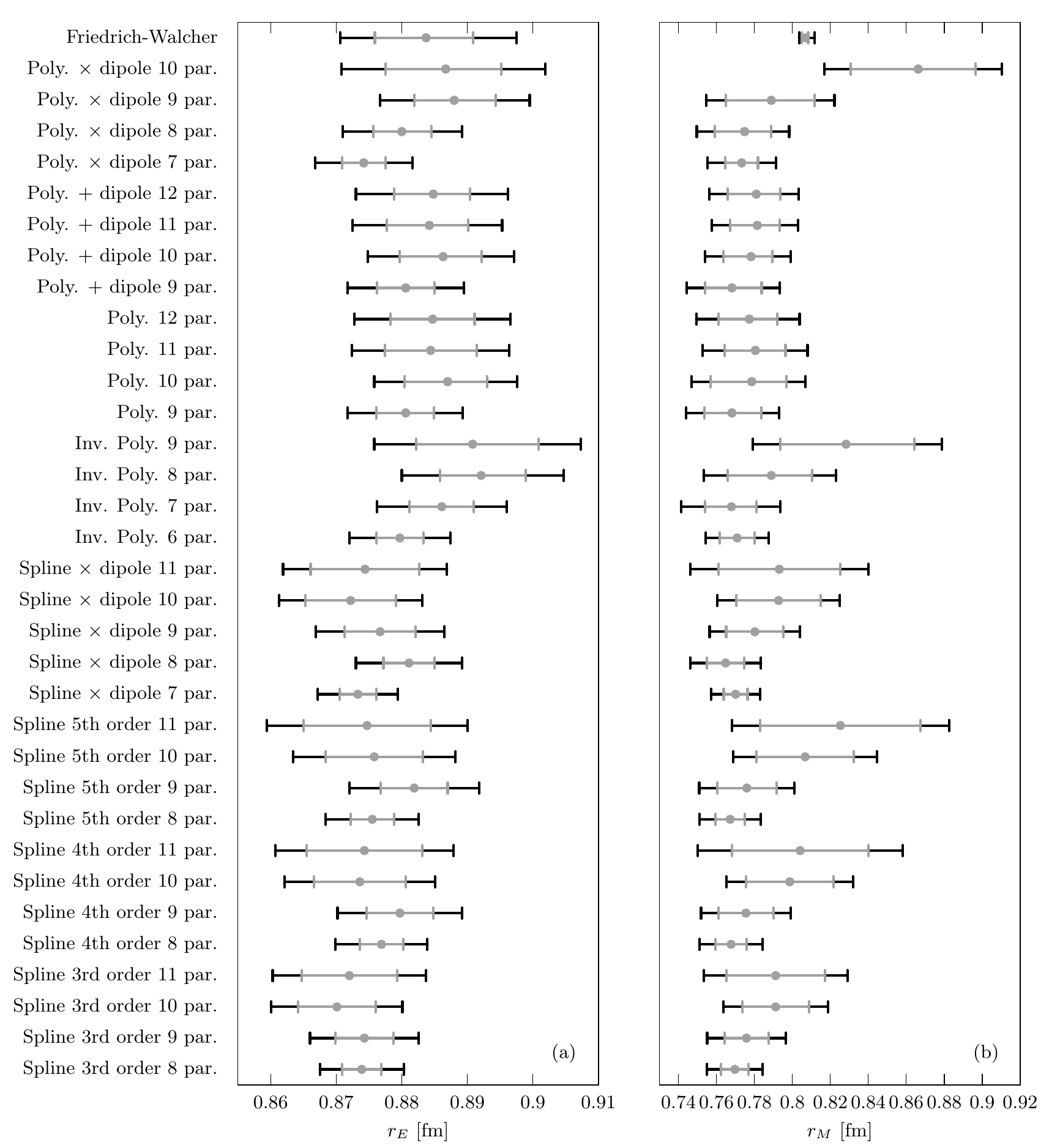}
   \end{center}
  \caption{\label{figradalle}The electric and magnetic rms radii as extracted with the different models. Gray: Statistical error; black: Linearly added systematic error.}
\end{figure*}

In order to estimate the model dependency for the extracted radii, the radii are determined with all models described before and for some variation in $N_p$. The results are shown in Fig.\  \ref{figradalle}.

The results for the charge radius fall somewhat apart into two groups
according to the model of the analysis, namely those determined with
the spline-based models and those from the polynomial-based
models. For each group, the electric and magnetic radius and their statistical and systematic error have been determined as the weighted average over the results from the single fits, where the weights are the linear sum of statistical and systematical error. The model error has been
calculated from the weighted variance of the values.

The final result is the (unweighted) arithmetic average of the values of the two groups. An additional error (labeled ``group''), accounting for the difference of the two groups, is attributed to the result. Since it cannot be assumed that this error is normal-distributed, it is taken as half of the difference of the two groups.

For most of the results below we applied the radiative
corrections described above, that is, with the Feshbach
Coulomb correction.
  In order to get a feeling for 
the effect of different corrections, we repeated the
fits with a TPE calculation using the approximation of
Ref.\ \cite{Borisyuk:2006uq}, yielding the values already published in
Ref.\ \cite{Bernauer:2011zz}, and also using the calculation from Refs.\ \cite{Arrington:2011dn,blundenpc,Blunden05}. 

\subsection{Electric radius}

As the average of the flexible models, we obtain
\begin{equation} 
\label{refinal}
r_E=0.879 (5)_\mathrm{stat.}(4)_\mathrm{syst.}(2)_\mathrm{model}(4)_\mathrm{group}.
\end{equation}
This value  is in complete agreement with the CODATA06
\cite{Mohr08} value of $0.8768 (69)$\ fm based mostly on atomic
measurements. It is also in complete accord with the old Mainz result
\cite{Simon80} when the Coulomb corrections
\cite{Rosenfelder99,Sick03} are applied. However, the results from
recent Lamb shift measurements on muonic hydrogen \cite{pohl,Antognini:1900ns} are
0.04\ fm smaller, i.e., 5 standard deviations from our result
(quadratically added errors) and almost 8 from the updated CODATA value from 2010 \cite{Mohr12} which combines our data and earlier scattering and atomic level measurements. 

Since this difference is unexplained yet, despite a multitude of efforts, we looked whether other ways of analysis of the scattering data would yield different results.

The Friedrich-Walcher model gives a slightly larger, but fully
compatible radius. 

For small $Q^2$, the contribution of the magnetic form factor to the cross section is so small that one can adopt one parametrization for $G_M$, subtract the magnetic contribution from the cross section and then fit the resulting $G_E$ at low $Q^2$ only using a simple model like a low-order polynomial.  This technique is similar to the method employed by Simon {\em et al.}\ \cite{Simon80}, where $G_M$ was set to $\mu_pG_E$ (scaling relation). In the present work, we apply it  to the 180-MeV data alone, using different parametrizations, different cut-off values in $Q^2$, and different $G_M$ models. The normalization was left floating, but the fit recovered the normalization given by the global fit on the 0.1\% level. This approach yields radii between 0.870 and 0.895 fm, with most values close to 0.880 fm, thus in excellent agreement with the final result of the global fit.

The two different TPE calculations we applied change the result by
significantly less than the statistical uncertainty alone.
In Ref.\ \cite{Arrington:2012dq}, Arrington compares several
recent calculations and finds all of them in good agreement for low-$Q^2$
and high $\varepsilon$. A full evaluation of the effect of all available calculations on the fit and the extracted radii is beyond the scope of this paper.

The inclusion of external data changes the value
marginally for the spline model with variable knot positions. The
Pad\'e model is not sufficiently flexible to achieve a comparably small $\chi^2_\mathrm{red}$.
Hence, it's extracted radius is unreliable and we refrain from quoting its result.

Table\ \ref{taberad} summarizes the electric radii determined with the different approaches together with the final result Eq.\ (\ref{refinal}). Despite all these efforts, we do not see a way to reconcile our result with those from muonic hydrogen. We do not expect that a future calculation of TPE corrections can reconcile our result with the muonic measurement completely, but we cannot rule out that such calculations may reduce the discrepancy. We want to note, however, that a large shift in the radius from TPE would in turn create tension with atomic measurements with electric hydrogen, albeit probably with less significance.

\begin{table}
\begin{center}
\begin{ruledtabular}
\begin{tabular}{l|c}Method& Electric radius $r_E$ in fm\\
\hline
\hline
Spline models (1) &$0.875(5)_\mathrm{stat.}(4)_\mathrm{syst.}(2)_\mathrm{model}$\\
Polynomial models (2) & $0.883(5)_\mathrm{stat.}(5)_\mathrm{syst.}(3)_\mathrm{model}$\\
Friedrich-Walcher & $0.884( ^{+7}_{-8})_\mathrm{stat.}(^{+7}_{-5})_\mathrm{syst.}$\\
\hline
\multicolumn {2}{l}{Spline with variable knots + external data:}\\
+ Rosenbluth data & $0.878$\\
+ all external data & $0.878$\\
\hline
{\bf Average of (1),(2)} &$0.879(5)_\mathrm{stat.}(4)_\mathrm{syst.}(2)_\mathrm{model}(4)_\mathrm{group}$\\
With TPE from \cite{Borisyuk:2006uq}&$0.876(5)_\mathrm{stat.}(4)_\mathrm{syst.}(2)_\mathrm{model}(5)_\mathrm{group}$\\
With TPE from \cite{Arrington:2011dn,blundenpc,Blunden05}&$0.875(5)_\mathrm{stat.}(4)_\mathrm{syst.}(2)_\mathrm{model}(5)_\mathrm{group}$\\
\end{tabular}
\end{ruledtabular}
\end{center}
\caption{\label{taberad}Results for the electric radius.}
\end{table}

\subsection{Magnetic radius \label{sec:V.C}}
Table \ref{tabmrad} gives an overview of the results for the magnetic radius of the proton. The statistical and systematic uncertainties are larger, since the radius is determines as an extrapolation for $Q^2\rightarrow 0$ where the cross section is less sensitive to magnetic scattering. Interestingly, the difference between splines and polynomials is much smaller than for the electric form factor. 
This gives some confidence that the wiggle in $G_M$ at small $Q^2$ (see Fig.\ \ref{fig_emem}) is not an artifact of the fitted models. The only experimental reason for this wiggle, which we could think of and not rule out directly, would be a systematic error with the angular rotation of the spectrometers around the target. From the precision of the setup of the spectrometer turntable \cite{Blomqvist:1998xn}, which we have rechecked,  and the checks with the overlapping angular settings of the spectrometer described above, in particular the measurements on the left and right side, we can exclude such an explanation.

On the other hand, the change is relatively large when applying the two different TPE corrections. This may be due to the somewhat unexpectedly large deviations of the TPE calculations from a linear behavior at small $\varepsilon$, i.e., larger scattering angles.
The Coulomb calculation from Refs.\ \cite{arringtonsick04,Arrington:2011kv}
does not show this behavior at small $\varepsilon$ and it has been
argued in Ref.\ \cite{Arrington:2012dq} that calculations beyond second Born
might be needed, reflecting the uncertainty in current TPE calculations. The phenomenological determination of TPE effects in this work may serve as a help for a better theoretical description.

Future results from direct measurements of the TPE effect are expected from several experiments \cite{vepp3,classtpe,olympus} for large $Q^2$ and from Ref.\ \cite{Gilman:2013fk} for low $Q^2$. They will help resolve this issue.

We observe that larger parameter numbers tend to produce larger magnetic radii. We do not believe that this stems from insufficient flexibility for the lower $N_p$, as a plateau in $\chi^2_\mathrm{red}$ is clearly already reached for the smaller parameter numbers. However, more flexible fits are more susceptible to follow statistical fluctuations. These possibly less reliable fits have larger uncertainties than fits with smaller parameter numbers. Hence, with the chosen weighted averaging, the impact of these fits is lessened. However, this choice and the choice of the parameter number are somewhat subjective. For comparison, an unweighted average would enlarge the radius by 0.008 fm.

Previous determinations from elastic electron scattering give a
significantly larger magnetic radius (see Ref.\ \cite{Ron:2011rd} and
references therein). However, since available data for $Q^2 < 0.2$
GeV$^2$ had large error bars and could not resolve the structure the
new data indicates, the validity of the extrapolation of these
previous determinations is questionable. The values for $G_E$ of Ref.\
\cite{Borkowski74} indicate a $\sim1\%$ deviation in the normalization
of the form factors---the extrapolation to $Q^2$ aims at 0.99.
Applying the same shift to the $G_M$ data of that paper, one might
even recognize the wiggle in this old data set.

The hyperfine splitting in electric hydrogen represents an alternative
method to determine $r_M$. The radius $r_M$ enters in the hyperfine
splitting via the Zemach radius $r_Z$. For the extraction, one has to
make an ansatz for the electric and magnetic form factor shape. The
corresponding analysis in Ref.\ \cite{volotka} was performed with the
standard dipole with $r_E=0.8750(68)$\ fm and $r_M$ was left as a fit
parameter. The measured value is $r_Z=1.045(16)$\ fm and the variation
resulted in $r_M=0.778(29)$\ fm in complete agreement with the value
we obtain. However, Carlson {\em et al.}\ \cite{Carlson:2008ke}
elaborated an update of this analysis by including better polarization
corrections, i.e., TPE effects, and more recent form factor
parametrizations. These parametrizations yield $r_Z=1.069$\ fm
\cite{Kelly:2004hm}, 1.091\ fm\cite{Arrington:2006hm}, and 1.089\ fm
\cite{Arrington07}. It is somewhat model dependent to convert these
$r_Z$ to $r_M$, but one gets as an indication $r_M=0.82(1)$\ fm,
0.86(1)\ fm, and 0.86(1)\ fm for the $r_Z$s above, respectively. Their
analysis did not yet take our new data into account and, hence, has to
be taken with the poor knowledge of $G_M$ at small $Q^2$ from the
previous measurements and, consequently, also of $r_Z$ in mind.

The most recent determination of the magnetic radius stems from the
laser spectroscopy of muonic hydrogen \cite{Antognini:1900ns}. From
the measurement of two transition frequencies the Zemach radius
$r_Z=1.082(37) \ \mathrm{fm}$ has been extracted and the charge radius
has been reevaluated as $r_E=0.84087(39)\ \mathrm{fm}$. Using the
analytical ansatz with the standard dipole the authors determine
$r_M=0.87(6)\ \mathrm{fm}$ and claim consistency with electron
scattering. However, assuming the dipole parametrization they used for
the extraction, these values predict a form-factor ratio at odds with that
of polarization experiments (e.g., Ref.\ \cite{Ron:2011rd}). One would need to
shift $r_M$ down to resolve this discrepancy, and then be at odds with
the scattering experiments they compared to, or assume a different
form-factor shape, which could in turn invalidate the extraction.
Electron scattering provides more information than just the radii,
and only if simultaneous agreement in additional moments of the charge
and magnetization distribution is reached, one can claim consistency.

The Zemach radius derived from the form factors presented in this
paper is $r_Z=1.045(4)$\ fm \cite{Distler:2010zq}, i.e., within the
error margin of the laser spectroscopy of muonic hydrogen and
identical to the value from normal hydrogen. Putting this Zemach
radius into the calculation of Carlson {\em et al.}, one gets
$r_M=0.777(10)$\ fm, in perfect agreement with the result from the elastic
electron scattering of this work.

\begin{table}
\begin{center}
\begin{ruledtabular}
\begin{tabular}{l|c}
Method& Magnetic radius $r_M$ in fm\\
\hline
\hline
Spline models (1) &$0.775(12)_\mathrm{stat.}(9)_\mathrm{syst.}(4)_\mathrm{model}$\\
Polynomial models (2) & $0.778(^{+14}_{-15})_\mathrm{stat.}(10)_\mathrm{syst.}(6)_\mathrm{model}$\\
Friedrich-Walcher & $0.807(2)_\mathrm{stat.}(^{+4}_{-1})_\mathrm{syst.}$\\
\hline
\multicolumn {2}{l}{Spline with variable knots + external data:}\\
+ Rosenbluth data & $0.772$\\
+ all external data & $0.769$\\
\hline
{\bf Average of (1),(2)} &$0.777(13)_\mathrm{stat.}(9)_\mathrm{syst.}(5)_\mathrm{model}(2)_\mathrm{group}$\\
With TPE from \cite{Borisyuk:2006uq}&$0.803(13)_\mathrm{stat.}(9)_\mathrm{syst.}(5)_\mathrm{model}(3)_\mathrm{group}$\\
With TPE from \cite{Arrington:2011dn,blundenpc,Blunden05}&$0.799(13)_\mathrm{stat.}(9)_\mathrm{syst.}(5)_\mathrm{model}(3)_\mathrm{group}$\\
\end{tabular}
\end{ruledtabular}
\end{center}
\caption{\label{tabmrad}Results for the magnetic radius.}
\end{table}

\section{Conclusions}
 This paper presents the details of the highly precise measurement of elastic electron scattering off the proton performed at MAMI, the first results of which have already been published in Ref.\ \cite{Bernauer:2010wm}. The analysis differs in some respects from the customary approach as follows:
\begin{itemize}
\item The Rosenbluth separation and classic error propagation are given up as they are unnecessary steps limiting the precision and the amount of information extracted from the data. Instead, we perform a direct fit of sufficiently flexible models for the form factors to the whole body of measurements, avoiding the unnecessary limitation for the Rosenbluth separation and the badly controlled correlation in the resulting form factors. However, we also show the consistency of the two methods. We extracted sensible confidence bands without any approximation to a linear behavior using Monte Carlo techniques and performed an extensive study of model dependency.
\item The calibration problem present in any determination of absolute cross sections has been overcome by fitting the normalization of sets of cross sections of which the relative measurement-to-measurement normalization was well under control by an explicit luminosity measurement using an extra spectrometer. The absolute normalization was fixed by the well-known form-factor values at $Q^2 = 0$. This procedure only needs a weak assumption on the smoothness of the form factors for the $Q^2=0$ limit. Our data has the furthest reach toward lowest $Q^2$ to date, minimizing the impact of this assumption. The use of a spectrometer as luminosity monitor to fix the relative normalization between individual measurements and the large overlap between our data sets makes a precise determination of the absolute normalization for all measurements possible.  
\item We have only applied the standard radiative corrections but not the hitherto debated two-photon-exchange contributions. However, an empirical form has been derived from the  inconsistency of the $G_E$ and $G_M$ data, extracted from measurements with polarized and unpolarized electrons, respectively,  which may be interpreted by radiative corrections as TPE or other physics (see also Ref.\ \cite{Bernauer:2011zz}).
\end{itemize}

The new method has been also applied to the world data set together with the data of this paper. The analysis represents a coherent summary of the present  knowledge of the form factor of the proton. 

From the slopes of the form factors for $Q^2\rightarrow 0$  we determined the electric and magnetic radii of the
proton. The values extracted here from the whole body of electron-scattering data are at variance with those determined recently with very high precision from muonic hydrogen. In spite of a multitude of efforts, there is no generally accepted explanation yet.

The discrepancy for the magnetic radius determined with the different methods is somewhat less dramatic. While its determination from
the hyperfine splitting in electric hydrogen requires the knowledge of the Zemach radius
 (which, in turn, needs the information from the electric and magnetic form factors),
the determination from electron scattering is hampered by the limited sensitivity of
 the cross section to magnetic scattering at low $Q^2$. In contrast to some (re-)analyses of
 the hyperfine splitting and to a new measurement on muonic hydrogen, we determine a
 magnetic radius for the proton which is substantially smaller than the electric radius.
 While this result fits well to the direct measurement of the ratio $G_E/G_M$
using polarized degrees of freedom at quite small $Q^2$, the muonic result does not.

The larger slope (with respect to the dipole) observed in $G_E$ gives rise to  a larger charge radius, compatible with older extractions.  While the data clearly exhibits this feature, the conflict with the muonic Lamb-shift measurements certainly warrants further study. The wiggles in $G_M$ are at the limit of significance and further measurements are needed for an independent verification. If the results of this paper would be confirmed confirmed and if these structures would survive other efforts for an explanation like the application of TPE,  they would hint at the existence of effective degrees of freedom which may be a yardstick feature  to be replicated by theory, for example in lattice calculations (we refer to Refs.\ \cite{Alexandrou:2013joa,Syritsyn:2009mx,Green:2013hja} for current results).
 
\acknowledgments{We thank the accelerator group of MAMI for its
excellent support. This work was supported in part by the Deutsche
Forschungsgemeinschaft with the Collaborative Research Center 443 and 1044, the Federal State of Rhineland-Palatinate and the French CEA and CNRS/IN2P3.}

\bibliography{thesisbib}{}

\newcommand{\noopsort}[1]{} \newcommand{\printfirst}[2]{#1}
  \newcommand{\singleletter}[1]{#1} \newcommand{\switchargs}[2]{#2#1}
\begin{thebibliography}{112}%
\makeatletter
\providecommand \@ifxundefined [1]{%
 \@ifx{#1\undefined}
}%
\providecommand \@ifnum [1]{%
 \ifnum #1\expandafter \@firstoftwo
 \else \expandafter \@secondoftwo
 \fi
}%
\providecommand \@ifx [1]{%
 \ifx #1\expandafter \@firstoftwo
 \else \expandafter \@secondoftwo
 \fi
}%
\providecommand \natexlab [1]{#1}%
\providecommand \enquote  [1]{``#1''}%
\providecommand \bibnamefont  [1]{#1}%
\providecommand \bibfnamefont [1]{#1}%
\providecommand \citenamefont [1]{#1}%
\providecommand \href@noop [0]{\@secondoftwo}%
\providecommand \href [0]{\begingroup \@sanitize@url \@href}%
\providecommand \@href[1]{\@@startlink{#1}\@@href}%
\providecommand \@@href[1]{\endgroup#1\@@endlink}%
\providecommand \@sanitize@url [0]{\catcode `\\12\catcode `\$12\catcode
  `\&12\catcode `\#12\catcode `\^12\catcode `\_12\catcode `\%12\relax}%
\providecommand \@@startlink[1]{}%
\providecommand \@@endlink[0]{}%
\providecommand \url  [0]{\begingroup\@sanitize@url \@url }%
\providecommand \@url [1]{\endgroup\@href {#1}{\urlprefix }}%
\providecommand \urlprefix  [0]{URL }%
\providecommand \Eprint [0]{\href }%
\providecommand \doibase [0]{http://dx.doi.org/}%
\providecommand \selectlanguage [0]{\@gobble}%
\providecommand \bibinfo  [0]{\@secondoftwo}%
\providecommand \bibfield  [0]{\@secondoftwo}%
\providecommand \translation [1]{[#1]}%
\providecommand \BibitemOpen [0]{}%
\providecommand \bibitemStop [0]{}%
\providecommand \bibitemNoStop [0]{.\EOS\space}%
\providecommand \EOS [0]{\spacefactor3000\relax}%
\providecommand \BibitemShut  [1]{\csname bibitem#1\endcsname}%
\let\auto@bib@innerbib\@empty
\bibitem [{\citenamefont {Vanderhaeghen}\ and\ \citenamefont
  {Walcher}(2010)}]{Vanderhaeghen:2010nd}%
  \BibitemOpen
  \bibfield  {author} {\bibinfo {author} {\bibfnamefont {M.}~\bibnamefont
  {Vanderhaeghen}}\ and\ \bibinfo {author} {\bibfnamefont
  {{\singleletter{Th}}.}~\bibnamefont {Walcher}},\ }\href {\doibase
  10.1080/10619127.2011.554757} {\bibfield  {journal} {\bibinfo  {journal}
  {Nuclear Phys. News}\ }\textbf {\bibinfo {volume} {21}},\ \bibinfo {pages}
  {14} (\bibinfo {year} {2010})},\ \Eprint {http://arxiv.org/abs/1008.4225}
  {1008.4225 [hep-ph]} \BibitemShut {NoStop}%
\bibitem [{\citenamefont {Friedrich}\ and\ \citenamefont
  {Walcher}(2003)}]{fw03}%
  \BibitemOpen
  \bibfield  {author} {\bibinfo {author} {\bibfnamefont {J.}~\bibnamefont
  {Friedrich}}\ and\ \bibinfo {author} {\bibfnamefont
  {{\singleletter{Th}}.}~\bibnamefont {Walcher}},\ }\href {\doibase
  10.1140/epja/i2003-10025-3} {\bibfield  {journal} {\bibinfo  {journal} {Eur.
  Phys. J.}\ }\textbf {\bibinfo {volume} {A17}},\ \bibinfo {pages} {607}
  (\bibinfo {year} {2003})},\ \Eprint {http://arxiv.org/abs/hep-ph/0303054}
  {hep-ph/0303054} \BibitemShut {NoStop}%
\bibitem [{\citenamefont {Arrington}(2004)}]{Arrington03}%
  \BibitemOpen
  \bibfield  {author} {\bibinfo {author} {\bibfnamefont {J.}~\bibnamefont
  {Arrington}},\ }\href {\doibase 10.1103/PhysRevC.69.022201} {\bibfield
  {journal} {\bibinfo  {journal} {Phys. Rev.}\ }\textbf {\bibinfo {volume}
  {C69}},\ \bibinfo {pages} {022201} (\bibinfo {year} {2004})},\ \Eprint
  {http://arxiv.org/abs/nucl-ex/0309011} {nucl-ex/0309011} \BibitemShut
  {NoStop}%
\bibitem [{\citenamefont {Arrington}\ \emph {et~al.}(2007)\citenamefont
  {Arrington}, \citenamefont {Melnitchouk},\ and\ \citenamefont
  {Tjon}}]{Arrington07}%
  \BibitemOpen
  \bibfield  {author} {\bibinfo {author} {\bibfnamefont {J.}~\bibnamefont
  {Arrington}}, \bibinfo {author} {\bibfnamefont {W.}~\bibnamefont
  {Melnitchouk}}, \ and\ \bibinfo {author} {\bibfnamefont {J.~A.}\ \bibnamefont
  {Tjon}},\ }\href {\doibase 10.1103/PhysRevC.76.035205} {\bibfield  {journal}
  {\bibinfo  {journal} {Phys. Rev.}\ }\textbf {\bibinfo {volume} {C76}},\
  \bibinfo {pages} {035205} (\bibinfo {year} {2007})},\ \Eprint
  {http://arxiv.org/abs/nucl-ex/0707.1861} {nucl-ex/0707.1861} \BibitemShut
  {NoStop}%
\bibitem [{\citenamefont {Venkat}\ \emph {et~al.}(2011)\citenamefont {Venkat},
  \citenamefont {Arrington}, \citenamefont {Miller},\ and\ \citenamefont
  {Zhan}}]{venkat11}%
  \BibitemOpen
  \bibfield  {author} {\bibinfo {author} {\bibfnamefont {S.}~\bibnamefont
  {Venkat}}, \bibinfo {author} {\bibfnamefont {J.}~\bibnamefont {Arrington}},
  \bibinfo {author} {\bibfnamefont {G.~A.}\ \bibnamefont {Miller}}, \ and\
  \bibinfo {author} {\bibfnamefont {X.}~\bibnamefont {Zhan}},\ }\href {\doibase
  10.1103/PhysRevC.83.015203} {\bibfield  {journal} {\bibinfo  {journal} {Phys.
  Rev. C}\ }\textbf {\bibinfo {volume} {83}},\ \bibinfo {pages} {015203}
  (\bibinfo {year} {2011})}\BibitemShut {NoStop}%
\bibitem [{\citenamefont {Kelly}(2004)}]{Kelly:2004hm}%
  \BibitemOpen
  \bibfield  {author} {\bibinfo {author} {\bibfnamefont {J.~J.}\ \bibnamefont
  {Kelly}},\ }\href {\doibase 10.1103/PhysRevC.70.068202} {\bibfield  {journal}
  {\bibinfo  {journal} {Phys. Rev.}\ }\textbf {\bibinfo {volume} {C70}},\
  \bibinfo {pages} {068202} (\bibinfo {year} {2004})}\BibitemShut {NoStop}%
\bibitem [{\citenamefont {Alberico}\ \emph {et~al.}(2009)\citenamefont
  {Alberico}, \citenamefont {Bilenky}, \citenamefont {Giunti},\ and\
  \citenamefont {Graczyk}}]{alberico09}%
  \BibitemOpen
  \bibfield  {author} {\bibinfo {author} {\bibfnamefont {W.~M.}\ \bibnamefont
  {Alberico}}, \bibinfo {author} {\bibfnamefont {S.~M.}\ \bibnamefont
  {Bilenky}}, \bibinfo {author} {\bibfnamefont {C.}~\bibnamefont {Giunti}}, \
  and\ \bibinfo {author} {\bibfnamefont {K.~M.}\ \bibnamefont {Graczyk}},\
  }\href {\doibase 10.1103/PhysRevC.79.065204} {\bibfield  {journal} {\bibinfo
  {journal} {Phys. Rev. C}\ }\textbf {\bibinfo {volume} {79}},\ \bibinfo
  {pages} {065204} (\bibinfo {year} {2009})}\BibitemShut {NoStop}%
\bibitem [{\citenamefont {Bernauer}\ \emph {et~al.}(2010)\citenamefont
  {Bernauer}, \citenamefont {Achenbach}, \citenamefont {Ayerbe~Gayoso},
  \citenamefont {B\"ohm}, \citenamefont {Bosnar} \emph
  {et~al.}}]{Bernauer:2010wm}%
  \BibitemOpen
  \bibfield  {author} {\bibinfo {author} {\bibfnamefont {J.~C.}\ \bibnamefont
  {Bernauer}}, \bibinfo {author} {\bibfnamefont {P.}~\bibnamefont {Achenbach}},
  \bibinfo {author} {\bibfnamefont {C.}~\bibnamefont {Ayerbe~Gayoso}}, \bibinfo
  {author} {\bibfnamefont {R.}~\bibnamefont {B\"ohm}}, \bibinfo {author}
  {\bibfnamefont {D.}~\bibnamefont {Bosnar}},  \emph {et~al.} (\bibinfo
  {collaboration} {A1 Collaboration}),\ }\href {\doibase
  10.1103/PhysRevLett.105.242001} {\bibfield  {journal} {\bibinfo  {journal}
  {Phys. Rev. Lett.}\ }\textbf {\bibinfo {volume} {105}},\ \bibinfo {pages}
  {242001} (\bibinfo {year} {2010})},\ \Eprint {http://arxiv.org/abs/1007.5076}
  {1007.5076 [nucl-ex]} \BibitemShut {NoStop}%
\bibitem [{\citenamefont {Blomqvist}\ \emph {et~al.}(1998)\citenamefont
  {Blomqvist}, \citenamefont {Boeglin}, \citenamefont {B\"ohm}, \citenamefont
  {Distler}, \citenamefont {Edelhoff} \emph {et~al.}}]{Blomqvist:1998xn}%
  \BibitemOpen
  \bibfield  {author} {\bibinfo {author} {\bibfnamefont {K.~I.}\ \bibnamefont
  {Blomqvist}}, \bibinfo {author} {\bibfnamefont {W.~U.}\ \bibnamefont
  {Boeglin}}, \bibinfo {author} {\bibfnamefont {R.}~\bibnamefont {B\"ohm}},
  \bibinfo {author} {\bibfnamefont {M.}~\bibnamefont {Distler}}, \bibinfo
  {author} {\bibfnamefont {R.}~\bibnamefont {Edelhoff}},  \emph {et~al.},\
  }\href {\doibase 10.1016/S0168-9002(97)01133-9} {\bibfield  {journal}
  {\bibinfo  {journal} {Nucl. Instrum. Meth.}\ }\textbf {\bibinfo {volume}
  {A403}},\ \bibinfo {pages} {263} (\bibinfo {year} {1998})}\BibitemShut
  {NoStop}%
\bibitem [{\citenamefont {Hofstadter}\ and\ \citenamefont
  {McAllister}(1955)}]{Hofstadter55}%
  \BibitemOpen
  \bibfield  {author} {\bibinfo {author} {\bibfnamefont {R.}~\bibnamefont
  {Hofstadter}}\ and\ \bibinfo {author} {\bibfnamefont {R.~W.}\ \bibnamefont
  {McAllister}},\ }\href {\doibase 10.1103/PhysRev.98.217} {\bibfield
  {journal} {\bibinfo  {journal} {Phys. Rev.}\ }\textbf {\bibinfo {volume}
  {98}},\ \bibinfo {pages} {217} (\bibinfo {year} {1955})}\BibitemShut
  {NoStop}%
\bibitem [{\citenamefont {Guichon}\ and\ \citenamefont
  {Vanderhaeghen}(2003)}]{Guichon03}%
  \BibitemOpen
  \bibfield  {author} {\bibinfo {author} {\bibfnamefont {P.~A.~M.}\
  \bibnamefont {Guichon}}\ and\ \bibinfo {author} {\bibfnamefont
  {M.}~\bibnamefont {Vanderhaeghen}},\ }\href {\doibase
  10.1103/PhysRevLett.91.142303} {\bibfield  {journal} {\bibinfo  {journal}
  {Phys. Rev. Lett.}\ }\textbf {\bibinfo {volume} {91}},\ \bibinfo {pages}
  {142303} (\bibinfo {year} {2003})},\ \Eprint
  {http://arxiv.org/abs/hep-ph/0306007} {hep-ph/0306007} \BibitemShut {NoStop}%
\bibitem [{\citenamefont {Pohl}\ \emph {et~al.}(2010)\citenamefont {Pohl},
  \citenamefont {Antognini}, \citenamefont {Nez}, \citenamefont {Amaro},
  \citenamefont {Biraben} \emph {et~al.}}]{pohl}%
  \BibitemOpen
  \bibfield  {author} {\bibinfo {author} {\bibfnamefont {R.}~\bibnamefont
  {Pohl}}, \bibinfo {author} {\bibfnamefont {A.}~\bibnamefont {Antognini}},
  \bibinfo {author} {\bibfnamefont {F.}~\bibnamefont {Nez}}, \bibinfo {author}
  {\bibfnamefont {F.~D.}\ \bibnamefont {Amaro}}, \bibinfo {author}
  {\bibfnamefont {F.}~\bibnamefont {Biraben}},  \emph {et~al.},\ }\href
  {\doibase 10.1038/nature09250} {\bibfield  {journal} {\bibinfo  {journal}
  {Nature}\ }\textbf {\bibinfo {volume} {466}},\ \bibinfo {pages} {213}
  (\bibinfo {year} {2010})}\BibitemShut {NoStop}%
\bibitem [{\citenamefont {Mohr}\ \emph {et~al.}(2008)\citenamefont {Mohr},
  \citenamefont {Taylor},\ and\ \citenamefont {Newell}}]{Mohr08}%
  \BibitemOpen
  \bibfield  {author} {\bibinfo {author} {\bibfnamefont {P.~J.}\ \bibnamefont
  {Mohr}}, \bibinfo {author} {\bibfnamefont {B.~N.}\ \bibnamefont {Taylor}}, \
  and\ \bibinfo {author} {\bibfnamefont {D.~B.}\ \bibnamefont {Newell}},\
  }\href {\doibase 10.1103/RevModPhys.80.633} {\bibfield  {journal} {\bibinfo
  {journal} {Rev. Mod. Phys.}\ }\textbf {\bibinfo {volume} {80}},\ \bibinfo
  {pages} {633} (\bibinfo {year} {2008})},\ \Eprint
  {http://arxiv.org/abs/physics.atom-ph/0801.0028} {physics.atom-ph/0801.0028}
  \BibitemShut {NoStop}%
\bibitem [{\citenamefont {Antognini}\ \emph {et~al.}(2013)\citenamefont
  {Antognini}, \citenamefont {Nez}, \citenamefont {Schuhmann}, \citenamefont
  {Amaro}, \citenamefont {Biraben} \emph {et~al.}}]{Antognini:1900ns}%
  \BibitemOpen
  \bibfield  {author} {\bibinfo {author} {\bibfnamefont {A.}~\bibnamefont
  {Antognini}}, \bibinfo {author} {\bibfnamefont {F.}~\bibnamefont {Nez}},
  \bibinfo {author} {\bibfnamefont {K.}~\bibnamefont {Schuhmann}}, \bibinfo
  {author} {\bibfnamefont {F.~D.}\ \bibnamefont {Amaro}}, \bibinfo {author}
  {\bibfnamefont {F.}~\bibnamefont {Biraben}},  \emph {et~al.},\ }\href
  {\doibase 10.1126/science.1230016} {\bibfield  {journal} {\bibinfo  {journal}
  {Science}\ }\textbf {\bibinfo {volume} {339}},\ \bibinfo {pages} {417}
  (\bibinfo {year} {2013})}\BibitemShut {NoStop}%
\bibitem [{\citenamefont {Volotka}\ \emph {et~al.}(2005)\citenamefont
  {Volotka}, \citenamefont {Shabaev}, \citenamefont {Plunien},\ and\
  \citenamefont {Soff}}]{volotka}%
  \BibitemOpen
  \bibfield  {author} {\bibinfo {author} {\bibfnamefont {A.~V.}\ \bibnamefont
  {Volotka}}, \bibinfo {author} {\bibfnamefont {V.~M.}\ \bibnamefont
  {Shabaev}}, \bibinfo {author} {\bibfnamefont {G.}~\bibnamefont {Plunien}}, \
  and\ \bibinfo {author} {\bibfnamefont {G.}~\bibnamefont {Soff}},\ }\href
  {\doibase 10.1140/epjd/e2005-00025-9} {\bibfield  {journal} {\bibinfo
  {journal} {Eur. Phys. J.}\ }\textbf {\bibinfo {volume} {D33}},\ \bibinfo
  {pages} {23} (\bibinfo {year} {2005})}\BibitemShut {NoStop}%
\bibitem [{\citenamefont {Ron}\ \emph {et~al.}(2011)\citenamefont {Ron},
  \citenamefont {Zhan}, \citenamefont {Glister}, \citenamefont {Lee},
  \citenamefont {Allada} \emph {et~al.}}]{Ron:2011rd}%
  \BibitemOpen
  \bibfield  {author} {\bibinfo {author} {\bibfnamefont {G.}~\bibnamefont
  {Ron}}, \bibinfo {author} {\bibfnamefont {X.}~\bibnamefont {Zhan}}, \bibinfo
  {author} {\bibfnamefont {J.}~\bibnamefont {Glister}}, \bibinfo {author}
  {\bibfnamefont {B.}~\bibnamefont {Lee}}, \bibinfo {author} {\bibfnamefont
  {K.}~\bibnamefont {Allada}},  \emph {et~al.} (\bibinfo {collaboration} {The
  Jefferson Lab Hall A Collaboration}),\ }\href {\doibase
  10.1103/PhysRevC.84.055204} {\bibfield  {journal} {\bibinfo  {journal} {Phys.
  Rev.}\ }\textbf {\bibinfo {volume} {C84}},\ \bibinfo {pages} {055204}
  (\bibinfo {year} {2011})},\ \Eprint {http://arxiv.org/abs/1103.5784}
  {1103.5784 [nucl-ex]} \BibitemShut {NoStop}%
\bibitem [{\citenamefont {Meziane}\ \emph {et~al.}(2011)\citenamefont
  {Meziane}, \citenamefont {Brash}, \citenamefont {Gilman}, \citenamefont
  {Jones}, \citenamefont {Luo} \emph {et~al.}}]{Meziane:2010xc}%
  \BibitemOpen
  \bibfield  {author} {\bibinfo {author} {\bibfnamefont {M.}~\bibnamefont
  {Meziane}}, \bibinfo {author} {\bibfnamefont {E.~J.}\ \bibnamefont {Brash}},
  \bibinfo {author} {\bibfnamefont {R.}~\bibnamefont {Gilman}}, \bibinfo
  {author} {\bibfnamefont {M.~K.}\ \bibnamefont {Jones}}, \bibinfo {author}
  {\bibfnamefont {W.}~\bibnamefont {Luo}},  \emph {et~al.} (\bibinfo
  {collaboration} {GEp2gamma Collaboration}),\ }\href {\doibase
  10.1103/PhysRevLett.106.132501} {\bibfield  {journal} {\bibinfo  {journal}
  {Phys. Rev. Lett.}\ }\textbf {\bibinfo {volume} {106}},\ \bibinfo {pages}
  {132501} (\bibinfo {year} {2011})},\ \Eprint {http://arxiv.org/abs/1012.0339}
  {1012.0339 [nucl-ex]} \BibitemShut {NoStop}%
\bibitem [{\citenamefont {Arrington}(2011)}]{Arrington:2011kv}%
  \BibitemOpen
  \bibfield  {author} {\bibinfo {author} {\bibfnamefont {J.}~\bibnamefont
  {Arrington}},\ }\href {\doibase 10.1103/PhysRevLett.107.119101} {\bibfield
  {journal} {\bibinfo  {journal} {Phys. Rev. Lett.}\ }\textbf {\bibinfo
  {volume} {107}},\ \bibinfo {pages} {119101} (\bibinfo {year} {2011})},\
  \Eprint {http://arxiv.org/abs/1108.3058} {1108.3058 [nucl-ex]} \BibitemShut
  {NoStop}%
\bibitem [{\citenamefont {Bernauer}\ \emph {et~al.}(2011)\citenamefont
  {Bernauer}, \citenamefont {Achenbach}, \citenamefont {Ayerbe~Gayoso},
  \citenamefont {B\"ohm}, \citenamefont {Bosnar} \emph
  {et~al.}}]{Bernauer:2011zz}%
  \BibitemOpen
  \bibfield  {author} {\bibinfo {author} {\bibfnamefont {J.~C.}\ \bibnamefont
  {Bernauer}}, \bibinfo {author} {\bibfnamefont {P.}~\bibnamefont {Achenbach}},
  \bibinfo {author} {\bibfnamefont {C.}~\bibnamefont {Ayerbe~Gayoso}}, \bibinfo
  {author} {\bibfnamefont {R.}~\bibnamefont {B\"ohm}}, \bibinfo {author}
  {\bibfnamefont {D.}~\bibnamefont {Bosnar}},  \emph {et~al.},\ }\href
  {\doibase 10.1103/PhysRevLett.107.119102} {\bibfield  {journal} {\bibinfo
  {journal} {Phys. Rev. Lett.}\ }\textbf {\bibinfo {volume} {107}},\ \bibinfo
  {pages} {119102} (\bibinfo {year} {2011})}\BibitemShut {NoStop}%
\bibitem [{\citenamefont {Meissner}(2007)}]{Meissner:2007tp}%
  \BibitemOpen
  \bibfield  {author} {\bibinfo {author} {\bibfnamefont {U.-G.}\ \bibnamefont
  {Meissner}},\ }\href {\doibase 10.1063/1.2734299} {\bibfield  {journal}
  {\bibinfo  {journal} {AIP Conf. Proc.}\ }\textbf {\bibinfo {volume} {904}},\
  \bibinfo {pages} {142} (\bibinfo {year} {2007})},\ \Eprint
  {http://arxiv.org/abs/nucl-th/0701094} {nucl-th/0701094 [nucl-th]}
  \BibitemShut {NoStop}%
\bibitem [{sup()}]{supp}%
  \BibitemOpen
  \href@noop {} {}\bibinfo {note} {Available at
  \url{http://journals.aps.org/prc/abstract/10.1103/PhysRevC.90.015206}}\BibitemShut
  {NoStop}%
\bibitem [{\citenamefont {Herminghaus}\ \emph {et~al.}(1976)\citenamefont
  {Herminghaus}, \citenamefont {Feder}, \citenamefont {Kaiser}, \citenamefont
  {Manz},\ and\ \citenamefont {Von Der~Schmitt}}]{Herminghaus76}%
  \BibitemOpen
  \bibfield  {author} {\bibinfo {author} {\bibfnamefont {H.}~\bibnamefont
  {Herminghaus}}, \bibinfo {author} {\bibfnamefont {A.}~\bibnamefont {Feder}},
  \bibinfo {author} {\bibfnamefont {K.~H.}\ \bibnamefont {Kaiser}}, \bibinfo
  {author} {\bibfnamefont {W.}~\bibnamefont {Manz}}, \ and\ \bibinfo {author}
  {\bibfnamefont {H.}~\bibnamefont {Von Der~Schmitt}},\ }\href {\doibase
  10.1016/0029-554X(76)90145-2} {\bibfield  {journal} {\bibinfo  {journal}
  {Nucl. Instrum. Methods}\ }\textbf {\bibinfo {volume} {138}},\ \bibinfo
  {pages} {1} (\bibinfo {year} {1976})}\BibitemShut {NoStop}%
\bibitem [{\citenamefont {Jankowiak}(2006)}]{Jankowiak:2006yc}%
  \BibitemOpen
  \bibfield  {author} {\bibinfo {author} {\bibfnamefont {A.}~\bibnamefont
  {Jankowiak}},\ }\href {\doibase 10.1140/epja/i2006-09-016-3} {\bibfield
  {journal} {\bibinfo  {journal} {Eur. Phys. J.}\ }\textbf {\bibinfo {volume}
  {A28S1}},\ \bibinfo {pages} {149} (\bibinfo {year} {2006})}\BibitemShut
  {NoStop}%
\bibitem [{\citenamefont {Kaiser}\ \emph {et~al.}(2008)\citenamefont {Kaiser},
  \citenamefont {Aulenbacher}, \citenamefont {Chubarov}, \citenamefont {Dehn},
  \citenamefont {Euteneuer} \emph {et~al.}}]{Kaiser:2008zz}%
  \BibitemOpen
  \bibfield  {author} {\bibinfo {author} {\bibfnamefont {K.-H.}\ \bibnamefont
  {Kaiser}}, \bibinfo {author} {\bibfnamefont {K.}~\bibnamefont {Aulenbacher}},
  \bibinfo {author} {\bibfnamefont {O.}~\bibnamefont {Chubarov}}, \bibinfo
  {author} {\bibfnamefont {M.}~\bibnamefont {Dehn}}, \bibinfo {author}
  {\bibfnamefont {H.}~\bibnamefont {Euteneuer}},  \emph {et~al.},\ }\href
  {\doibase 10.1016/j.nima.2008.05.018} {\bibfield  {journal} {\bibinfo
  {journal} {Nucl. Instrum. Methods}\ }\textbf {\bibinfo {volume} {A593}},\
  \bibinfo {pages} {159} (\bibinfo {year} {2008})}\BibitemShut {NoStop}%
\bibitem [{\citenamefont {Bernauer}(2004)}]{Jcb04}%
  \BibitemOpen
  \bibfield  {author} {\bibinfo {author} {\bibfnamefont {J.~C.}\ \bibnamefont
  {Bernauer}},\ }\emph {\bibinfo {title} {{Vorbereitung zur hochpr{\"a}zisen
  Messung des elektrischen und magnetischen Formfaktors von Protonen}}},\
  \href@noop {} {Master's thesis},\ \bibinfo  {school} {Institut f{\"u}r
  Kernphysik der Universit{\"a}t Mainz} (\bibinfo {year} {2004})\BibitemShut
  {NoStop}%
\bibitem [{\citenamefont {Dehn}(2010)}]{dehn}%
  \BibitemOpen
  \bibfield  {author} {\bibinfo {author} {\bibfnamefont {M.}~\bibnamefont
  {Dehn}},\ }\href@noop {} {\enquote {\bibinfo {title} {{A1 beam position
  stabilization}},}\ } (\bibinfo {year} {2010}),\ \bibinfo {note} {(private
  communication)}\BibitemShut {NoStop}%
\bibitem [{\citenamefont {Yennie}\ \emph {et~al.}(1957)\citenamefont {Yennie},
  \citenamefont {L\'evy},\ and\ \citenamefont {Ravenhall}}]{yennie57}%
  \BibitemOpen
  \bibfield  {author} {\bibinfo {author} {\bibfnamefont {D.~R.}\ \bibnamefont
  {Yennie}}, \bibinfo {author} {\bibfnamefont {M.~M.}\ \bibnamefont {L\'evy}},
  \ and\ \bibinfo {author} {\bibfnamefont {D.~G.}\ \bibnamefont {Ravenhall}},\
  }\href {\doibase 10.1103/RevModPhys.29.144} {\bibfield  {journal} {\bibinfo
  {journal} {Rev. Mod. Phys.}\ }\textbf {\bibinfo {volume} {29}},\ \bibinfo
  {pages} {144} (\bibinfo {year} {1957})}\BibitemShut {NoStop}%
\bibitem [{\citenamefont {Ernst}\ \emph {et~al.}(1960)\citenamefont {Ernst},
  \citenamefont {Sachs},\ and\ \citenamefont {Wali}}]{Ernst}%
  \BibitemOpen
  \bibfield  {author} {\bibinfo {author} {\bibfnamefont {F.~J.}\ \bibnamefont
  {Ernst}}, \bibinfo {author} {\bibfnamefont {R.~G.}\ \bibnamefont {Sachs}}, \
  and\ \bibinfo {author} {\bibfnamefont {K.~C.}\ \bibnamefont {Wali}},\ }\href
  {\doibase 10.1103/PhysRev.119.1105} {\bibfield  {journal} {\bibinfo
  {journal} {Phys. Rev.}\ }\textbf {\bibinfo {volume} {119}},\ \bibinfo {pages}
  {1105} (\bibinfo {year} {1960})}\BibitemShut {NoStop}%
\bibitem [{\citenamefont {Sachs}(1962)}]{Sachs}%
  \BibitemOpen
  \bibfield  {author} {\bibinfo {author} {\bibfnamefont {R.~G.}\ \bibnamefont
  {Sachs}},\ }\href {\doibase 10.1103/PhysRev.126.2256} {\bibfield  {journal}
  {\bibinfo  {journal} {Phys. Rev.}\ }\textbf {\bibinfo {volume} {126}},\
  \bibinfo {pages} {2256} (\bibinfo {year} {1962})}\BibitemShut {NoStop}%
\bibitem [{\citenamefont {Rosenbluth}(1950)}]{Rosenbluth}%
  \BibitemOpen
  \bibfield  {author} {\bibinfo {author} {\bibfnamefont {M.~N.}\ \bibnamefont
  {Rosenbluth}},\ }\href {\doibase 10.1103/PhysRev.79.615} {\bibfield
  {journal} {\bibinfo  {journal} {Phys. Rev.}\ }\textbf {\bibinfo {volume}
  {79}},\ \bibinfo {pages} {615} (\bibinfo {year} {1950})}\BibitemShut
  {NoStop}%
\bibitem [{\citenamefont {Maximon}\ and\ \citenamefont
  {Tjon}(2000)}]{Maximon2000}%
  \BibitemOpen
  \bibfield  {author} {\bibinfo {author} {\bibfnamefont {L.~C.}\ \bibnamefont
  {Maximon}}\ and\ \bibinfo {author} {\bibfnamefont {J.~A.}\ \bibnamefont
  {Tjon}},\ }\href {\doibase 10.1103/PhysRevC.62.054320} {\bibfield  {journal}
  {\bibinfo  {journal} {Phys. Rev.}\ }\textbf {\bibinfo {volume} {C62}},\
  \bibinfo {pages} {054320} (\bibinfo {year} {2000})},\ \Eprint
  {http://arxiv.org/abs/nucl-th/0002058} {nucl-th/0002058} \BibitemShut
  {NoStop}%
\bibitem [{\citenamefont {Vanderhaeghen}\ \emph {et~al.}(2000)\citenamefont
  {Vanderhaeghen}, \citenamefont {Friedrich}, \citenamefont {Lhuillier},
  \citenamefont {Marchand}, \citenamefont {Van~Hoorebeke},\ and\ \citenamefont
  {Van~de Wiele}}]{Vanderhaeghen2000}%
  \BibitemOpen
  \bibfield  {author} {\bibinfo {author} {\bibfnamefont {M.}~\bibnamefont
  {Vanderhaeghen}}, \bibinfo {author} {\bibfnamefont {J.~M.}\ \bibnamefont
  {Friedrich}}, \bibinfo {author} {\bibfnamefont {D.}~\bibnamefont
  {Lhuillier}}, \bibinfo {author} {\bibfnamefont {D.}~\bibnamefont {Marchand}},
  \bibinfo {author} {\bibfnamefont {L.}~\bibnamefont {Van~Hoorebeke}}, \ and\
  \bibinfo {author} {\bibfnamefont {J.}~\bibnamefont {Van~de Wiele}},\ }\href
  {\doibase 10.1103/PhysRevC.62.025501} {\bibfield  {journal} {\bibinfo
  {journal} {Phys. Rev.}\ }\textbf {\bibinfo {volume} {C62}},\ \bibinfo {pages}
  {025501} (\bibinfo {year} {2000})},\ \Eprint
  {http://arxiv.org/abs/hep-ph/0001100} {hep-ph/0001100} \BibitemShut {NoStop}%
\bibitem [{\citenamefont {Bloch}\ and\ \citenamefont
  {Nordsieck}(1937)}]{Bloch1937}%
  \BibitemOpen
  \bibfield  {author} {\bibinfo {author} {\bibfnamefont {F.}~\bibnamefont
  {Bloch}}\ and\ \bibinfo {author} {\bibfnamefont {A.}~\bibnamefont
  {Nordsieck}},\ }\href {\doibase 10.1103/PhysRev.52.54} {\bibfield  {journal}
  {\bibinfo  {journal} {Phys. Rev.}\ }\textbf {\bibinfo {volume} {52}},\
  \bibinfo {pages} {54} (\bibinfo {year} {1937})}\BibitemShut {NoStop}%
\bibitem [{\citenamefont {Jauch}\ and\ \citenamefont
  {Rohrlich}(1954)}]{Jauch1954}%
  \BibitemOpen
  \bibfield  {author} {\bibinfo {author} {\bibfnamefont {J.~M.}\ \bibnamefont
  {Jauch}}\ and\ \bibinfo {author} {\bibfnamefont {F.}~\bibnamefont
  {Rohrlich}},\ }\href@noop {} {\bibfield  {journal} {\bibinfo  {journal}
  {Helvet, Phys. Acta}\ }\textbf {\bibinfo {volume} {27}},\ \bibinfo {pages}
  {613} (\bibinfo {year} {1954})}\BibitemShut {NoStop}%
\bibitem [{\citenamefont {Friedrich}(2000)}]{friedjm00}%
  \BibitemOpen
  \bibfield  {author} {\bibinfo {author} {\bibfnamefont {J.~M.}\ \bibnamefont
  {Friedrich}},\ }\emph {\bibinfo {title} {{Messung der Virtuellen
  Comptonstreuung an MAMI zur Bestimmung Generalisierter Polarisierbarkeiten
  des Protons}}},\ \href@noop {} {Ph.D. thesis},\ \bibinfo  {school} {Johannes
  Gutenberg-Universit{\"a}t Mainz} (\bibinfo {year} {2000})\BibitemShut
  {NoStop}%
\bibitem [{\citenamefont {Yennie}\ \emph {et~al.}(1961)\citenamefont {Yennie},
  \citenamefont {Frautschi},\ and\ \citenamefont {Suura}}]{Yennie1961}%
  \BibitemOpen
  \bibfield  {author} {\bibinfo {author} {\bibfnamefont {D.~R.}\ \bibnamefont
  {Yennie}}, \bibinfo {author} {\bibfnamefont {S.~C.}\ \bibnamefont
  {Frautschi}}, \ and\ \bibinfo {author} {\bibfnamefont {H.}~\bibnamefont
  {Suura}},\ }\href {\doibase 10.1016/0003-4916(61)90151-8} {\bibfield
  {journal} {\bibinfo  {journal} {Ann. Phys.}\ }\textbf {\bibinfo {volume}
  {13}},\ \bibinfo {pages} {379} (\bibinfo {year} {1961})}\BibitemShut
  {NoStop}%
\bibitem [{\citenamefont {Rosenfelder}(2000)}]{Rosenfelder99}%
  \BibitemOpen
  \bibfield  {author} {\bibinfo {author} {\bibfnamefont {R.}~\bibnamefont
  {Rosenfelder}},\ }\href {\doibase 10.1016/S0370-2693(00)00316-6} {\bibfield
  {journal} {\bibinfo  {journal} {Phys. Lett.}\ }\textbf {\bibinfo {volume}
  {B479}},\ \bibinfo {pages} {381} (\bibinfo {year} {2000})},\ \Eprint
  {http://arxiv.org/abs/nucl-th/9912031} {nucl-th/9912031} \BibitemShut
  {NoStop}%
\bibitem [{\citenamefont {McKinley}\ and\ \citenamefont
  {Feshbach}(1948)}]{McKinley:1948zz}%
  \BibitemOpen
  \bibfield  {author} {\bibinfo {author} {\bibfnamefont {W.~A.}\ \bibnamefont
  {McKinley}}\ and\ \bibinfo {author} {\bibfnamefont {H.}~\bibnamefont
  {Feshbach}},\ }\href {\doibase 10.1103/PhysRev.74.1759} {\bibfield  {journal}
  {\bibinfo  {journal} {Phys. Rev.}\ }\textbf {\bibinfo {volume} {74}},\
  \bibinfo {pages} {1759} (\bibinfo {year} {1948})}\BibitemShut {NoStop}%
\bibitem [{\citenamefont {Tsai}(1961)}]{tsai61}%
  \BibitemOpen
  \bibfield  {author} {\bibinfo {author} {\bibfnamefont {Y.-S.}\ \bibnamefont
  {Tsai}},\ }\href {\doibase 10.1103/PhysRev.122.1898} {\bibfield  {journal}
  {\bibinfo  {journal} {Phys. Rev.}\ }\textbf {\bibinfo {volume} {122}},\
  \bibinfo {pages} {1898} (\bibinfo {year} {1961})}\BibitemShut {NoStop}%
\bibitem [{\citenamefont {{Jover Ma\~{n}as}}(2003)}]{Jovas03}%
  \BibitemOpen
  \bibfield  {author} {\bibinfo {author} {\bibfnamefont {G.~V.}\ \bibnamefont
  {{Jover Ma\~{n}as}}},\ }\emph {\bibinfo {title} {{Simulation of Double
  Polarization Variables in Virtual Compton Scattering}}},\ \href@noop {}
  {Master's thesis},\ \bibinfo  {school} {Institut f{\"u}r Kernphysik der
  Universit{\"a}t Mainz} (\bibinfo {year} {2003})\BibitemShut {NoStop}%
\bibitem [{\citenamefont {Press}\ \emph {et~al.}(1992)\citenamefont {Press},
  \citenamefont {Teukolsky}, \citenamefont {Vetterling},\ and\ \citenamefont
  {Flannery}}]{NR}%
  \BibitemOpen
  \bibfield  {author} {\bibinfo {author} {\bibfnamefont {W.}~\bibnamefont
  {Press}}, \bibinfo {author} {\bibfnamefont {S.}~\bibnamefont {Teukolsky}},
  \bibinfo {author} {\bibfnamefont {W.}~\bibnamefont {Vetterling}}, \ and\
  \bibinfo {author} {\bibfnamefont {B.}~\bibnamefont {Flannery}},\ }\href@noop
  {} {\emph {\bibinfo {title} {{Numerical Recipes in C}}}},\ \bibinfo {edition}
  {2nd}\ ed.\ (\bibinfo  {publisher} {Cambridge University Press},\ \bibinfo
  {address} {Cambridge, UK},\ \bibinfo {year} {1992})\BibitemShut {NoStop}%
\bibitem [{\citenamefont {Bernauer}(2010)}]{bernauerphd}%
  \BibitemOpen
  \bibfield  {author} {\bibinfo {author} {\bibfnamefont {J.~C.}\ \bibnamefont
  {Bernauer}},\ }\emph {\bibinfo {title} {{Measurement of the elastic
  electron-proton cross section and separation of the electric and magnetic
  form factor in the $Q^2$ range from 0.004 to 1~$ ( \mathrm{GeV}/c)^{2}$}}},\
  \href@noop {} {Ph.D. thesis},\ \bibinfo  {school} {Johannes
  Gutenberg-Universit{\"a}t Mainz} (\bibinfo {year} {2010})\BibitemShut
  {NoStop}%
\bibitem [{\citenamefont {Crawford}\ \emph {et~al.}(2007)\citenamefont
  {Crawford}, \citenamefont {Sindile}, \citenamefont {Akdogan}, \citenamefont
  {Alarcon}, \citenamefont {Bertozzi} \emph {et~al.}}]{Crawford07}%
  \BibitemOpen
  \bibfield  {author} {\bibinfo {author} {\bibfnamefont {C.~B.}\ \bibnamefont
  {Crawford}}, \bibinfo {author} {\bibfnamefont {A.}~\bibnamefont {Sindile}},
  \bibinfo {author} {\bibfnamefont {T.}~\bibnamefont {Akdogan}}, \bibinfo
  {author} {\bibfnamefont {R.}~\bibnamefont {Alarcon}}, \bibinfo {author}
  {\bibfnamefont {W.}~\bibnamefont {Bertozzi}},  \emph {et~al.},\ }\href
  {\doibase 10.1103/PhysRevLett.98.052301} {\bibfield  {journal} {\bibinfo
  {journal} {Phys. Rev. Lett.}\ }\textbf {\bibinfo {volume} {98}},\ \bibinfo
  {pages} {052301} (\bibinfo {year} {2007})}\BibitemShut {NoStop}%
\bibitem [{\citenamefont {Gayou}\ \emph {et~al.}(2001)\citenamefont {Gayou},
  \citenamefont {Wijesooriya}, \citenamefont {Afanasev}, \citenamefont
  {Amarian}, \citenamefont {Aniol} \emph {et~al.}}]{Gayou01}%
  \BibitemOpen
  \bibfield  {author} {\bibinfo {author} {\bibfnamefont {O.}~\bibnamefont
  {Gayou}}, \bibinfo {author} {\bibfnamefont {K.}~\bibnamefont {Wijesooriya}},
  \bibinfo {author} {\bibfnamefont {A.}~\bibnamefont {Afanasev}}, \bibinfo
  {author} {\bibfnamefont {M.}~\bibnamefont {Amarian}}, \bibinfo {author}
  {\bibfnamefont {K.}~\bibnamefont {Aniol}},  \emph {et~al.},\ }\href {\doibase
  10.1103/PhysRevC.64.038202} {\bibfield  {journal} {\bibinfo  {journal} {Phys.
  Rev.}\ }\textbf {\bibinfo {volume} {C64}},\ \bibinfo {pages} {038202}
  (\bibinfo {year} {2001})}\BibitemShut {NoStop}%
\bibitem [{\citenamefont {Gayou}\ \emph {et~al.}(2002)\citenamefont {Gayou},
  \citenamefont {Aniol}, \citenamefont {Averett}, \citenamefont {Benmokhtar},
  \citenamefont {Bertozzi} \emph {et~al.}}]{Gayou02}%
  \BibitemOpen
  \bibfield  {author} {\bibinfo {author} {\bibfnamefont {O.}~\bibnamefont
  {Gayou}}, \bibinfo {author} {\bibfnamefont {K.~A.}\ \bibnamefont {Aniol}},
  \bibinfo {author} {\bibfnamefont {T.}~\bibnamefont {Averett}}, \bibinfo
  {author} {\bibfnamefont {F.}~\bibnamefont {Benmokhtar}}, \bibinfo {author}
  {\bibfnamefont {W.}~\bibnamefont {Bertozzi}},  \emph {et~al.} (\bibinfo
  {collaboration} {Jefferson Lab Hall A Collaboration}),\ }\href {\doibase
  10.1103/PhysRevLett.88.092301} {\bibfield  {journal} {\bibinfo  {journal}
  {Phys. Rev. Lett.}\ }\textbf {\bibinfo {volume} {88}},\ \bibinfo {pages}
  {092301} (\bibinfo {year} {2002})}\BibitemShut {NoStop}%
\bibitem [{\citenamefont {Puckett}\ \emph {et~al.}(2012)\citenamefont
  {Puckett}, \citenamefont {Brash}, \citenamefont {Gayou}, \citenamefont
  {Jones}, \citenamefont {Pentchev} \emph {et~al.}}]{Puckett:2011xg}%
  \BibitemOpen
  \bibfield  {author} {\bibinfo {author} {\bibfnamefont {A.~J.~R.}\
  \bibnamefont {Puckett}}, \bibinfo {author} {\bibfnamefont {E.~J.}\
  \bibnamefont {Brash}}, \bibinfo {author} {\bibfnamefont {O.}~\bibnamefont
  {Gayou}}, \bibinfo {author} {\bibfnamefont {M.~K.}\ \bibnamefont {Jones}},
  \bibinfo {author} {\bibfnamefont {L.}~\bibnamefont {Pentchev}},  \emph
  {et~al.},\ }\href@noop {} {\bibfield  {journal} {\bibinfo  {journal} {Phys.
  Rev.}\ }\textbf {\bibinfo {volume} {C85}},\ \bibinfo {pages} {045203}
  (\bibinfo {year} {2012})},\ \Eprint {http://arxiv.org/abs/1102.5737}
  {1102.5737 [nucl-ex]} \BibitemShut {NoStop}%
\bibitem [{\citenamefont {Jones}\ \emph {et~al.}(2000)\citenamefont {Jones},
  \citenamefont {Aniol}, \citenamefont {Baker}, \citenamefont {Berthot},
  \citenamefont {Bertin} \emph {et~al.}}]{Jones00}%
  \BibitemOpen
  \bibfield  {author} {\bibinfo {author} {\bibfnamefont {M.~K.}\ \bibnamefont
  {Jones}}, \bibinfo {author} {\bibfnamefont {K.~A.}\ \bibnamefont {Aniol}},
  \bibinfo {author} {\bibfnamefont {F.~T.}\ \bibnamefont {Baker}}, \bibinfo
  {author} {\bibfnamefont {J.}~\bibnamefont {Berthot}}, \bibinfo {author}
  {\bibfnamefont {P.~Y.}\ \bibnamefont {Bertin}},  \emph {et~al.},\ }\href
  {\doibase 10.1103/PhysRevLett.84.1398} {\bibfield  {journal} {\bibinfo
  {journal} {Phys. Rev. Lett.}\ }\textbf {\bibinfo {volume} {84}},\ \bibinfo
  {pages} {1398} (\bibinfo {year} {2000})}\BibitemShut {NoStop}%
\bibitem [{\citenamefont {Jones}\ \emph {et~al.}(2006)\citenamefont {Jones},
  \citenamefont {Aghalaryan}, \citenamefont {Ahmidouch}, \citenamefont
  {Asaturyan}, \citenamefont {Bloch} \emph {et~al.}}]{Jones06}%
  \BibitemOpen
  \bibfield  {author} {\bibinfo {author} {\bibfnamefont {M.~K.}\ \bibnamefont
  {Jones}}, \bibinfo {author} {\bibfnamefont {A.}~\bibnamefont {Aghalaryan}},
  \bibinfo {author} {\bibfnamefont {A.}~\bibnamefont {Ahmidouch}}, \bibinfo
  {author} {\bibfnamefont {R.}~\bibnamefont {Asaturyan}}, \bibinfo {author}
  {\bibfnamefont {F.}~\bibnamefont {Bloch}},  \emph {et~al.} (\bibinfo
  {collaboration} {Resonance Spin Structure Collaboration}),\ }\href {\doibase
  10.1103/PhysRevC.74.035201} {\bibfield  {journal} {\bibinfo  {journal} {Phys.
  Rev.}\ }\textbf {\bibinfo {volume} {C74}},\ \bibinfo {pages} {035201}
  (\bibinfo {year} {2006})}\BibitemShut {NoStop}%
\bibitem [{\citenamefont {MacLachlan}\ \emph {et~al.}(2006)\citenamefont
  {MacLachlan}, \citenamefont {Aghalaryan}, \citenamefont {Ahmidouch},
  \citenamefont {Anderson}, \citenamefont {Asaturyan} \emph
  {et~al.}}]{MacLachlan06}%
  \BibitemOpen
  \bibfield  {author} {\bibinfo {author} {\bibfnamefont {G.}~\bibnamefont
  {MacLachlan}}, \bibinfo {author} {\bibfnamefont {A.}~\bibnamefont
  {Aghalaryan}}, \bibinfo {author} {\bibfnamefont {A.}~\bibnamefont
  {Ahmidouch}}, \bibinfo {author} {\bibfnamefont {B.~D.}\ \bibnamefont
  {Anderson}}, \bibinfo {author} {\bibfnamefont {R.}~\bibnamefont {Asaturyan}},
   \emph {et~al.},\ }\href {\doibase 10.1016/j.nuclphysa.2005.09.012}
  {\bibfield  {journal} {\bibinfo  {journal} {Nuclear Phys.}\ }\textbf
  {\bibinfo {volume} {A764}},\ \bibinfo {pages} {261 } (\bibinfo {year}
  {2006})}\BibitemShut {NoStop}%
\bibitem [{\citenamefont {Milbrath}\ \emph {et~al.}(1998)\citenamefont
  {Milbrath}, \citenamefont {McIntyre}, \citenamefont {Armstrong},
  \citenamefont {Barkhuff}, \citenamefont {Bertozzi} \emph
  {et~al.}}]{Milbrath98}%
  \BibitemOpen
  \bibfield  {author} {\bibinfo {author} {\bibfnamefont {B.~D.}\ \bibnamefont
  {Milbrath}}, \bibinfo {author} {\bibfnamefont {J.~I.}\ \bibnamefont
  {McIntyre}}, \bibinfo {author} {\bibfnamefont {C.~S.}\ \bibnamefont
  {Armstrong}}, \bibinfo {author} {\bibfnamefont {D.~H.}\ \bibnamefont
  {Barkhuff}}, \bibinfo {author} {\bibfnamefont {W.}~\bibnamefont {Bertozzi}},
  \emph {et~al.},\ }\href {\doibase 10.1103/PhysRevLett.80.452} {\bibfield
  {journal} {\bibinfo  {journal} {Phys. Rev. Lett.}\ }\textbf {\bibinfo
  {volume} {80}},\ \bibinfo {pages} {452} (\bibinfo {year} {1998})}\BibitemShut
  {NoStop}%
\bibitem [{\citenamefont {Pospischil}\ \emph {et~al.}(2001)\citenamefont
  {Pospischil}, \citenamefont {Bartsch}, \citenamefont {Baumann}, \citenamefont
  {B{\"o}hm}, \citenamefont {Bohinc} \emph {et~al.}}]{Pospischil01}%
  \BibitemOpen
  \bibfield  {author} {\bibinfo {author} {\bibfnamefont {T.}~\bibnamefont
  {Pospischil}}, \bibinfo {author} {\bibfnamefont {P.}~\bibnamefont {Bartsch}},
  \bibinfo {author} {\bibfnamefont {D.}~\bibnamefont {Baumann}}, \bibinfo
  {author} {\bibfnamefont {R.}~\bibnamefont {B{\"o}hm}}, \bibinfo {author}
  {\bibfnamefont {K.}~\bibnamefont {Bohinc}},  \emph {et~al.} (\bibinfo
  {collaboration} {A1 Collaboration}),\ }\href {\doibase 10.1007/s100500170046}
  {\bibfield  {journal} {\bibinfo  {journal} {Eur. Phys. J.}\ }\textbf
  {\bibinfo {volume} {A12}},\ \bibinfo {pages} {125} (\bibinfo {year}
  {2001})}\BibitemShut {NoStop}%
\bibitem [{\citenamefont {Puckett}\ \emph {et~al.}(2010)\citenamefont
  {Puckett}, \citenamefont {Brash}, \citenamefont {Jones}, \citenamefont {Luo},
  \citenamefont {Meziane} \emph {et~al.}}]{Puckett10}%
  \BibitemOpen
  \bibfield  {author} {\bibinfo {author} {\bibfnamefont {A.~J.~R.}\
  \bibnamefont {Puckett}}, \bibinfo {author} {\bibfnamefont {E.~J.}\
  \bibnamefont {Brash}}, \bibinfo {author} {\bibfnamefont {M.~K.}\ \bibnamefont
  {Jones}}, \bibinfo {author} {\bibfnamefont {W.}~\bibnamefont {Luo}}, \bibinfo
  {author} {\bibfnamefont {M.}~\bibnamefont {Meziane}},  \emph {et~al.},\
  }\href {\doibase 10.1103/PhysRevLett.104.242301} {\bibfield  {journal}
  {\bibinfo  {journal} {Phys. Rev. Lett.}\ }\textbf {\bibinfo {volume} {104}},\
  \bibinfo {pages} {242301} (\bibinfo {year} {2010})}\BibitemShut {NoStop}%
\bibitem [{\citenamefont {Punjabi}\ \emph {et~al.}(2005)\citenamefont
  {Punjabi}, \citenamefont {Perdrisat}, \citenamefont {Aniol}, \citenamefont
  {Baker}, \citenamefont {Berthot} \emph {et~al.}}]{Punjabi05}%
  \BibitemOpen
  \bibfield  {author} {\bibinfo {author} {\bibfnamefont {V.}~\bibnamefont
  {Punjabi}}, \bibinfo {author} {\bibfnamefont {C.~F.}\ \bibnamefont
  {Perdrisat}}, \bibinfo {author} {\bibfnamefont {K.~A.}\ \bibnamefont
  {Aniol}}, \bibinfo {author} {\bibfnamefont {F.~T.}\ \bibnamefont {Baker}},
  \bibinfo {author} {\bibfnamefont {J.}~\bibnamefont {Berthot}},  \emph
  {et~al.},\ }\href {\doibase 10.1103/PhysRevC.71.055202} {\bibfield  {journal}
  {\bibinfo  {journal} {Phys. Rev.}\ }\textbf {\bibinfo {volume} {C71}},\
  \bibinfo {pages} {055202} (\bibinfo {year} {2005})}\BibitemShut {NoStop}%
\bibitem [{\citenamefont {Ron}\ \emph {et~al.}(2007)\citenamefont {Ron},
  \citenamefont {Glister}, \citenamefont {Lee}, \citenamefont {Allada},
  \citenamefont {Armstrong} \emph {et~al.}}]{Ron07}%
  \BibitemOpen
  \bibfield  {author} {\bibinfo {author} {\bibfnamefont {G.}~\bibnamefont
  {Ron}}, \bibinfo {author} {\bibfnamefont {J.}~\bibnamefont {Glister}},
  \bibinfo {author} {\bibfnamefont {B.}~\bibnamefont {Lee}}, \bibinfo {author}
  {\bibfnamefont {K.}~\bibnamefont {Allada}}, \bibinfo {author} {\bibfnamefont
  {W.}~\bibnamefont {Armstrong}},  \emph {et~al.},\ }\href {\doibase
  10.1103/PhysRevLett.99.202002} {\bibfield  {journal} {\bibinfo  {journal}
  {Phys. Rev. Lett.}\ }\textbf {\bibinfo {volume} {99}},\ \bibinfo {pages}
  {202002} (\bibinfo {year} {2007})},\ \Eprint
  {http://arxiv.org/abs/nucl-ex/0706.0128} {nucl-ex/0706.0128} \BibitemShut
  {NoStop}%
\bibitem [{\citenamefont {Zhan}\ \emph {et~al.}(2011)\citenamefont {Zhan},
  \citenamefont {Allada}, \citenamefont {Armstrong}, \citenamefont {Arrington},
  \citenamefont {Bertozzi} \emph {et~al.}}]{Zhan11}%
  \BibitemOpen
  \bibfield  {author} {\bibinfo {author} {\bibfnamefont {X.}~\bibnamefont
  {Zhan}}, \bibinfo {author} {\bibfnamefont {K.}~\bibnamefont {Allada}},
  \bibinfo {author} {\bibfnamefont {D.~S.}\ \bibnamefont {Armstrong}}, \bibinfo
  {author} {\bibfnamefont {J.}~\bibnamefont {Arrington}}, \bibinfo {author}
  {\bibfnamefont {W.}~\bibnamefont {Bertozzi}},  \emph {et~al.},\ }\href
  {\doibase 10.1016/j.physletb.2011.10.002} {\bibfield  {journal} {\bibinfo
  {journal} {Phys. Lett.}\ }\textbf {\bibinfo {volume} {B705}},\ \bibinfo
  {pages} {59} (\bibinfo {year} {2011})},\ \Eprint
  {http://arxiv.org/abs/1102.0318} {1102.0318 [nucl-ex]} \BibitemShut {NoStop}%
\bibitem [{\citenamefont {Christy}\ \emph {et~al.}(2004)\citenamefont
  {Christy}, \citenamefont {Ahmidouch}, \citenamefont {Armstrong},
  \citenamefont {Arrington}, \citenamefont {Asaturyan} \emph
  {et~al.}}]{Christy04}%
  \BibitemOpen
  \bibfield  {author} {\bibinfo {author} {\bibfnamefont {M.~E.}\ \bibnamefont
  {Christy}}, \bibinfo {author} {\bibfnamefont {A.}~\bibnamefont {Ahmidouch}},
  \bibinfo {author} {\bibfnamefont {C.~S.}\ \bibnamefont {Armstrong}}, \bibinfo
  {author} {\bibfnamefont {J.}~\bibnamefont {Arrington}}, \bibinfo {author}
  {\bibfnamefont {R.}~\bibnamefont {Asaturyan}},  \emph {et~al.},\ }\href
  {\doibase 10.1103/PhysRevC.70.015206} {\bibfield  {journal} {\bibinfo
  {journal} {Phys. Rev.}\ }\textbf {\bibinfo {volume} {C70}},\ \bibinfo {pages}
  {015206} (\bibinfo {year} {2004})}\BibitemShut {NoStop}%
\bibitem [{\citenamefont {Janssens}\ \emph {et~al.}(1966)\citenamefont
  {Janssens}, \citenamefont {Hofstadter}, \citenamefont {Hughes},\ and\
  \citenamefont {Yearian}}]{Janssens65}%
  \BibitemOpen
  \bibfield  {author} {\bibinfo {author} {\bibfnamefont {T.}~\bibnamefont
  {Janssens}}, \bibinfo {author} {\bibfnamefont {R.}~\bibnamefont
  {Hofstadter}}, \bibinfo {author} {\bibfnamefont {E.~B.}\ \bibnamefont
  {Hughes}}, \ and\ \bibinfo {author} {\bibfnamefont {M.~R.}\ \bibnamefont
  {Yearian}},\ }\href {\doibase 10.1103/PhysRev.142.922} {\bibfield  {journal}
  {\bibinfo  {journal} {Phys. Rev.}\ }\textbf {\bibinfo {volume} {142}},\
  \bibinfo {pages} {922} (\bibinfo {year} {1966})}\BibitemShut {NoStop}%
\bibitem [{\citenamefont {Qattan}\ \emph {et~al.}(2005)\citenamefont {Qattan},
  \citenamefont {Arrington}, \citenamefont {Segel}, \citenamefont {Zheng},
  \citenamefont {Aniol} \emph {et~al.}}]{Qattan}%
  \BibitemOpen
  \bibfield  {author} {\bibinfo {author} {\bibfnamefont {I.~A.}\ \bibnamefont
  {Qattan}}, \bibinfo {author} {\bibfnamefont {J.}~\bibnamefont {Arrington}},
  \bibinfo {author} {\bibfnamefont {R.~E.}\ \bibnamefont {Segel}}, \bibinfo
  {author} {\bibfnamefont {X.}~\bibnamefont {Zheng}}, \bibinfo {author}
  {\bibfnamefont {K.}~\bibnamefont {Aniol}},  \emph {et~al.},\ }\href {\doibase
  10.1103/PhysRevLett.94.142301} {\bibfield  {journal} {\bibinfo  {journal}
  {Phys. Rev. Lett.}\ }\textbf {\bibinfo {volume} {94}},\ \bibinfo {pages}
  {142301} (\bibinfo {year} {2005})}\BibitemShut {NoStop}%
\bibitem [{\citenamefont {Sill}\ \emph {et~al.}(1993)\citenamefont {Sill},
  \citenamefont {Arnold}, \citenamefont {Bosted}, \citenamefont {Chang},
  \citenamefont {Gomez} \emph {et~al.}}]{Sill93}%
  \BibitemOpen
  \bibfield  {author} {\bibinfo {author} {\bibfnamefont {A.~F.}\ \bibnamefont
  {Sill}}, \bibinfo {author} {\bibfnamefont {R.~G.}\ \bibnamefont {Arnold}},
  \bibinfo {author} {\bibfnamefont {P.~E.}\ \bibnamefont {Bosted}}, \bibinfo
  {author} {\bibfnamefont {C.~C.}\ \bibnamefont {Chang}}, \bibinfo {author}
  {\bibfnamefont {J.}~\bibnamefont {Gomez}},  \emph {et~al.},\ }\href {\doibase
  10.1103/PhysRevD.48.29} {\bibfield  {journal} {\bibinfo  {journal} {Phys.
  Rev.}\ }\textbf {\bibinfo {volume} {D48}},\ \bibinfo {pages} {29} (\bibinfo
  {year} {1993})}\BibitemShut {NoStop}%
\bibitem [{\citenamefont {Simon}\ \emph {et~al.}(1980)\citenamefont {Simon},
  \citenamefont {Schmitt}, \citenamefont {Borkowski},\ and\ \citenamefont
  {Walther}}]{Simon80}%
  \BibitemOpen
  \bibfield  {author} {\bibinfo {author} {\bibfnamefont {G.~G.}\ \bibnamefont
  {Simon}}, \bibinfo {author} {\bibfnamefont {C.}~\bibnamefont {Schmitt}},
  \bibinfo {author} {\bibfnamefont {F.}~\bibnamefont {Borkowski}}, \ and\
  \bibinfo {author} {\bibfnamefont {V.~H.}\ \bibnamefont {Walther}},\ }\href
  {\doibase 10.1016/0375-9474(80)90104-9} {\bibfield  {journal} {\bibinfo
  {journal} {Nucl. Phys.}\ }\textbf {\bibinfo {volume} {A333}},\ \bibinfo
  {pages} {381} (\bibinfo {year} {1980})}\BibitemShut {NoStop}%
\bibitem [{\citenamefont {Walker}\ \emph {et~al.}(1994)\citenamefont {Walker},
  \citenamefont {Filippone}, \citenamefont {Jourdan}, \citenamefont {Milner},
  \citenamefont {McKeown} \emph {et~al.}}]{Walker}%
  \BibitemOpen
  \bibfield  {author} {\bibinfo {author} {\bibfnamefont {R.~C.}\ \bibnamefont
  {Walker}}, \bibinfo {author} {\bibfnamefont {B.~W.}\ \bibnamefont
  {Filippone}}, \bibinfo {author} {\bibfnamefont {J.}~\bibnamefont {Jourdan}},
  \bibinfo {author} {\bibfnamefont {R.}~\bibnamefont {Milner}}, \bibinfo
  {author} {\bibfnamefont {R.}~\bibnamefont {McKeown}},  \emph {et~al.},\
  }\href {\doibase 10.1103/PhysRevD.49.5671} {\bibfield  {journal} {\bibinfo
  {journal} {Phys. Rev.}\ }\textbf {\bibinfo {volume} {D49}},\ \bibinfo {pages}
  {5671} (\bibinfo {year} {1994})}\BibitemShut {NoStop}%
\bibitem [{\citenamefont {Andivahis}\ \emph {et~al.}(1994)\citenamefont
  {Andivahis}, \citenamefont {Bosted}, \citenamefont {Lung}, \citenamefont
  {Stuart}, \citenamefont {Alster} \emph {et~al.}}]{Andivahis}%
  \BibitemOpen
  \bibfield  {author} {\bibinfo {author} {\bibfnamefont {L.}~\bibnamefont
  {Andivahis}}, \bibinfo {author} {\bibfnamefont {P.~E.}\ \bibnamefont
  {Bosted}}, \bibinfo {author} {\bibfnamefont {A.}~\bibnamefont {Lung}},
  \bibinfo {author} {\bibfnamefont {L.~M.}\ \bibnamefont {Stuart}}, \bibinfo
  {author} {\bibfnamefont {J.}~\bibnamefont {Alster}},  \emph {et~al.},\ }\href
  {\doibase 10.1103/PhysRevD.50.5491} {\bibfield  {journal} {\bibinfo
  {journal} {Phys. Rev.}\ }\textbf {\bibinfo {volume} {D50}},\ \bibinfo {pages}
  {5491} (\bibinfo {year} {1994})}\BibitemShut {NoStop}%
\bibitem [{\citenamefont {Borkowski}\ \emph {et~al.}(1975)\citenamefont
  {Borkowski}, \citenamefont {Peuser}, \citenamefont {Simon}, \citenamefont
  {Walther},\ and\ \citenamefont {Wendling}}]{Borkowski74}%
  \BibitemOpen
  \bibfield  {author} {\bibinfo {author} {\bibfnamefont {F.}~\bibnamefont
  {Borkowski}}, \bibinfo {author} {\bibfnamefont {P.}~\bibnamefont {Peuser}},
  \bibinfo {author} {\bibfnamefont {G.~G.}\ \bibnamefont {Simon}}, \bibinfo
  {author} {\bibfnamefont {V.~H.}\ \bibnamefont {Walther}}, \ and\ \bibinfo
  {author} {\bibfnamefont {R.~D.}\ \bibnamefont {Wendling}},\ }\href {\doibase
  10.1016/0550-3213(75)90514-3} {\bibfield  {journal} {\bibinfo  {journal}
  {Nucl. Phys.}\ }\textbf {\bibinfo {volume} {B93}},\ \bibinfo {pages} {461}
  (\bibinfo {year} {1975})}\BibitemShut {NoStop}%
\bibitem [{\citenamefont {Borkowski}\ \emph {et~al.}(1974)\citenamefont
  {Borkowski}, \citenamefont {Peuser}, \citenamefont {Simon}, \citenamefont
  {Walther},\ and\ \citenamefont {Wendling}}]{Borkowski74-2}%
  \BibitemOpen
  \bibfield  {author} {\bibinfo {author} {\bibfnamefont {F.}~\bibnamefont
  {Borkowski}}, \bibinfo {author} {\bibfnamefont {P.}~\bibnamefont {Peuser}},
  \bibinfo {author} {\bibfnamefont {G.~G.}\ \bibnamefont {Simon}}, \bibinfo
  {author} {\bibfnamefont {V.~H.}\ \bibnamefont {Walther}}, \ and\ \bibinfo
  {author} {\bibfnamefont {R.~D.}\ \bibnamefont {Wendling}},\ }\href {\doibase
  10.1016/0375-9474(74)90392-3} {\bibfield  {journal} {\bibinfo  {journal}
  {Nucl. Phys.}\ }\textbf {\bibinfo {volume} {A222}},\ \bibinfo {pages} {269}
  (\bibinfo {year} {1974})}\BibitemShut {NoStop}%
\bibitem [{\citenamefont {Goitein}\ \emph {et~al.}(1970)\citenamefont
  {Goitein}, \citenamefont {Budnitz}, \citenamefont {Carroll}, \citenamefont
  {Chen}, \citenamefont {Dunning}, \citenamefont {Hanson}, \citenamefont
  {Imrie}, \citenamefont {Mistretta},\ and\ \citenamefont
  {Wilson}}]{Goitein70}%
  \BibitemOpen
  \bibfield  {author} {\bibinfo {author} {\bibfnamefont {M.}~\bibnamefont
  {Goitein}}, \bibinfo {author} {\bibfnamefont {R.~J.}\ \bibnamefont
  {Budnitz}}, \bibinfo {author} {\bibfnamefont {L.}~\bibnamefont {Carroll}},
  \bibinfo {author} {\bibfnamefont {J.~R.}\ \bibnamefont {Chen}}, \bibinfo
  {author} {\bibfnamefont {J.~R.}\ \bibnamefont {Dunning}}, \bibinfo {author}
  {\bibfnamefont {K.}~\bibnamefont {Hanson}}, \bibinfo {author} {\bibfnamefont
  {D.~C.}\ \bibnamefont {Imrie}}, \bibinfo {author} {\bibfnamefont
  {C.}~\bibnamefont {Mistretta}}, \ and\ \bibinfo {author} {\bibfnamefont
  {R.}~\bibnamefont {Wilson}},\ }\href {\doibase 10.1103/PhysRevD.1.2449}
  {\bibfield  {journal} {\bibinfo  {journal} {Phys. Rev.}\ }\textbf {\bibinfo
  {volume} {D1}},\ \bibinfo {pages} {2449} (\bibinfo {year}
  {1970})}\BibitemShut {NoStop}%
\bibitem [{\citenamefont {Litt}\ \emph {et~al.}(1970)\citenamefont {Litt},
  \citenamefont {Buschhorn}, \citenamefont {Coward}, \citenamefont
  {Destaebler}, \citenamefont {Mo} \emph {et~al.}}]{Litt69}%
  \BibitemOpen
  \bibfield  {author} {\bibinfo {author} {\bibfnamefont {J.}~\bibnamefont
  {Litt}}, \bibinfo {author} {\bibfnamefont {G.}~\bibnamefont {Buschhorn}},
  \bibinfo {author} {\bibfnamefont {D.~H.}\ \bibnamefont {Coward}}, \bibinfo
  {author} {\bibfnamefont {H.}~\bibnamefont {Destaebler}}, \bibinfo {author}
  {\bibfnamefont {L.~W.}\ \bibnamefont {Mo}},  \emph {et~al.},\ }\href
  {\doibase 10.1016/0370-2693(70)90015-8} {\bibfield  {journal} {\bibinfo
  {journal} {Phys. Lett.}\ }\textbf {\bibinfo {volume} {B31}},\ \bibinfo
  {pages} {40} (\bibinfo {year} {1970})}\BibitemShut {NoStop}%
\bibitem [{\citenamefont {Price}\ \emph {et~al.}(1971)\citenamefont {Price},
  \citenamefont {Dunning}, \citenamefont {Goitein}, \citenamefont {Hanson},
  \citenamefont {Kirk},\ and\ \citenamefont {Wilson}}]{Price71}%
  \BibitemOpen
  \bibfield  {author} {\bibinfo {author} {\bibfnamefont {L.~E.}\ \bibnamefont
  {Price}}, \bibinfo {author} {\bibfnamefont {J.~R.}\ \bibnamefont {Dunning}},
  \bibinfo {author} {\bibfnamefont {M.}~\bibnamefont {Goitein}}, \bibinfo
  {author} {\bibfnamefont {K.}~\bibnamefont {Hanson}}, \bibinfo {author}
  {\bibfnamefont {T.}~\bibnamefont {Kirk}}, \ and\ \bibinfo {author}
  {\bibfnamefont {R.}~\bibnamefont {Wilson}},\ }\href {\doibase
  10.1103/PhysRevD.4.45} {\bibfield  {journal} {\bibinfo  {journal} {Phys.
  Rev.}\ }\textbf {\bibinfo {volume} {D4}},\ \bibinfo {pages} {45} (\bibinfo
  {year} {1971})}\BibitemShut {NoStop}%
\bibitem [{\citenamefont {Bosted}\ \emph {et~al.}(1990)\citenamefont {Bosted},
  \citenamefont {Katramatou}, \citenamefont {Arnold}, \citenamefont {Benton},
  \citenamefont {Clogher} \emph {et~al.}}]{Bosted90}%
  \BibitemOpen
  \bibfield  {author} {\bibinfo {author} {\bibfnamefont {P.~E.}\ \bibnamefont
  {Bosted}}, \bibinfo {author} {\bibfnamefont {A.~T.}\ \bibnamefont
  {Katramatou}}, \bibinfo {author} {\bibfnamefont {R.~G.}\ \bibnamefont
  {Arnold}}, \bibinfo {author} {\bibfnamefont {D.}~\bibnamefont {Benton}},
  \bibinfo {author} {\bibfnamefont {L.}~\bibnamefont {Clogher}},  \emph
  {et~al.},\ }\href {\doibase 10.1103/PhysRevC.42.38} {\bibfield  {journal}
  {\bibinfo  {journal} {Phys. Rev.}\ }\textbf {\bibinfo {volume} {C42}},\
  \bibinfo {pages} {38} (\bibinfo {year} {1990})}\BibitemShut {NoStop}%
\bibitem [{\citenamefont {Rock}\ \emph {et~al.}(1992)\citenamefont {Rock},
  \citenamefont {Arnold}, \citenamefont {Bosted}, \citenamefont {Chertok},
  \citenamefont {Mecking}, \citenamefont {Schmidt}, \citenamefont {Szalata},
  \citenamefont {York},\ and\ \citenamefont {Zdarko}}]{Rock92}%
  \BibitemOpen
  \bibfield  {author} {\bibinfo {author} {\bibfnamefont {S.}~\bibnamefont
  {Rock}}, \bibinfo {author} {\bibfnamefont {R.~G.}\ \bibnamefont {Arnold}},
  \bibinfo {author} {\bibfnamefont {P.~E.}\ \bibnamefont {Bosted}}, \bibinfo
  {author} {\bibfnamefont {B.~T.}\ \bibnamefont {Chertok}}, \bibinfo {author}
  {\bibfnamefont {B.~A.}\ \bibnamefont {Mecking}}, \bibinfo {author}
  {\bibfnamefont {I.}~\bibnamefont {Schmidt}}, \bibinfo {author} {\bibfnamefont
  {Z.~M.}\ \bibnamefont {Szalata}}, \bibinfo {author} {\bibfnamefont {R.~C.}\
  \bibnamefont {York}}, \ and\ \bibinfo {author} {\bibfnamefont
  {R.}~\bibnamefont {Zdarko}},\ }\href {\doibase 10.1103/PhysRevD.46.24}
  {\bibfield  {journal} {\bibinfo  {journal} {Phys. Rev.}\ }\textbf {\bibinfo
  {volume} {D46}},\ \bibinfo {pages} {24} (\bibinfo {year} {1992})}\BibitemShut
  {NoStop}%
\bibitem [{\citenamefont {Stein}\ \emph {et~al.}(1975)\citenamefont {Stein},
  \citenamefont {Atwood}, \citenamefont {Bloom}, \citenamefont {Cottrell},
  \citenamefont {DeStaebler}, \citenamefont {Jordan}, \citenamefont {Piel},
  \citenamefont {Prescott}, \citenamefont {Siemann},\ and\ \citenamefont
  {Taylor}}]{Stein75}%
  \BibitemOpen
  \bibfield  {author} {\bibinfo {author} {\bibfnamefont {S.}~\bibnamefont
  {Stein}}, \bibinfo {author} {\bibfnamefont {W.~B.}\ \bibnamefont {Atwood}},
  \bibinfo {author} {\bibfnamefont {E.~D.}\ \bibnamefont {Bloom}}, \bibinfo
  {author} {\bibfnamefont {R.~L.~A.}\ \bibnamefont {Cottrell}}, \bibinfo
  {author} {\bibfnamefont {H.}~\bibnamefont {DeStaebler}}, \bibinfo {author}
  {\bibfnamefont {C.~L.}\ \bibnamefont {Jordan}}, \bibinfo {author}
  {\bibfnamefont {H.~G.}\ \bibnamefont {Piel}}, \bibinfo {author}
  {\bibfnamefont {C.~Y.}\ \bibnamefont {Prescott}}, \bibinfo {author}
  {\bibfnamefont {R.}~\bibnamefont {Siemann}}, \ and\ \bibinfo {author}
  {\bibfnamefont {R.~E.}\ \bibnamefont {Taylor}},\ }\href {\doibase
  10.1103/PhysRevD.12.1884} {\bibfield  {journal} {\bibinfo  {journal} {Phys.
  Rev.}\ }\textbf {\bibinfo {volume} {D12}},\ \bibinfo {pages} {1884} (\bibinfo
  {year} {1975})}\BibitemShut {NoStop}%
\bibitem [{\citenamefont {Mo}\ and\ \citenamefont {Tsai}(1969)}]{MoTsai}%
  \BibitemOpen
  \bibfield  {author} {\bibinfo {author} {\bibfnamefont {L.~W.}\ \bibnamefont
  {Mo}}\ and\ \bibinfo {author} {\bibfnamefont {Y.-S.}\ \bibnamefont {Tsai}},\
  }\href {\doibase 10.1103/RevModPhys.41.205} {\bibfield  {journal} {\bibinfo
  {journal} {Rev. Mod. Phys.}\ }\textbf {\bibinfo {volume} {41}},\ \bibinfo
  {pages} {205} (\bibinfo {year} {1969})}\BibitemShut {NoStop}%
\bibitem [{\citenamefont {Meister}\ and\ \citenamefont
  {Yennie}(1963)}]{MeisterYennie}%
  \BibitemOpen
  \bibfield  {author} {\bibinfo {author} {\bibfnamefont {N.}~\bibnamefont
  {Meister}}\ and\ \bibinfo {author} {\bibfnamefont {D.~R.}\ \bibnamefont
  {Yennie}},\ }\href {\doibase 10.1103/PhysRev.130.1210} {\bibfield  {journal}
  {\bibinfo  {journal} {Phys. Rev.}\ }\textbf {\bibinfo {volume} {130}},\
  \bibinfo {pages} {1210} (\bibinfo {year} {1963})}\BibitemShut {NoStop}%
\bibitem [{\citenamefont {Hand}\ \emph {et~al.}(1963)\citenamefont {Hand},
  \citenamefont {Miller},\ and\ \citenamefont {Wilson}}]{hand63}%
  \BibitemOpen
  \bibfield  {author} {\bibinfo {author} {\bibfnamefont {L.~N.}\ \bibnamefont
  {Hand}}, \bibinfo {author} {\bibfnamefont {D.~G.}\ \bibnamefont {Miller}}, \
  and\ \bibinfo {author} {\bibfnamefont {R.}~\bibnamefont {Wilson}},\ }\href
  {\doibase 10.1103/RevModPhys.35.335} {\bibfield  {journal} {\bibinfo
  {journal} {Rev. Mod. Phys.}\ }\textbf {\bibinfo {volume} {35}},\ \bibinfo
  {pages} {335} (\bibinfo {year} {1963})}\BibitemShut {NoStop}%
\bibitem [{\citenamefont {Povh}\ \emph {et~al.}(2004)\citenamefont {Povh},
  \citenamefont {Rith}, \citenamefont {Scholz},\ and\ \citenamefont
  {Zetsche}}]{povh04}%
  \BibitemOpen
  \bibfield  {author} {\bibinfo {author} {\bibfnamefont {B.}~\bibnamefont
  {Povh}}, \bibinfo {author} {\bibfnamefont {K.}~\bibnamefont {Rith}}, \bibinfo
  {author} {\bibfnamefont {C.}~\bibnamefont {Scholz}}, \ and\ \bibinfo {author}
  {\bibfnamefont {F.}~\bibnamefont {Zetsche}},\ }\href@noop {} {\emph {\bibinfo
  {title} {Teilchen und Kerne. Eine Einf{\"u}hrung in die physikalischen
  Konzepte}}},\ \bibinfo {edition} {sixth}\ ed.\ (\bibinfo  {publisher}
  {Springer-Verlag Berlin},\ \bibinfo {year} {2004})\BibitemShut {NoStop}%
\bibitem [{\citenamefont {Sick}(1974)}]{Sick74}%
  \BibitemOpen
  \bibfield  {author} {\bibinfo {author} {\bibfnamefont {I.}~\bibnamefont
  {Sick}},\ }\href {\doibase DOI: 10.1016/0375-9474(74)90039-6} {\bibfield
  {journal} {\bibinfo  {journal} {Nuclear Physics}\ }\textbf {\bibinfo {volume}
  {A218}},\ \bibinfo {pages} {509 } (\bibinfo {year} {1974})}\BibitemShut
  {NoStop}%
\bibitem [{\citenamefont {Sick}(2003)}]{Sick03}%
  \BibitemOpen
  \bibfield  {author} {\bibinfo {author} {\bibfnamefont {I.}~\bibnamefont
  {Sick}},\ }\href {\doibase 10.1016/j.physletb.2003.09.092} {\bibfield
  {journal} {\bibinfo  {journal} {Phys. Lett.}\ }\textbf {\bibinfo {volume}
  {B576}},\ \bibinfo {pages} {62} (\bibinfo {year} {2003})},\ \Eprint
  {http://arxiv.org/abs/nucl-ex/0310008} {nucl-ex/0310008} \BibitemShut
  {NoStop}%
\bibitem [{\citenamefont {Gari}\ and\ \citenamefont
  {Kr{\"u}mpelmann}(1992)}]{Gari92}%
  \BibitemOpen
  \bibfield  {author} {\bibinfo {author} {\bibfnamefont {M.~F.}\ \bibnamefont
  {Gari}}\ and\ \bibinfo {author} {\bibfnamefont {W.}~\bibnamefont
  {Kr{\"u}mpelmann}},\ }\href {\doibase 10.1016/0370-2693(92)90516-7}
  {\bibfield  {journal} {\bibinfo  {journal} {Phys. Lett.}\ }\textbf {\bibinfo
  {volume} {B274}},\ \bibinfo {pages} {159} (\bibinfo {year} {1992})},\
  \bibinfo {note} {erratum-ibid.{\bf B282}:483,1992}\BibitemShut {NoStop}%
\bibitem [{\citenamefont {Lomon}(2001)}]{Lomon01}%
  \BibitemOpen
  \bibfield  {author} {\bibinfo {author} {\bibfnamefont {E.~L.}\ \bibnamefont
  {Lomon}},\ }\href {\doibase 10.1103/PhysRevC.64.035204} {\bibfield  {journal}
  {\bibinfo  {journal} {Phys. Rev.}\ }\textbf {\bibinfo {volume} {C64}},\
  \bibinfo {pages} {035204} (\bibinfo {year} {2001})},\ \Eprint
  {http://arxiv.org/abs/nucl-th/0104039} {nucl-th/0104039} \BibitemShut
  {NoStop}%
\bibitem [{\citenamefont {Lomon}(2002)}]{Lomon02}%
  \BibitemOpen
  \bibfield  {author} {\bibinfo {author} {\bibfnamefont {E.~L.}\ \bibnamefont
  {Lomon}},\ }\href {\doibase 10.1103/PhysRevC.66.045501} {\bibfield  {journal}
  {\bibinfo  {journal} {Phys. Rev.}\ }\textbf {\bibinfo {volume} {C66}},\
  \bibinfo {pages} {045501} (\bibinfo {year} {2002})},\ \Eprint
  {http://arxiv.org/abs/nucl-th/0203081} {nucl-th/0203081} \BibitemShut
  {NoStop}%
\bibitem [{\citenamefont {Lomon}(2006)}]{Lomon06}%
  \BibitemOpen
  \bibfield  {author} {\bibinfo {author} {\bibfnamefont {E.~L.}\ \bibnamefont
  {Lomon}},\ }\href@noop {} {\  (\bibinfo {year} {2006})},\ \Eprint
  {http://arxiv.org/abs/nucl-th/0609020} {nucl-th/0609020} \BibitemShut
  {NoStop}%
\bibitem [{\citenamefont {Bauer}\ \emph {et~al.}(2012)\citenamefont {Bauer},
  \citenamefont {Bernauer},\ and\ \citenamefont {Scherer}}]{Bauer:2012pv}%
  \BibitemOpen
  \bibfield  {author} {\bibinfo {author} {\bibfnamefont {T.}~\bibnamefont
  {Bauer}}, \bibinfo {author} {\bibfnamefont {J.~C.}\ \bibnamefont {Bernauer}},
  \ and\ \bibinfo {author} {\bibfnamefont {S.}~\bibnamefont {Scherer}},\ }\href
  {\doibase 10.1103/PhysRevC.86.065206} {\bibfield  {journal} {\bibinfo
  {journal} {Phys.Rev.}\ }\textbf {\bibinfo {volume} {C86}},\ \bibinfo {pages}
  {065206} (\bibinfo {year} {2012})},\ \Eprint {http://arxiv.org/abs/1209.3872}
  {1209.3872 [nucl-th]} \BibitemShut {NoStop}%
\bibitem [{\citenamefont {Lorenz}\ \emph {et~al.}(2012)\citenamefont {Lorenz},
  \citenamefont {Hammer},\ and\ \citenamefont {Meissner}}]{Lorenz:2012tm}%
  \BibitemOpen
  \bibfield  {author} {\bibinfo {author} {\bibfnamefont {I.}~\bibnamefont
  {Lorenz}}, \bibinfo {author} {\bibfnamefont {H.-W.}\ \bibnamefont {Hammer}},
  \ and\ \bibinfo {author} {\bibfnamefont {U.-G.}\ \bibnamefont {Meissner}},\
  }\href {\doibase 10.1140/epja/i2012-12151-1} {\bibfield  {journal} {\bibinfo
  {journal} {Eur.Phys.J.}\ }\textbf {\bibinfo {volume} {A48}},\ \bibinfo
  {pages} {151} (\bibinfo {year} {2012})},\ \Eprint
  {http://arxiv.org/abs/1205.6628} {1205.6628 [hep-ph]} \BibitemShut {NoStop}%
\bibitem [{\citenamefont {Guttmann}\ \emph {et~al.}(2011)\citenamefont
  {Guttmann}, \citenamefont {Kivel}, \citenamefont {Meziane},\ and\
  \citenamefont {Vanderhaeghen}}]{Guttmann:2010au}%
  \BibitemOpen
  \bibfield  {author} {\bibinfo {author} {\bibfnamefont {J.}~\bibnamefont
  {Guttmann}}, \bibinfo {author} {\bibfnamefont {N.}~\bibnamefont {Kivel}},
  \bibinfo {author} {\bibfnamefont {M.}~\bibnamefont {Meziane}}, \ and\
  \bibinfo {author} {\bibfnamefont {M.}~\bibnamefont {Vanderhaeghen}},\ }\href
  {\doibase 10.1140/epja/i2011-11077-4} {\bibfield  {journal} {\bibinfo
  {journal} {The European Physical Journal A}\ }\textbf {\bibinfo {volume}
  {47}},\ \bibinfo {eid} {77} (\bibinfo {year} {2011}),\
  10.1140/epja/i2011-11077-4}\BibitemShut {NoStop}%
\bibitem [{\citenamefont {Borisyuk}\ and\ \citenamefont
  {Kobushkin}(2011)}]{Borisyuk:2010ep}%
  \BibitemOpen
  \bibfield  {author} {\bibinfo {author} {\bibfnamefont {D.}~\bibnamefont
  {Borisyuk}}\ and\ \bibinfo {author} {\bibfnamefont {A.}~\bibnamefont
  {Kobushkin}},\ }\href {\doibase 10.1103/PhysRevD.83.057501} {\bibfield
  {journal} {\bibinfo  {journal} {Phys. Rev.}\ }\textbf {\bibinfo {volume}
  {D83}},\ \bibinfo {pages} {057501} (\bibinfo {year} {2011})},\ \Eprint
  {http://arxiv.org/abs/1012.3746} {1012.3746 [hep-ph]} \BibitemShut {NoStop}%
\bibitem [{\citenamefont {Blunden}\ \emph {et~al.}(2003)\citenamefont
  {Blunden}, \citenamefont {Melnitchouk},\ and\ \citenamefont
  {Tjon}}]{Blunden:2003sp}%
  \BibitemOpen
  \bibfield  {author} {\bibinfo {author} {\bibfnamefont {P.~G.}\ \bibnamefont
  {Blunden}}, \bibinfo {author} {\bibfnamefont {W.}~\bibnamefont
  {Melnitchouk}}, \ and\ \bibinfo {author} {\bibfnamefont {J.~A.}\ \bibnamefont
  {Tjon}},\ }\href {\doibase 10.1103/PhysRevLett.91.142304} {\bibfield
  {journal} {\bibinfo  {journal} {Phys. Rev. Lett.}\ }\textbf {\bibinfo
  {volume} {91}},\ \bibinfo {pages} {142304} (\bibinfo {year} {2003})},\
  \Eprint {http://arxiv.org/abs/nucl-th/0306076} {nucl-th/0306076 [nucl-th]}
  \BibitemShut {NoStop}%
\bibitem [{\citenamefont {Chen}\ \emph {et~al.}(2007)\citenamefont {Chen},
  \citenamefont {Kao},\ and\ \citenamefont {Yang}}]{Chen07}%
  \BibitemOpen
  \bibfield  {author} {\bibinfo {author} {\bibfnamefont {Y.~C.}\ \bibnamefont
  {Chen}}, \bibinfo {author} {\bibfnamefont {C.~W.}\ \bibnamefont {Kao}}, \
  and\ \bibinfo {author} {\bibfnamefont {S.~N.}\ \bibnamefont {Yang}},\ }\href
  {\doibase http://dx.doi.org/10.1016/j.physletb.2007.07.044} {\bibfield
  {journal} {\bibinfo  {journal} {Phys. Lett. B}\ }\textbf {\bibinfo {volume}
  {652}},\ \bibinfo {pages} {269 } (\bibinfo {year} {2007})}\BibitemShut
  {NoStop}%
\bibitem [{\citenamefont {Berger}\ \emph {et~al.}(1971)\citenamefont {Berger},
  \citenamefont {Burkert}, \citenamefont {Knop}, \citenamefont {Langenbeck},\
  and\ \citenamefont {Rith}}]{Berger71}%
  \BibitemOpen
  \bibfield  {author} {\bibinfo {author} {\bibfnamefont {C.}~\bibnamefont
  {Berger}}, \bibinfo {author} {\bibfnamefont {V.}~\bibnamefont {Burkert}},
  \bibinfo {author} {\bibfnamefont {G.}~\bibnamefont {Knop}}, \bibinfo {author}
  {\bibfnamefont {B.}~\bibnamefont {Langenbeck}}, \ and\ \bibinfo {author}
  {\bibfnamefont {K.}~\bibnamefont {Rith}},\ }\href {\doibase
  10.1016/0370-2693(71)90448-5} {\bibfield  {journal} {\bibinfo  {journal}
  {Phys. Lett.}\ }\textbf {\bibinfo {volume} {B35}},\ \bibinfo {pages} {87}
  (\bibinfo {year} {1971})}\BibitemShut {NoStop}%
\bibitem [{\citenamefont {Hanson}\ \emph {et~al.}(1973)\citenamefont {Hanson},
  \citenamefont {Dunning}, \citenamefont {Goitein}, \citenamefont {Kirk},
  \citenamefont {Price},\ and\ \citenamefont {Wilson}}]{Hanson73}%
  \BibitemOpen
  \bibfield  {author} {\bibinfo {author} {\bibfnamefont {K.~M.}\ \bibnamefont
  {Hanson}}, \bibinfo {author} {\bibfnamefont {J.~R.}\ \bibnamefont {Dunning}},
  \bibinfo {author} {\bibfnamefont {M.}~\bibnamefont {Goitein}}, \bibinfo
  {author} {\bibfnamefont {T.}~\bibnamefont {Kirk}}, \bibinfo {author}
  {\bibfnamefont {L.~E.}\ \bibnamefont {Price}}, \ and\ \bibinfo {author}
  {\bibfnamefont {R.}~\bibnamefont {Wilson}},\ }\href {\doibase
  10.1103/PhysRevD.8.753} {\bibfield  {journal} {\bibinfo  {journal} {Phys.
  Rev.}\ }\textbf {\bibinfo {volume} {D8}},\ \bibinfo {pages} {753} (\bibinfo
  {year} {1973})}\BibitemShut {NoStop}%
\bibitem [{\citenamefont {Bartel}\ \emph {et~al.}(1973)\citenamefont {Bartel},
  \citenamefont {B{\"u}sser}, \citenamefont {Dix}, \citenamefont {Felst},
  \citenamefont {Harms}, \citenamefont {Krehbiel}, \citenamefont {Kuhlmann},
  \citenamefont {McElroy}, \citenamefont {Meyer},\ and\ \citenamefont
  {Weber}}]{Bartel73}%
  \BibitemOpen
  \bibfield  {author} {\bibinfo {author} {\bibfnamefont {W.}~\bibnamefont
  {Bartel}}, \bibinfo {author} {\bibfnamefont {F.-W.}\ \bibnamefont
  {B{\"u}sser}}, \bibinfo {author} {\bibfnamefont {W.-R.}\ \bibnamefont {Dix}},
  \bibinfo {author} {\bibfnamefont {R.}~\bibnamefont {Felst}}, \bibinfo
  {author} {\bibfnamefont {D.}~\bibnamefont {Harms}}, \bibinfo {author}
  {\bibfnamefont {H.}~\bibnamefont {Krehbiel}}, \bibinfo {author}
  {\bibfnamefont {P.~E.}\ \bibnamefont {Kuhlmann}}, \bibinfo {author}
  {\bibfnamefont {J.}~\bibnamefont {McElroy}}, \bibinfo {author} {\bibfnamefont
  {J.}~\bibnamefont {Meyer}}, \ and\ \bibinfo {author} {\bibfnamefont
  {G.}~\bibnamefont {Weber}},\ }\href {\doibase 10.1016/0550-3213(73)90594-4}
  {\bibfield  {journal} {\bibinfo  {journal} {Nucl. Phys.}\ }\textbf {\bibinfo
  {volume} {B58}},\ \bibinfo {pages} {429} (\bibinfo {year}
  {1973})}\BibitemShut {NoStop}%
\bibitem [{\citenamefont {Pospischil}\ \emph {et~al.}(2002)\citenamefont
  {Pospischil}, \citenamefont {Bartsch}, \citenamefont {Baumann}, \citenamefont
  {B{\"o}hm}, \citenamefont {Bohinc} \emph {et~al.}}]{Pospischil:2000pu}%
  \BibitemOpen
  \bibfield  {author} {\bibinfo {author} {\bibfnamefont {T.}~\bibnamefont
  {Pospischil}}, \bibinfo {author} {\bibfnamefont {P.}~\bibnamefont {Bartsch}},
  \bibinfo {author} {\bibfnamefont {D.}~\bibnamefont {Baumann}}, \bibinfo
  {author} {\bibfnamefont {R.}~\bibnamefont {B{\"o}hm}}, \bibinfo {author}
  {\bibfnamefont {K.}~\bibnamefont {Bohinc}},  \emph {et~al.} (\bibinfo
  {collaboration} {A1 Collaboration}),\ }\href {\doibase
  10.1016/S0168-9002(01)01955-6} {\bibfield  {journal} {\bibinfo  {journal}
  {Nucl. Instrum. Methods}\ }\textbf {\bibinfo {volume} {A483}},\ \bibinfo
  {pages} {713} (\bibinfo {year} {2002})},\ \Eprint
  {http://arxiv.org/abs/nucl-ex/0010007} {nucl-ex/0010007} \BibitemShut
  {NoStop}%
\bibitem [{\citenamefont {Dieterich}\ \emph {et~al.}(2001)\citenamefont
  {Dieterich}, \citenamefont {Bartsch}, \citenamefont {Baumann}, \citenamefont
  {Bermuth}, \citenamefont {Bohinc} \emph {et~al.}}]{Dieterich01}%
  \BibitemOpen
  \bibfield  {author} {\bibinfo {author} {\bibfnamefont {S.}~\bibnamefont
  {Dieterich}}, \bibinfo {author} {\bibfnamefont {P.}~\bibnamefont {Bartsch}},
  \bibinfo {author} {\bibfnamefont {D.}~\bibnamefont {Baumann}}, \bibinfo
  {author} {\bibfnamefont {J.}~\bibnamefont {Bermuth}}, \bibinfo {author}
  {\bibfnamefont {K.}~\bibnamefont {Bohinc}},  \emph {et~al.} (\bibinfo
  {collaboration} {A1 Collaboration}),\ }\href {\doibase DOI:
  10.1016/S0370-2693(01)00052-1} {\bibfield  {journal} {\bibinfo  {journal}
  {Phys. Lett.}\ }\textbf {\bibinfo {volume} {B500}},\ \bibinfo {pages} {47 }
  (\bibinfo {year} {2001})}\BibitemShut {NoStop}%
\bibitem [{\citenamefont {Murphy}\ \emph {et~al.}(1974)\citenamefont {Murphy},
  \citenamefont {Shin},\ and\ \citenamefont {Skopik}}]{Murphy74}%
  \BibitemOpen
  \bibfield  {author} {\bibinfo {author} {\bibfnamefont {J.~J.}\ \bibnamefont
  {Murphy}}, \bibinfo {author} {\bibfnamefont {Y.~M.}\ \bibnamefont {Shin}}, \
  and\ \bibinfo {author} {\bibfnamefont {D.~M.}\ \bibnamefont {Skopik}},\
  }\href {\doibase 10.1103/PhysRevC.9.2125} {\bibfield  {journal} {\bibinfo
  {journal} {Phys. Rev.}\ }\textbf {\bibinfo {volume} {C9}},\ \bibinfo {pages}
  {2125} (\bibinfo {year} {1974})}\BibitemShut {NoStop}%
\bibitem [{\citenamefont {Ledwig}\ \emph {et~al.}(2012)\citenamefont {Ledwig},
  \citenamefont {Martin-Camalich}, \citenamefont {Pascalutsa},\ and\
  \citenamefont {Vanderhaeghen}}]{Ledwig:2011cx}%
  \BibitemOpen
  \bibfield  {author} {\bibinfo {author} {\bibfnamefont {T.}~\bibnamefont
  {Ledwig}}, \bibinfo {author} {\bibfnamefont {J.}~\bibnamefont
  {Martin-Camalich}}, \bibinfo {author} {\bibfnamefont {V.}~\bibnamefont
  {Pascalutsa}}, \ and\ \bibinfo {author} {\bibfnamefont {M.}~\bibnamefont
  {Vanderhaeghen}},\ }\href {\doibase 10.1103/PhysRevD.85.034013} {\bibfield
  {journal} {\bibinfo  {journal} {Phys. Rev.}\ }\textbf {\bibinfo {volume}
  {D85}},\ \bibinfo {pages} {034013} (\bibinfo {year} {2012})},\ \Eprint
  {http://arxiv.org/abs/1108.2523} {1108.2523 [hep-ph]} \BibitemShut {NoStop}%
\bibitem [{\citenamefont {Lyman}\ \emph {et~al.}(1951)\citenamefont {Lyman},
  \citenamefont {Hanson},\ and\ \citenamefont {Scott}}]{Hanson:1951zz}%
  \BibitemOpen
  \bibfield  {author} {\bibinfo {author} {\bibfnamefont {E.~M.}\ \bibnamefont
  {Lyman}}, \bibinfo {author} {\bibfnamefont {A.~O.}\ \bibnamefont {Hanson}}, \
  and\ \bibinfo {author} {\bibfnamefont {M.~B.}\ \bibnamefont {Scott}},\ }\href
  {\doibase 10.1103/PhysRev.84.626} {\bibfield  {journal} {\bibinfo  {journal}
  {Phys. Rev.}\ }\textbf {\bibinfo {volume} {84}},\ \bibinfo {pages} {626}
  (\bibinfo {year} {1951})}\BibitemShut {NoStop}%
\bibitem [{\citenamefont {James}(2006)}]{james}%
  \BibitemOpen
  \bibfield  {author} {\bibinfo {author} {\bibfnamefont {F.}~\bibnamefont
  {James}},\ }\href@noop {} {\emph {\bibinfo {title} {{Statistical Methods in
  Experimental Physics}}}},\ \bibinfo {edition} {2nd}\ ed.\ (\bibinfo
  {publisher} {World Scientific, Singapore},\ \bibinfo {year}
  {2006})\BibitemShut {NoStop}%
\bibitem [{\citenamefont {Borisyuk}\ and\ \citenamefont
  {Kobushkin}(2007)}]{Borisyuk:2006uq}%
  \BibitemOpen
  \bibfield  {author} {\bibinfo {author} {\bibfnamefont {D.}~\bibnamefont
  {Borisyuk}}\ and\ \bibinfo {author} {\bibfnamefont {A.}~\bibnamefont
  {Kobushkin}},\ }\href {\doibase 10.1103/PhysRevC.75.038202} {\bibfield
  {journal} {\bibinfo  {journal} {Phys. Rev.}\ }\textbf {\bibinfo {volume}
  {C75}},\ \bibinfo {pages} {038202} (\bibinfo {year} {2007})},\ \Eprint
  {http://arxiv.org/abs/nucl-th/0612104} {nucl-th/0612104 [nucl-th]}
  \BibitemShut {NoStop}%
\bibitem [{\citenamefont {Blunden}\ \emph {et~al.}(2005)\citenamefont
  {Blunden}, \citenamefont {Melnitchouk},\ and\ \citenamefont
  {Tjon}}]{Blunden05}%
  \BibitemOpen
  \bibfield  {author} {\bibinfo {author} {\bibfnamefont {P.~G.}\ \bibnamefont
  {Blunden}}, \bibinfo {author} {\bibfnamefont {W.}~\bibnamefont
  {Melnitchouk}}, \ and\ \bibinfo {author} {\bibfnamefont {J.~A.}\ \bibnamefont
  {Tjon}},\ }\href {\doibase 10.1103/PhysRevC.72.034612} {\bibfield  {journal}
  {\bibinfo  {journal} {Phys. Rev. C}\ }\textbf {\bibinfo {volume} {72}},\
  \bibinfo {pages} {034612} (\bibinfo {year} {2005})},\ \Eprint
  {http://arxiv.org/abs/nucl-th/0506039} {nucl-th/0506039 [nucl-th]}
  \BibitemShut {NoStop}%
\bibitem [{\citenamefont {Arrington}\ \emph {et~al.}(2011)\citenamefont
  {Arrington}, \citenamefont {Blunden},\ and\ \citenamefont
  {Melnitchouk}}]{Arrington:2011dn}%
  \BibitemOpen
  \bibfield  {author} {\bibinfo {author} {\bibfnamefont {J.}~\bibnamefont
  {Arrington}}, \bibinfo {author} {\bibfnamefont {P.~G.}\ \bibnamefont
  {Blunden}}, \ and\ \bibinfo {author} {\bibfnamefont {W.}~\bibnamefont
  {Melnitchouk}},\ }\href {\doibase 10.1016/j.ppnp.2011.07.003} {\bibfield
  {journal} {\bibinfo  {journal} {Prog. Part. Nucl. Phys.}\ }\textbf {\bibinfo
  {volume} {66}},\ \bibinfo {pages} {782} (\bibinfo {year} {2011})},\ \Eprint
  {http://arxiv.org/abs/1105.0951} {1105.0951 [nucl-th]} \BibitemShut {NoStop}%
\bibitem [{\citenamefont {Blunden}()}]{blundenpc}%
  \BibitemOpen
  \bibfield  {author} {\bibinfo {author} {\bibfnamefont {P.~G.}\ \bibnamefont
  {Blunden}},\ }\href@noop {} {}\bibinfo {note} {(private
  communication)}\BibitemShut {NoStop}%
\bibitem [{\citenamefont {Mohr}\ \emph {et~al.}(2012)\citenamefont {Mohr},
  \citenamefont {Taylor},\ and\ \citenamefont {Newell}}]{Mohr12}%
  \BibitemOpen
  \bibfield  {author} {\bibinfo {author} {\bibfnamefont {P.~J.}\ \bibnamefont
  {Mohr}}, \bibinfo {author} {\bibfnamefont {B.~N.}\ \bibnamefont {Taylor}}, \
  and\ \bibinfo {author} {\bibfnamefont {D.~B.}\ \bibnamefont {Newell}},\
  }\href {\doibase 10.1103/RevModPhys.84.1527} {\bibfield  {journal} {\bibinfo
  {journal} {Rev. Mod. Phys.}\ }\textbf {\bibinfo {volume} {84}},\ \bibinfo
  {pages} {1527} (\bibinfo {year} {2012})}\BibitemShut {NoStop}%
\bibitem [{\citenamefont {Arrington}(2013)}]{Arrington:2012dq}%
  \BibitemOpen
  \bibfield  {author} {\bibinfo {author} {\bibfnamefont {J.}~\bibnamefont
  {Arrington}},\ }\href {\doibase 10.1088/0954-3899/40/11/115003} {\bibfield
  {journal} {\bibinfo  {journal} {J.Phys.}\ }\textbf {\bibinfo {volume}
  {G40}},\ \bibinfo {pages} {115003} (\bibinfo {year} {2013})},\ \Eprint
  {http://arxiv.org/abs/1210.2677} {1210.2677 [nucl-ex]} \BibitemShut {NoStop}%
\bibitem [{\citenamefont {Arrington}\ and\ \citenamefont
  {Sick}(2004)}]{arringtonsick04}%
  \BibitemOpen
  \bibfield  {author} {\bibinfo {author} {\bibfnamefont {J.}~\bibnamefont
  {Arrington}}\ and\ \bibinfo {author} {\bibfnamefont {I.}~\bibnamefont
  {Sick}},\ }\href {\doibase 10.1103/PhysRevC.70.028203} {\bibfield  {journal}
  {\bibinfo  {journal} {Phys. Rev. C}\ }\textbf {\bibinfo {volume} {70}},\
  \bibinfo {pages} {028203} (\bibinfo {year} {2004})},\ \Eprint
  {http://arxiv.org/abs/nucl-ex/0406014} {nucl-ex/0406014 [nucl-ex]}
  \BibitemShut {NoStop}%
\bibitem [{\citenamefont {Gramolin}\ \emph {et~al.}(2012)\citenamefont
  {Gramolin}, \citenamefont {Arrington}, \citenamefont {Barkov}, \citenamefont
  {Dmitriev}, \citenamefont {Gauzshtein} \emph {et~al.}}]{vepp3}%
  \BibitemOpen
  \bibfield  {author} {\bibinfo {author} {\bibfnamefont {A.~V.}\ \bibnamefont
  {Gramolin}}, \bibinfo {author} {\bibfnamefont {J.}~\bibnamefont {Arrington}},
  \bibinfo {author} {\bibfnamefont {L.~M.}\ \bibnamefont {Barkov}}, \bibinfo
  {author} {\bibfnamefont {V.~F.}\ \bibnamefont {Dmitriev}}, \bibinfo {author}
  {\bibfnamefont {V.~V.}\ \bibnamefont {Gauzshtein}},  \emph {et~al.},\ }\href
  {\doibase 10.1016/j.nuclphysbps.2012.02.045} {\bibfield  {journal} {\bibinfo
  {journal} {Nucl. Phys. Proc. Suppl.}\ }\textbf {\bibinfo {volume}
  {225-227}},\ \bibinfo {pages} {216} (\bibinfo {year} {2012})},\ \Eprint
  {http://arxiv.org/abs/1112.5369} {1112.5369 [nucl-ex]} \BibitemShut {NoStop}%
\bibitem [{\citenamefont {Arrington}\ \emph {et~al.}()\citenamefont
  {Arrington}, \citenamefont {Zheng}, \citenamefont {Klein}, \citenamefont
  {Sober}, \citenamefont {Joo} \emph {et~al.}}]{classtpe}%
  \BibitemOpen
  \bibfield  {author} {\bibinfo {author} {\bibfnamefont {J.}~\bibnamefont
  {Arrington}}, \bibinfo {author} {\bibfnamefont {X.}~\bibnamefont {Zheng}},
  \bibinfo {author} {\bibfnamefont {F.}~\bibnamefont {Klein}}, \bibinfo
  {author} {\bibfnamefont {D.}~\bibnamefont {Sober}}, \bibinfo {author}
  {\bibfnamefont {K.}~\bibnamefont {Joo}},  \emph {et~al.},\ }\href
  {http://www.jlab.org/exp_prog/experiments/E-04-116.html} {\emph {\bibinfo
  {title} {E04-116: Beyond the Born Approximation: A Precise Comparison of
  Positron-Proton and Electron-Proton Elastic Scattering in CLAS}}},\ \bibinfo
  {type} {Tech. Rep.}\ (\bibinfo  {institution} {Jefferson
  Laboratory})\BibitemShut {NoStop}%
\bibitem [{\citenamefont {Milner}\ \emph {et~al.}(2014)\citenamefont {Milner},
  \citenamefont {Hasell}, \citenamefont {Kohl}, \citenamefont {Schneekloth}
  \emph {et~al.}}]{olympus}%
  \BibitemOpen
  \bibfield  {author} {\bibinfo {author} {\bibfnamefont {R.}~\bibnamefont
  {Milner}}, \bibinfo {author} {\bibfnamefont {D.}~\bibnamefont {Hasell}},
  \bibinfo {author} {\bibfnamefont {M.}~\bibnamefont {Kohl}}, \bibinfo {author}
  {\bibfnamefont {U.}~\bibnamefont {Schneekloth}},  \emph {et~al.},\ }\href
  {\doibase 10.1016/j.nima.2013.12.035} {\bibfield  {journal} {\bibinfo
  {journal} {Nucl. Instrum. Methods}\ }\textbf {\bibinfo {volume} {741}},\
  \bibinfo {pages} {1 } (\bibinfo {year} {2014})},\ \Eprint
  {http://arxiv.org/abs/1312.1730} {1312.1730 [physics.ins-det]} \BibitemShut
  {NoStop}%
\bibitem [{\citenamefont {Gilman}\ \emph {et~al.}(2013)\citenamefont {Gilman},
  \citenamefont {Downie}, \citenamefont {Ron}, \citenamefont {Afanasev},
  \citenamefont {Arrington} \emph {et~al.}}]{Gilman:2013fk}%
  \BibitemOpen
  \bibfield  {author} {\bibinfo {author} {\bibfnamefont {R.}~\bibnamefont
  {Gilman}}, \bibinfo {author} {\bibfnamefont {E.~J.}\ \bibnamefont {Downie}},
  \bibinfo {author} {\bibfnamefont {G.}~\bibnamefont {Ron}}, \bibinfo {author}
  {\bibfnamefont {A.}~\bibnamefont {Afanasev}}, \bibinfo {author}
  {\bibfnamefont {J.}~\bibnamefont {Arrington}},  \emph {et~al.} (\bibinfo
  {collaboration} {MUSE Collaboration}),\ }\href@noop {} {\  (\bibinfo {year}
  {2013})},\ \Eprint {http://arxiv.org/abs/1303.2160} {1303.2160 [nucl-ex]}
  \BibitemShut {NoStop}%
\bibitem [{\citenamefont {Carlson}\ \emph {et~al.}(2008)\citenamefont
  {Carlson}, \citenamefont {Nazaryan},\ and\ \citenamefont
  {Griffioen}}]{Carlson:2008ke}%
  \BibitemOpen
  \bibfield  {author} {\bibinfo {author} {\bibfnamefont {C.~E.}\ \bibnamefont
  {Carlson}}, \bibinfo {author} {\bibfnamefont {V.}~\bibnamefont {Nazaryan}}, \
  and\ \bibinfo {author} {\bibfnamefont {K.}~\bibnamefont {Griffioen}},\ }\href
  {\doibase 10.1103/PhysRevA.78.022517} {\bibfield  {journal} {\bibinfo
  {journal} {Phys. Rev.}\ }\textbf {\bibinfo {volume} {A78}},\ \bibinfo {pages}
  {022517} (\bibinfo {year} {2008})},\ \Eprint {http://arxiv.org/abs/0805.2603}
  {0805.2603 [physics.atom-ph]} \BibitemShut {NoStop}%
\bibitem [{\citenamefont {Arrington}\ and\ \citenamefont
  {Sick}(2007)}]{Arrington:2006hm}%
  \BibitemOpen
  \bibfield  {author} {\bibinfo {author} {\bibfnamefont {J.}~\bibnamefont
  {Arrington}}\ and\ \bibinfo {author} {\bibfnamefont {I.}~\bibnamefont
  {Sick}},\ }\href {\doibase 10.1103/PhysRevC.76.035201} {\bibfield  {journal}
  {\bibinfo  {journal} {Phys. Rev.}\ }\textbf {\bibinfo {volume} {C76}},\
  \bibinfo {pages} {035201} (\bibinfo {year} {2007})},\ \Eprint
  {http://arxiv.org/abs/nucl-th/0612079} {nucl-th/0612079 [nucl-th]}
  \BibitemShut {NoStop}%
\bibitem [{\citenamefont {Distler}\ \emph {et~al.}(2011)\citenamefont
  {Distler}, \citenamefont {Bernauer},\ and\ \citenamefont
  {Walcher}}]{Distler:2010zq}%
  \BibitemOpen
  \bibfield  {author} {\bibinfo {author} {\bibfnamefont {M.~O.}\ \bibnamefont
  {Distler}}, \bibinfo {author} {\bibfnamefont {J.~C.}\ \bibnamefont
  {Bernauer}}, \ and\ \bibinfo {author} {\bibfnamefont
  {{\singleletter{Th}}.}~\bibnamefont {Walcher}},\ }\href {\doibase
  10.1016/j.physletb.2010.12.067} {\bibfield  {journal} {\bibinfo  {journal}
  {Phys. Lett.}\ }\textbf {\bibinfo {volume} {B696}},\ \bibinfo {pages} {343}
  (\bibinfo {year} {2011})},\ \Eprint {http://arxiv.org/abs/1011.1861}
  {1011.1861 [nucl-th]} \BibitemShut {NoStop}%
\bibitem [{\citenamefont {Alexandrou}\ \emph {et~al.}(2013)\citenamefont
  {Alexandrou}, \citenamefont {Constantinou}, \citenamefont {Dinter},
  \citenamefont {Drach}, \citenamefont {Jansen} \emph
  {et~al.}}]{Alexandrou:2013joa}%
  \BibitemOpen
  \bibfield  {author} {\bibinfo {author} {\bibfnamefont {C.}~\bibnamefont
  {Alexandrou}}, \bibinfo {author} {\bibfnamefont {M.}~\bibnamefont
  {Constantinou}}, \bibinfo {author} {\bibfnamefont {S.}~\bibnamefont
  {Dinter}}, \bibinfo {author} {\bibfnamefont {V.}~\bibnamefont {Drach}},
  \bibinfo {author} {\bibfnamefont {K.}~\bibnamefont {Jansen}},  \emph
  {et~al.},\ }\href {\doibase 10.1103/PhysRevD.88.014509} {\bibfield  {journal}
  {\bibinfo  {journal} {Phys.Rev.}\ }\textbf {\bibinfo {volume} {D88}},\
  \bibinfo {pages} {014509} (\bibinfo {year} {2013})},\ \Eprint
  {http://arxiv.org/abs/1303.5979} {1303.5979 [hep-lat]} \BibitemShut {NoStop}%
\bibitem [{\citenamefont {Syritsyn}\ \emph {et~al.}(2010)\citenamefont
  {Syritsyn}, \citenamefont {Bratt}, \citenamefont {Lin}, \citenamefont
  {Meyer}, \citenamefont {Negele} \emph {et~al.}}]{Syritsyn:2009mx}%
  \BibitemOpen
  \bibfield  {author} {\bibinfo {author} {\bibfnamefont {S.}~\bibnamefont
  {Syritsyn}}, \bibinfo {author} {\bibfnamefont {J.}~\bibnamefont {Bratt}},
  \bibinfo {author} {\bibfnamefont {M.}~\bibnamefont {Lin}}, \bibinfo {author}
  {\bibfnamefont {H.}~\bibnamefont {Meyer}}, \bibinfo {author} {\bibfnamefont
  {J.}~\bibnamefont {Negele}},  \emph {et~al.},\ }\href {\doibase
  10.1103/PhysRevD.81.034507} {\bibfield  {journal} {\bibinfo  {journal}
  {Phys.Rev.}\ }\textbf {\bibinfo {volume} {D81}},\ \bibinfo {pages} {034507}
  (\bibinfo {year} {2010})},\ \Eprint {http://arxiv.org/abs/0907.4194}
  {0907.4194 [hep-lat]} \BibitemShut {NoStop}%
\bibitem [{\citenamefont {Green}\ \emph {et~al.}(2013)\citenamefont {Green},
  \citenamefont {Engelhardt}, \citenamefont {Krieg}, \citenamefont {Meinel},
  \citenamefont {Negele} \emph {et~al.}}]{Green:2013hja}%
  \BibitemOpen
  \bibfield  {author} {\bibinfo {author} {\bibfnamefont {J.}~\bibnamefont
  {Green}}, \bibinfo {author} {\bibfnamefont {M.}~\bibnamefont {Engelhardt}},
  \bibinfo {author} {\bibfnamefont {S.}~\bibnamefont {Krieg}}, \bibinfo
  {author} {\bibfnamefont {S.}~\bibnamefont {Meinel}}, \bibinfo {author}
  {\bibfnamefont {J.}~\bibnamefont {Negele}},  \emph {et~al.},\ }\href@noop {}
  {\  (\bibinfo {year} {2013})},\ \Eprint {http://arxiv.org/abs/1310.7043}
  {1310.7043 [hep-lat]} \BibitemShut {NoStop}%
\end{thebibliography}%

\end{document}